\documentclass[twocolumn]{aastex6}

\usepackage{amsmath}
\usepackage{graphicx}
\usepackage{array,longtable}
\usepackage{multirow}
\usepackage{subfigure}
\usepackage{epstopdf}

\shorttitle{SPH simulations of the IGC}
\shortauthors{Becerra et al.}

\def\roma{1}
\def\icra{2}
\def\losalamos{3}
\def\rio{4}

\begin{document}

\title{SPH simulations of the induced gravitational collapse scenario of long gamma-ray bursts associated with supernovae}

\author{L.~Becerra\altaffilmark{\roma,\icra},
                C.~L.~Ellinger\altaffilmark{\losalamos}, 
				C.~L.~Fryer\altaffilmark{\losalamos}, 
				J.~A.~Rueda\altaffilmark{\roma,\icra,\rio}, 
				R.~Ruffini\altaffilmark{\roma,\icra,\rio}
                }

\altaffiltext{\roma}{Dipartimento di Fisica and ICRA, 
                     Sapienza Universit\`a di Roma, 
                     P.le Aldo Moro 5, 
                     I--00185 Rome, 
                     Italy}
                     
\altaffiltext{\icra}{ICRANet, 
                     P.zza della Repubblica 10, 
                     I--65122 Pescara, 
                     Italy} 
																	
\altaffiltext{\losalamos}{CCS-2, Los Alamos National Laboratory, Los Alamos, NM 87545}

\altaffiltext{\rio}{ICRANet-Rio, 
                     Centro Brasileiro de Pesquisas F\'isicas, 
                     Rua Dr. Xavier Sigaud 150, 
                     22290--180 Rio de Janeiro, 
                     Brazil}

\begin{abstract}
We present the first three-dimensional (3D) smoothed-particle-hydrodynamics (SPH) simulations of the induced gravitational collapse (IGC) scenario of long-duration gamma-ray bursts (GRBs) associated with supernovae (SNe). We simulate the SN explosion of a carbon-oxygen core (CO$_{\rm core}$) forming a binary system with a neutron star (NS) companion. We follow the evolution of the SN ejecta, including their morphological structure, subjected to the gravitational field of both the new NS ($\nu$NS) formed at the center of the SN, and the one of the NS companion. We compute the accretion rate of the SN ejecta onto the NS companion as well as onto the $\nu$NS from SN matter fallback. We determine the fate of the binary system for a wide parameter space including different CO$_{\rm core}$ {and NS companion} masses, orbital periods and SN explosion geometry and energies. We {identify}, for selected NS nuclear equations-of-state, {the binary parameters leading the NS companion, by hypercritical accretion,} either to the mass-shedding limit, or to the secular axisymmetric instability for gravitational collapse to a black hole (BH), or to a more massive, fast rotating, stable NS. We also assess whether the binary remains or not gravitationally bound after the SN explosion, hence exploring the space of binary and SN explosion parameters leading to $\nu$NS-NS and $\nu$NS-BH binaries. The consequences of our results for the modeling of long GRBs{, i.e. X-ray flashes (XRFs) and binary-driven hypernovae (BdHNe),} are discussed.
\end{abstract}
\maketitle

\section{Introduction}

The induced gravitational collapse (IGC) concept was initially introduced to explain the temporal {and spatial} coincidence of long gamma-ray bursts (GRBs) with energies $E_{\rm iso}>10^{52}$~erg and type Ic supernovae (SNe). In a binary system composed of a carbon-oxygen (CO$_{\rm core}$) and a neutron star (NS), the explosion of the CO$_{\rm core}$ triggers the hypercritical accretion onto the NS. For the most compact binary systems, the accretion rate onto the NS is such that it can reach the critical mass against gravitational collapse and forming a black hole (BH) with consequence emission of a GRB. These systems have been called binary-driven hypernovae \citep[BdHNe; see][]{2014A&A...565L..10R}. Later, the IGC paradigm was extended to X-ray flashes (XRFs) with energies $E_{\rm iso}<10^{52}$~erg that occur when the accretion rate onto the NS is lower and so it does not induce its gravitational collapse to a BH but instead leads to a massive, fast rotating, stable NS. For further theoretical and observational details on these two subclasses of long GRBs, we refer the reader to \citet{2016ApJ...832..136R}, and references therein.

The physical picture of the IGC was first proposed in \citet{2001ApJ...555L.117R}, formally formulated in \citet{2012ApJ...758L...7R}, and then applied for the first time for the explanation of GRB 090618 in \citet{2012A&A...548L...5I}. \citet{2012ApJ...758L...7R} presented analytical estimates of the accretion rate and the possible fate of the accreting NS binary companion. This first simple model assumed: 1) a pre-SN uniform density profile following an homologous expansion and 2) approximately constant masses of the NS ($\approx 1.4~M_\odot$) and the pre-SN core ($\approx 4$--$8~M_\odot$).

The first 1-dimensional (1D) simulations of the IGC process were presented in \citet{2014ApJ...793L..36F}. These simulations included: 1) detailed SN explosions of the CO$_{\rm core}$ obtained from a 1D core-collapse SN code \citep{1999ApJ...516..892F}; 2) hydrodynamic details of the hypercritical accretion process; 3) the evolution of the SN ejecta falling into the Bondi-Hoyle accretion region all the way up to its incorporation into the NS surface. Following the Bondi-Hoyle formalism, they estimated accretion rates exceeding $10^{-3}~M_\odot$~s$^{-1}$, making highly possible that the NS reaches the critical mass and the BH formation.

{Already from these works, it was clear that the main binary properties that decide the occurrence of the IGC process are: 1) the orbital period, $P$, 2) the SN ejecta velocity, $v_{\rm ej}$, and 3) the initial NS companion mass. The shorter the period and the slower the ejecta the higher the Bondi-Hoyle accretion rate and the more massive the NS companion the less mass is needed to induce its gravitational collapse to a BH. Both $P$ and $v_{\rm ej}$ have a direct effect in the Bondi-Hoyle accretion rate via the gravitational capture radius but they have also an indirect role via the density of the accreted matter. Since the ejecta decompress during their expansion until the NS gravitational capture radius position, there is the obvious effect that the shorter the $P$ and the slower the $v_{\rm ej}$, the higher the density at the accretion site hence the higher the accretion rate. Clearly, the effect of $v_{\rm ej}$ can be seen as the effect of the SN kinetic/explosion energy since the stronger the explosion the higher the kinetic energy and the higher the expansion velocity and viceversa.}

{There were still additional effects needed to be included in the model to have a more general picture.} \citet{2015ApJ...812..100B} improved the analytical model relaxing assumptions made in \cite{2012ApJ...758L...7R}. The ejecta density profile was adopted as a power-law in radius and its evolution with time homologous. The amount of angular momentum transported by the ejecta entering the Bondi-Hoyle region and how much of it can be transferred to the NS was also estimated. It was there shown that the ejecta have enough angular momentum to circularize around the NS forming a disk-like structure and accretes on short timescales. {Bearing in mind the above mentioned effect of $P$ on the accretion rate,} it was there determined the critical binary period below which, i.e.~$P\lesssim P_{\rm max}$, the NS accretes enough mass and angular momentum such that its gravitational collapse to a BH is induced. {Conversely, for $P\gtrsim P_{\rm max}$, the NS does not gain enough mass and angular momentum to form a BH but just becomes a more massive NS.}

Later, we presented in \citet{2016ApJ...833..107B} a first attempt of a smoothed-particle-hydrodynamics (SPH)-like simulation of the SN ejecta expansion under the gravitational field of the NS companion. Specifically, we described the SN matter formed by point-like particles and modeled the initial power-law density profile of the CO$_{\rm core}$ by populating the inner layer with more particles and defined the initial conditions of the SN ejecta assuming an homologous velocity distribution in free expansion; i.e.~$v\propto r$. The particles trajectory were computed by solving the Newtonian equations of motion including the effects of the gravitational field of the NS companion. {In these simulations} we assumed a circular motion of the NS around the SN center, implemented the changes in the NS gravitational mass owing to the accretion process via the Bondi-Hoyle formalism following \citet{2015ApJ...812..100B} and, accordingly, removed from the system the particles falling within the Bondi-Hoyle surface. The accretion process was shown to proceed hypercritically thanks to the copious neutrino emission near the NS surface, which produces neutrino luminosities of up to $10^{52}$~erg~s$^{-1}$ and mean neutrino energies of 20~MeV. A detailed analysis of the fundamental neutrino emission properties in XRFs and BdHNe was presented in \citet{2018ApJ...852..120B}.

{\citet{2016ApJ...833..107B} made a first wide exploration of the binary parameters for the occurrence of the IGC process, hence of the systems leading to XRFs ($P\gtrsim P_{\rm max}$) and BdHNe ($P\lesssim P_{\rm max}$), as well as of the dependence of $P_{\rm max}$ on the CO$_{\rm core}$ and NS companion mass. In addition, these simulations produced a first visualization of the SN ejecta morphology. Indeed,} we showed in \citet{2016ApJ...833..107B} how the structure of the SN ejecta, becoming asymmetric by the presence of the accreting NS companion, becomes crucial for the inference of observable signatures in the GRB afterglow. {The specific example of XRF 060218 was there examined to show the asymmetry effect in the X-ray emission and to show the feedback of the accretion energy injected into the SN ejecta on its optical emission.}

{It became clear that the SN ejecta morphology and the feedback of a GRB emission onto the SN could be relevant also in BdHNe.} In \citet{2018ApJ...852...53R} we showed that the GRB $e^+e-$ plasma, expanding at relativistic velocities from the newborn BH site, engulfs different amounts of mass along different directions owing to the asymmetries developed in SN density profile, leading to different dynamics and consequently to different signatures for different viewing angles. The agreement of such a scenario with the observed emission from the X-ray flares in the BdHN early afterglow was there shown (see also section~\ref{sec:6} below for a discussion on this topic). The SN ejecta geometry affects as well the GeV emission observed in BdHNe \citep{2018arXiv180305476R}.

{\citet{2018ApJ...852...53R} also showed the relevance of the binary effects and the SN morphology on a most fundamental phenomenon: the GRB exploding within the SN impacts on it affecting its dynamics by transferring energy and momentum and finally transforming the ordinary SN into a hypernova (HN). Therefore, this model predicts that broad-lined SNe or HNe are not born as such but instead they are the outcome of the GRB impact on the SN. We have given evidence of such a SN-HN transition in a BdHN by identifying the moment of its occurrence in the case of GRB 151027A \citet{2017arXiv171205001R}.}

All these results point to the necessity of a detailed knowledge of the physical properties of the SN ejecta and in general of the binary system, in the 3D space and as a function of time, for the accurate inference of the consequences on the X, gamma-ray emission in {XRFs and} BdHNe and also on the GeV emission in BdHNe.

In view of the above we present here the first 3D hydrodynamic simulations of the IGC scenario. We have used the SPH technique as developed in the SNSPH code \citep{2006ApJ...643..292F}. The SNSPH is a tree-based, parallel code that has undergone rigorous testing and has been applied to study a wide variety of astrophysical problems \citep[see, e.g.,][]{2002ApJ...574L..65F,2006ApJ...640..891Y,2008ASPC..391..221D,2017ApJ...846L..15B}.  
The simulation starts from the moment at which the SN shock front reaches the CO$_{\rm core}$ external radius and, besides to calculate the accretion rate onto the NS companion, we also follow the evolution of the binary parameters (e.g. the binary separation, period, eccentricity) in order to determine if the final configuration becomes disrupted or not. This implies that we have introduced as well the gravitational effects of the remnant neutron star, the $\nu$NS formed at the center of the SN explosion, allowing us to calculate also the accretion onto it via matter fallback.

This article is organized as follows. In section~\ref{sec:2} we describe the main aspects of the SNSPH code \citep{2006ApJ...643..292F} and the algorithm applied to simulate the accretion process. In section~\ref{sec:3} we give the details on the construction of the initial binary configuration. Section~\ref{sec:4} shows the results of the simulations. We have covered a wide range of initial conditions for the binary system, i.e. we have varied  the CO$_{\rm core}$ progenitors,  the binary initial separation, SN total energy {and the initial NS mass companion}. {In Section~\ref{sec:5} we analyses if the binary system is disrupted or not by the mass loss due to the SN explosion of the CO$_{\rm core}$ and in Section~\ref{sec:6}} we compute the evolution of the binary and determine whether the stars' gravitational collapse is possible. In section~\ref{sec:7} we discuss the consequences on our results. {Specifically, in Section~\ref{sec:7.1} we analyze in depth the main parameters of the system that decide the fate of the NS companion, then we discuss, in Section~\ref{sec:7.2}, how these conditions could be realized in a consistent binary evolutionary path. Section~\ref{sec:7.3} contains estimates of the occurrence rate of these systems and Section~\ref{sec:7.4} outlines the consequences on the explanation of the GRB prompt and afterglow emission, as well as the prediction of new observables. } 
 Finally, in section~\ref{sec:8} we present our conclusions and perspectives for future work. In Appendix~\ref{app:1} we present convergence tests of the numerical simulations.

\section{SPH simulation}
\label{sec:2}

We use the 3D Lagrangian hydrodynamic code SNSPH \citep{2006ApJ...643..292F} to model the evolution of the binary system after the CO$_{\rm core}$ collapses and the  SN explosion occurs. The code follows the prescription of the SPH formalism in \citet{1990nmns.work..269B}. Basically, the fluid is divided by $N$ particles with  determined position, $\vec{r}_i$, mass, $m_i$, and smooth length, $h_i$. Physical quantities for each particle  are calculated through an interpolation of the form:
\begin{equation}
  A_i(\vec{r}_{i})=\sum_{j}A_j\left( \frac{m_j}{\rho_j} \right)W(|\vec{r}_{ij}|,h_{ij}),
	\label{eq:smoothV}
\end{equation}
where  $|\vec{r}_{ij}|=|\vec{r}_i-\vec{r}_j|$, $W$ is the smoothing kernel (that is equal zero if $r>2h$)  and $h_{ij}=(h_i+h_j)/2$ is the symmetric smooth length between particles $i$ and $j$. The code allows to evolve the smooth length with time as $dh_i/dt=-1/3(h_i/\rho_i)(d\rho_i/dt)$.

Then, the hydrodynamical equations of conservation of linear momentum and energy are written as:
\begin{align}
  \frac{d\vec{v}_i}{dt}&=-\sum_{j=1}^{N}m_j\left( \frac{P_i}{\rho_i^2}+\frac{P_j}{\rho_j^2}+\Pi_{ij}\right)\vec{\nabla}_i\,W(r,h_{ij}) + \vec{f}_{\rm g}\, ; \label{eq:SPHeqs_2} \\
	\frac{du_i}{dt}&=\sum_{j=1}^{N}m_j\left( \frac{P_i}{\rho_i^2}+\frac{1}{2}\Pi_{ij} \right)(\vec{v}_i-\vec{v}_j)\cdot \vec{\nabla}_i\,W(r,h_{ij})\, ,
	\label{eq:SPHeqs_1}
\end{align}
where $\vec{v}_i$, $P_i$, $\rho_i$ and $u_i$ are  the particle velocity,  pressure,  density and  internal energy, respectively. In order to handle shocks, an artificial viscosity term is introduced through $\Pi_{ij}$ as in \citet{1992ARA&A..30..543M,2005RPPh...68.1703M}:
\begin{equation}
  \Pi_{ij}=\left\{ \begin{array}{ccc}
    \frac{-\alpha(c_{i}+c_{j})\mu_{ij}+0.5 \beta\mu^2_{ij}}{\rho_{i}+\rho_{j}} &\mathrm{if} & \vec{v}_{ij}\cdot\vec{r}_{ij}<0 \\
    0 & & \mathrm{otherwise}
  \end{array}
    \right.,
  \label{eq:artvisc}
\end{equation}
where $c_{i}$ is the sound speed, $\vec{v}_{ij}=\vec{v}_i-\vec{v}_{j}$ and
{
\begin{equation}
  \mu_{ij}=\frac{h_{ij}\vec{v}_{ij}\cdot\vec{r}_{ij}}{|\vec{r}_{ij}|^2+\epsilon h_{ij}^2}\,.
  \label{eq:mu_artvisc}
\end{equation}
}
The form for the viscosity of Equation~(\ref{eq:artvisc}) can be interpreted as an bulk and von Neuman-Richtmyer viscosity parametrized by $\alpha$ and $\beta$ viscosity coefficients. In the simulations we adopt, as usual, $\alpha=1.0$ and $\beta=2.0$.

The last term of Equation~(\ref{eq:SPHeqs_2}) refers to the fluid self-gravity force. The particles are organized in a hashed oct-tree and the gravitational force is evaluated using the multipole acceptability criterion (MAC) described in \citet{WarrenSalmon1993,1995CoPhC..87..266W}.

Finally, the equation of state adopted treats the ions as a perfect-gas and takes into account the radiation pressure:
\begin{equation}
	P=\frac{1}{3}aT^4 + n_{\rm ion}\kappa T\, ;
	\label{eq:EOS_P}
\end{equation}
\begin{equation}
	u=aT^4 + \frac{3}{2}n_{\rm ion}\kappa T\, ,
	\label{eq:EOS_U}
\end{equation}
where $n_{\rm ion}$ is the number density of ions, $T$, the temperature and $a$ the radiation constant.

\subsection{Accretion Algorithm}\label{sec:2.1}

In the simulation, the remnant of the CO$_{\rm core}$ and the NS companion are model as two point masses  that only interact gravitationally with the other particles  and between them. Their equation of motion  will be:
\begin{equation}
	\frac{d\vec{v}_s}{dt}=\sum_{j=1}^{N}\frac{Gm_j}{|\vec{r}_s-\vec{r}_j|^3}\left(\vec{r}_j-\vec{r}_s\right)+\frac{GM_{s'}}{|\vec{r}_s-\vec{r}_{s'}|^3}\left(\vec{r}_{s'}-\vec{r}_s\right)\, ,
	\label{eq:Acc_sink}
\end{equation}
where subindex $s$ and $s'$ make reference to the stars. In the same way, each particle of the fluid will feel an additional force from the stars gravitational field:
\begin{equation}
  \vec{f}_{s,i}=\frac{GM_s}{|\vec{r}_s-\vec{r}_i|^3}\left(\vec{r}_s-\vec{r}_i\right)\, .
  \label{eq:acc_sink}
\end{equation}
The stars accrete a particle $j$ from the SN ejecta if the following conditions are fulfilled \citep{1995MNRAS.277..362B}:
\begin{enumerate}
	\item The particle is inside the star accretion radius, i.e.:
		$$|\vec{r}_j-\vec{r}_s|<R_{\rm j,acc}\, .$$
	\item The gravitational potential energy of the particle in the field of the star is larger than its kinetic energy, i.e.:
		$$\frac{GM_sm_j}{|\vec{r}_j-\vec{r}_s|}>\frac{1}{2}m_j|\vec{v}_j-\vec{v}_s|^2\, .$$
	\item The angular momentum of the particle relative to the star is less than the one it would have in a Keplerian orbit at $R_{\rm j,acc}$, i.e.:
		$$|(\vec{r}_j-\vec{r}_s)\times(\vec{v}_j-\vec{v}_s)|<\sqrt{GM_sR_{\rm j,acc}}\,.$$
\end{enumerate}
The accretion radius for a particle $j$ is defined as:
\begin{equation}
	R_{\rm j,acc}={\rm min}\left(\, \xi\,\frac{2GM_s}{v_{js}^2+c_{j}^2}\,,\,h_j \,\right),
	\label{eq:CaptureRadius}
\end{equation}
and we adopt in the simulations $\xi=0.05$--$0.1$ (see below for further details on this parameter).

These conditions are evaluated at the beginning of every time step. The particles that fulfill them are removed from the simulation and we update the properties of the star as:
\begin{eqnarray}
	M_{\rm s,new} &=& M_{\rm s}+\sum_{j}m_j \, ;\\
	\vec{v}_{\rm s,new} &=& \cfrac{M_s\vec{v}_s+\sum_j m_j\vec{v}_j}{M_{\rm s,new}}\, ;\\
	L_{\rm s,new} &=& M_{\rm s,new}\cfrac{L_s\vec{v}_s+\sum_j m_j(\vec{r}_{s,j}\times\vec{v}_{s,j})}{M_{\rm s}}\, .
	\label{eq:updateStar}
\end{eqnarray}
The above sum is over the particles accreted during the corresponding time step.

\section{Initial setup}
\label{sec:3}

Our calculations include a suite of pre-SN progenitors with zero-age-main- sequence (ZAMS) masses ranging from 15 to 40~M$_\odot$ obtained via the KEPLER code \citep{heger10}. The SN explosions are simulated with the 1D core-collapse code \citep{1999ApJ...516..892F} and the multi-parameter prescriptions introduced in \citet{2017arXiv171203415F} to mimic the supernova engine: the energy deposition rate and duration, the size of the convection cell above the base of the proto-NS and the time after bounce when the convective engine starts. These parameters are designed to include the uncertainties in the convection-enhanced supernova engine \citep[see][for details]{herant94,fryer07,murphy13}.

When the shock front reaches the edge of the CO$_{\rm core}$, the configuration is mapped into a 3D SPH-configuration of about 1 million particles with variable mass. This is done using the weight Voronoi tessellation (WVT) as described in \citet{2015PASA...32...48D}. 

The SPH configuration of the SN ejecta is constructed in the rotating reference frame of the progenitor star. In order to translate it to the center of mass reference frame of the initial binary system (CO$_{\rm core}$ + NS), the position and velocities of the particles are modified as follow:
\begin{eqnarray}
	\vec{r}_{i,\rm new}&=& \mathbb{R}\vec{r}_i-\vec{r}_{\rm CO}\, ;\\
	\vec{v}_{i,\rm new}&=& \mathbb{R}\vec{v}_i -\vec{r}_i\times\vec{\Omega}_{\rm orb} -\vec{v}_{\rm CO}\, ,
	\label{eq:RotationTrans}
\end{eqnarray}
where $\mathbb{R}$ is a rotation matrix, $\vec{r}_{\rm CO}$ and $\vec{v}_{\rm CO}$ are the position and velocity of the CO$_{\rm core}$ before the explosion and $\Omega_{\rm orb}$ is the binary orbital angular velocity, that is determined once the orbital separation and star masses are established:
\begin{equation}
	\Omega_{\rm orb}=\sqrt{\frac{G(M_{\rm CO}+M_{\rm ns})}{a_{\rm orb}^3}}\, ,
	\label{eq:Omegabinary}
\end{equation}
with $M_{\rm CO}$ being the CO$_{\rm core}$ total mass, $M_{\rm NS}$ the NS mass and $a_{\rm orb}$ the binary separation. The equatorial plane of the binary corresponds to the x-y plane, then the initial position of the stars (NS and $\nu$NS) are on the x-axis and its initial motion is counter-clockwise.

The minimum binary period that the system can have is given by the condition that the compactness of the CO$_{\rm core}$ is such that there is no Roche lobe overflow before the SN explosion. Then, the minimum binary separation is determined by \citep{1983ApJ...268..368E}:
\begin{equation}
	\frac{R_{\rm star}}{a_{\rm orb}}=\frac{0.49q^{2/3}}{0.6q^{2/3}+\ln\left( 1+q^{1/3} \right)},
	\label{eq:RocheLobe}
\end{equation}
with $q=M_{\rm CO}/M_{\rm ns}$.

\begin{figure}
	\centering
    \includegraphics[scale=0.6]{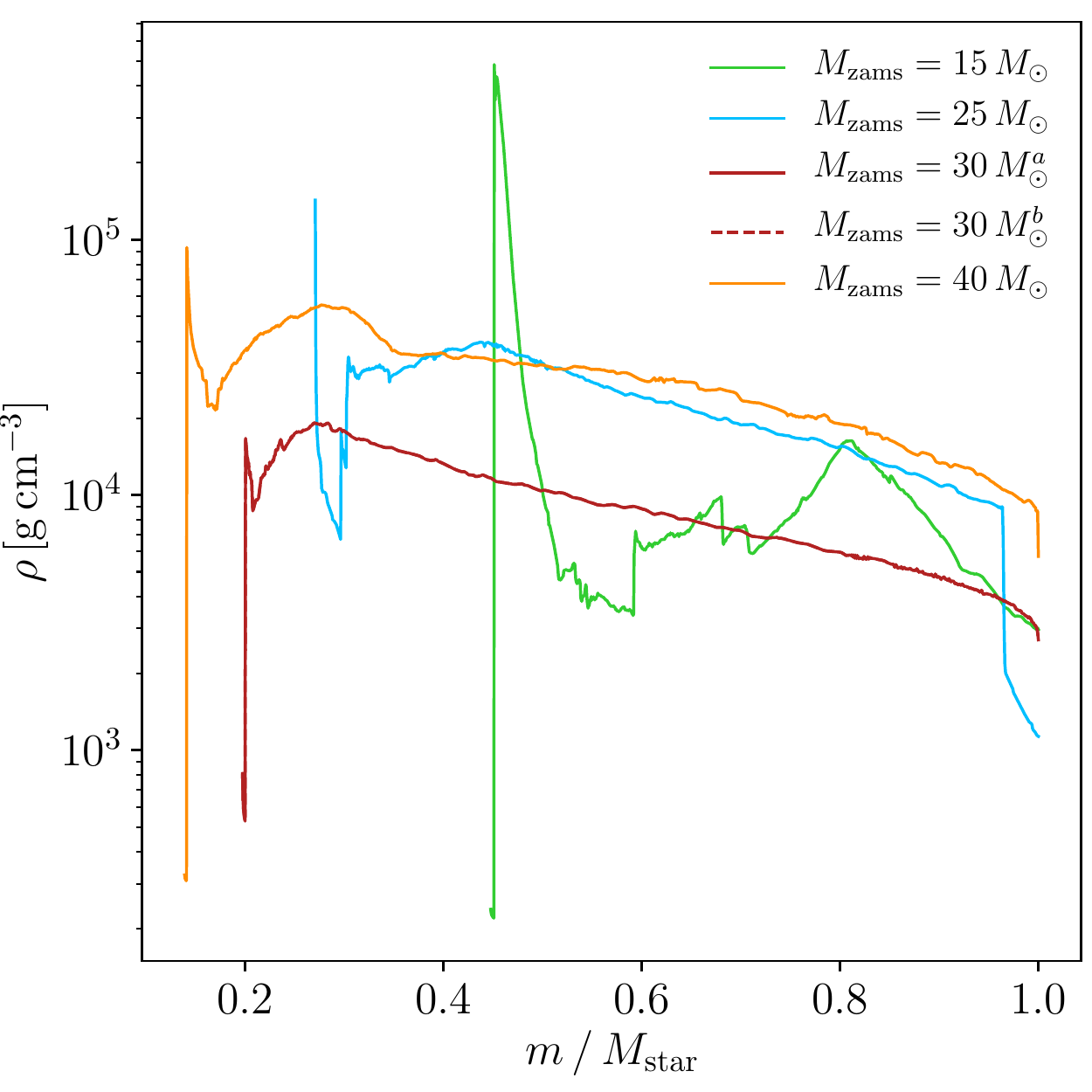}
	\caption{Density profile of the SN ejecta when the shock has reached the carbon-oxygen edge for the $M_{\rm zams}=15$, $25$, $30$ and $40\, M_\odot$ progenitors (see table~\ref{tab:ProgSN}). {The density is given as a function of the variable $m/M_{\rm star}$, namely the mass coordinate, $m$, normalized to the total mass of the CO$_{\rm core}$, $M_{\rm star}$}. At this moment, the 1D simulation is mapped into a 3D SPH-configuration.}
	\label{fig:Progrho}
\end{figure}
\begin{table*}
\centering
\caption{Properties of the CO$_{\rm core}$ progenitors.}
\setlength{\tabcolsep}{5pt}
\begin{tabular}{c|ccccccc}
  \hline
  \hline
  $M_\mathrm{ZAMS}$ & $M_\mathrm{rem} $ & $M_\mathrm{ej}$ &      $R_{\rm core}$        &      $R_{\rm star}$       &  $v_{\rm star,0}$ &   $E_\mathrm{grav}$         &     $m_j$   \\
   $(\,M_\odot\,)$  & $(\,M_{\odot}\,)$ & $(\,M_\odot\,)$ & $(\,10^8\, \mathrm{cm}\,)$ & $(\,10^9\,\mathrm{cm}\,)$ & $(\, 10^8 {\rm cm/s}\,)$ &$(\,10^{51}\, \mathrm{erg}\,)$ & $(\,10^{-6}\, M_\odot\,)$  \\
   \hline
    $15$ & $1.30$ & $1.606$ & $8.648$ & $5.156$ & $9.75$ & $0.2149$& $0.2-4.4$ \\
    $25$ & $1.85$ & $4.995$ & $2.141$ & $5.855$ & $5.43$ &$1.5797$& $2.2-11.4$ \\
    $30$\footnote{$E_{\rm sn}=1.09\times 10^{52}\, {\rm erg}$} & $1.75$ & $7.140$ & $28.33$ & $7.830$ & $8.78$ &$1.7916$& $1.9-58.9$ \\
    $30$\footnote{$E_{\rm sn}=2.19\times 10^{51}\, {\rm erg}$} & $1.75$ & $7.140$ & $13.84$& $7.751$ & $5.21$ &$1.5131$ & $1.9-58.9$ \\
    $40$ & $1.85$ & $11.50$ & $19.47$  &$6.529$ & $6.58$ &$4.4305$& $2.3-72.3$ \\
\hline
\hline
\end{tabular}
\tablecomments{Each progenitor was evolved with the KEPLER stellar evolution code \citep{heger10} and then was exploded artificially using the 1D core-collapse code presented in \citet{1999ApJ...516..892F}.}
\label{tab:ProgSN}
\end{table*}

Since we are interested in identifying the favorable conditions for which the NS companion can accrete enough mass and collapse into a BH, we will explore different initial conditions for the system. We have worked with four progenitor for the CO$_{\rm core}$ with different ZAMS mass: $M_{\rm zams}= 15$, $25$, $30$ and $40\, M_{\odot}$. In Table~\ref{tab:ProgSN} we present the main proprieties of each of these progenitors at the mapping moment: the SN mass ejected, $M_{\rm ej}$, the {the gravitational mass of the remnant star}, $M_{\rm rem}$ {(i.e.~the CO$_{\rm core}$ mass before its collapse is $M_{\rm star} = M_{\rm ej}+ M_{\rm rem}$)}, the SN ejecta innermost radius, $R_{\rm core}$, the CO$_{\rm core}$ radius when the collapse happens, $R_{\rm star}$, and the forward shock velocity, $v_{\rm star,0}$. In the last two columns of Table~\ref{tab:ProgSN} we specify the gravitational energy of the star, $E_{\rm grav}$, and the maximum and minimum masses of the SPH particles, $m_j$. Figure~\ref{fig:Progrho} shows the density profile of each progenitor at the moment of the mapping to the 3D SPH configuration (when the shock front of the explosion has reached the star surface). We have two models for the $30\, M_\odot$ progenitor, each with different SN explosion energy.

It is important to notice that we are working with progenitors that were evolved as isolated stars, i.e. without taking into account that they are part of a binary system. However, as it was indicated in \citet{2014ApJ...793L..36F}, there is a $3$--$4$ order of magnitude pressure jump between the CO$_{\rm core}$ and helium layer, this means that the star will not expand significantly when the helium layer is removed: the CO$_{\rm core}$ can be $1.5$--$2$ times larger \citep{2010ApJ...719.1445M}, then the minimum period of the system might increase by a factor of $1.8$--$2.8$.

The final fate of the system also will depend on the characteristics of the SN explosion. We run simulations varying the explosion energy of the SN. Rather than produce a broad range of explosion energies, (as we did with the $30\,M_\odot$ progenitor), we scaled the kinetic and internal energy of the particles behind the forward shock by a factor $\eta$. In this way, the internal structure of the progenitor does not change, just the velocity and the temperature.

In order to study the effect of an asymmetric SN expansion \citep[see, e.g.,][]{2012ARNPS..62..407J} we adopt a single-lobe prescription \citep{2005ApJ...635..487H} following \citet{2003apj...594..390h,2006ApJ...640..891Y}. Namely, the explosion is modified to a conical geometry parametrized by $\Theta$, the opening angle of the cone, and $f$, the ratio of the velocities between the particles inside and outside the cone. The velocities of the SPH particles behind the forward shock are then modified as:
\begin{equation}
	V_{\rm in-cone}=f\left[ \frac{1-f^2}{2}\cos \Theta +\frac{1+f^2}{2} \right]^{-1/2}V_{\rm symm}\, ;
	\label{eq:AssyVel_in}
\end{equation}
\begin{equation}
	V_{\rm out-cone}=\left[ \frac{1-f^2}{2}\cos \Theta +\frac{1+f^2}{2} \right]^{-1/2}V_{\rm symm}\,,
	\label{eq:AssyVel_out}
\end{equation}
where $V_{\rm symm}$ is the radial velocity of the original explosion. This prescription conserves the kinetic energy of the symmetry explosion, and to conserve the total energy of the supernova we scale the particles internal energy in the same way.

In the next section, we present the results of our SPH-simulations.
\begin{figure*}
  \centering
  \includegraphics[scale=0.33]{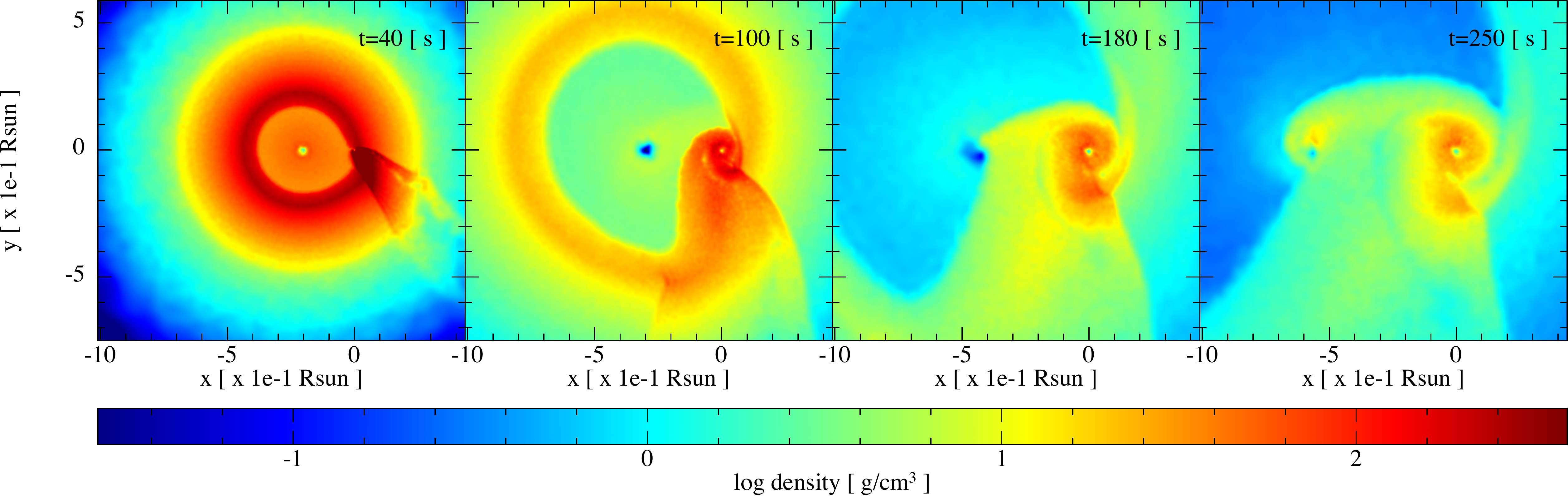}
 \includegraphics[scale=0.33]{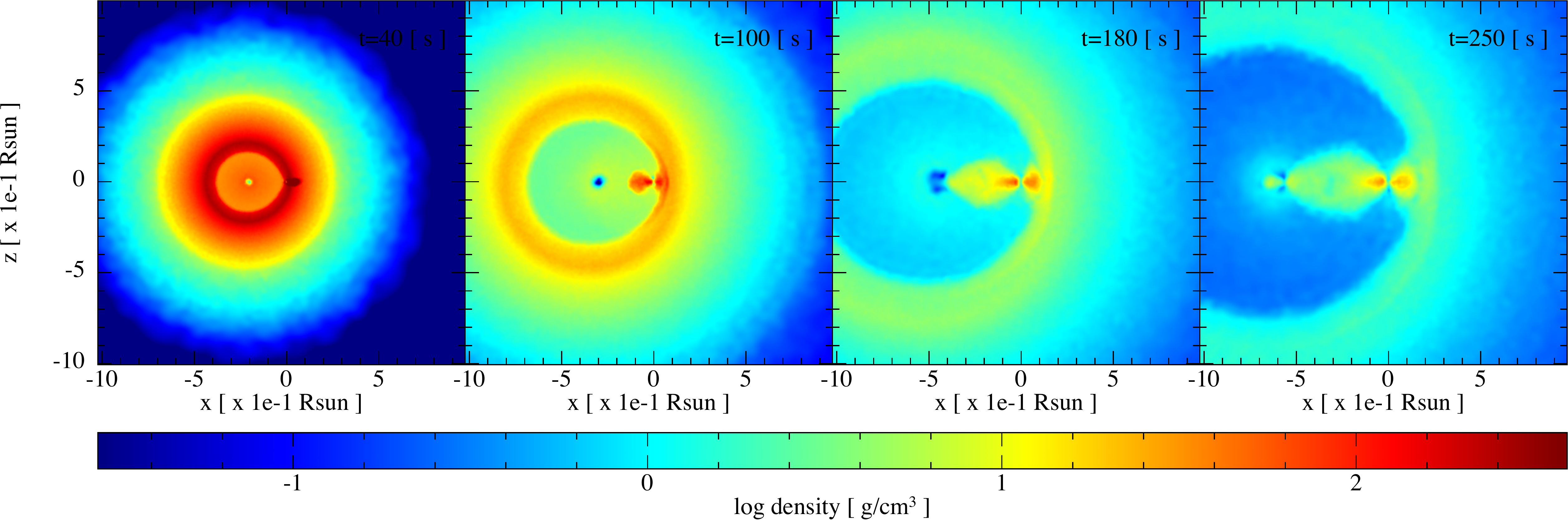}
 \caption{Snapshots of the SPH simulation of the IGC scenario. The initial binary system is formed by  a CO$_{\rm core}$, which progenitor is a $M_{\rm zams}=25\,M_\odot$, and a $2\, M_\odot$ NS with an initial orbital period of approximately $5$~min (Model 25M1p1e of table~\ref{tab:inconmodels2}).  The upper panel shows the mass density on the binary equatorial plane, at different times of the simulation, while the lower panel corresponds to the plane orthogonal to the binary equatorial plane. The reference system was rotated and translated in a way that the x-axis is along the line that joins the binary stars and the origin of the reference system is at the NS position. At $t=40$~seconds (first frame from left to right), it can be seen that the particles captured by the NS have formed a kind of tail behind it, then this particles star to circularize around the NS and a kind of thick disk is observed at $t=100$~s (second frame from left to right). The material captured by the gravitational field of the NS companion is also  attracted by the $\nu$NS and start to be accreted by it as can be seen at $t=180$~s (third frame). After around one initial orbital period, at $t=250$~s, around the both stars have been form a kind of disk structure. The $\nu$NS is along the x-axis at: $-2.02$, $-2.92$, $-3.73$ and $-5.64$ for $t=40$, $100$, $180$ and $250$~s, respectively. \\
  {\sc Note---}This figure and all the snapshots figures where done with the {\sc SNsplash} visualization program \citep{2011ascl.soft03004P}.}
  \label{fig:25Mzams2Mns}
\end{figure*}
%

\section{Results}
\label{sec:4}
%
\begin{figure}
  \centering
  \includegraphics[width=0.9\hsize]{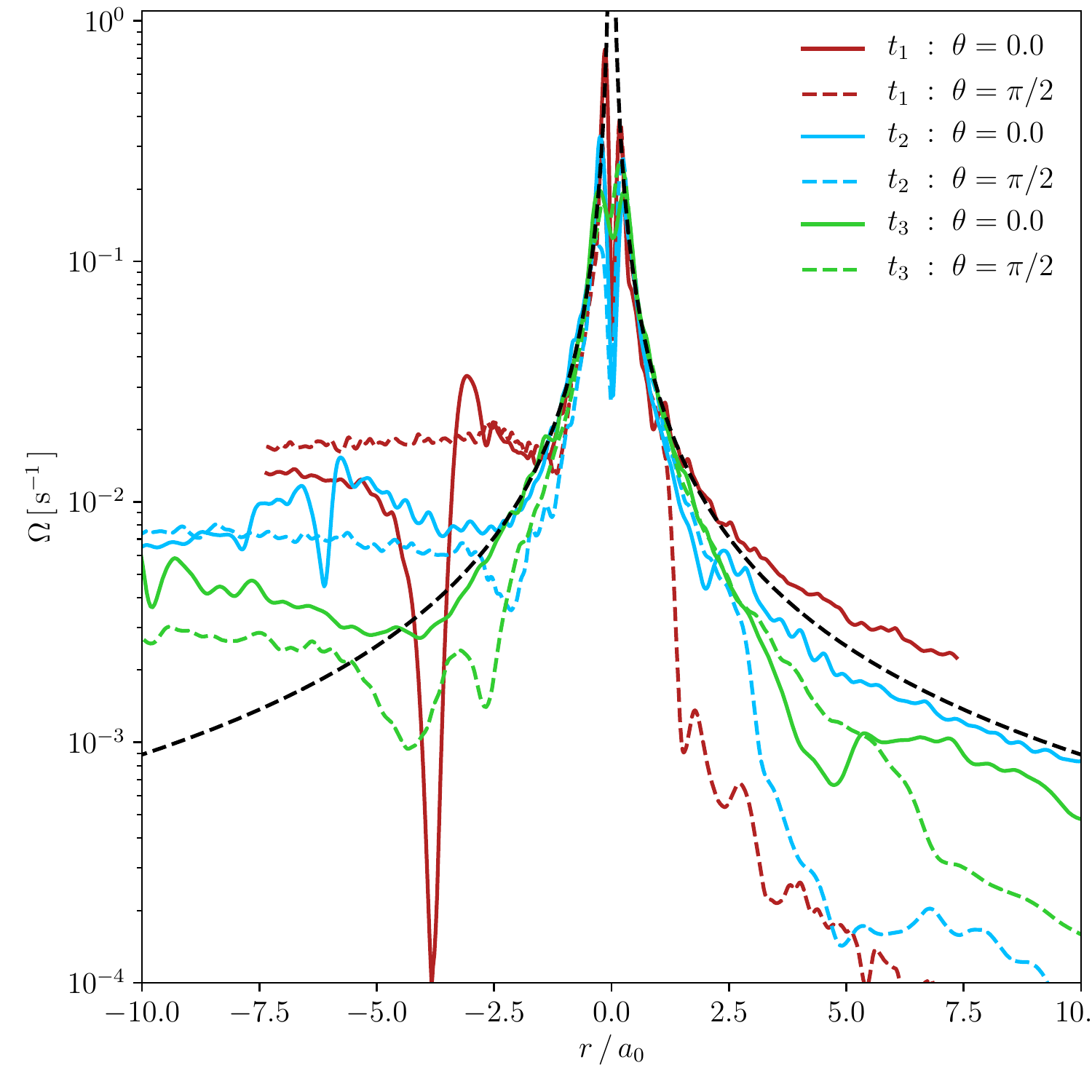}
  \caption{Angular velocity profiles of the SN material closed to the NS companion at three different times: $t_1=140.0$~s, $t_2=280.0$~s, $t_3=599.9$~s. The label $\theta=0$ corresponds to the line that joins the binary stars and the label $\theta=\pi/2$ is the line perpendicular to the latter and lies on the equatorial binary system. Close to the NS, the angular velocity approaches the Keplerian angular velocity (black line). This suggests that, before being accreted, the particles have enough angular momentum to circularize around the star a form a kind of disk structure. The minimum of the solid line in the $-x$ direction indicates the position of the $\nu$NS: at $3.01$, $5.04$ and $10.49$ for $t_1$, $t_2$ and $t_3$ respectively.}
  \label{fig:Omega25Mzams}
\vspace{0.5\floatsep}
  \centering
  \includegraphics[width=0.9\hsize]{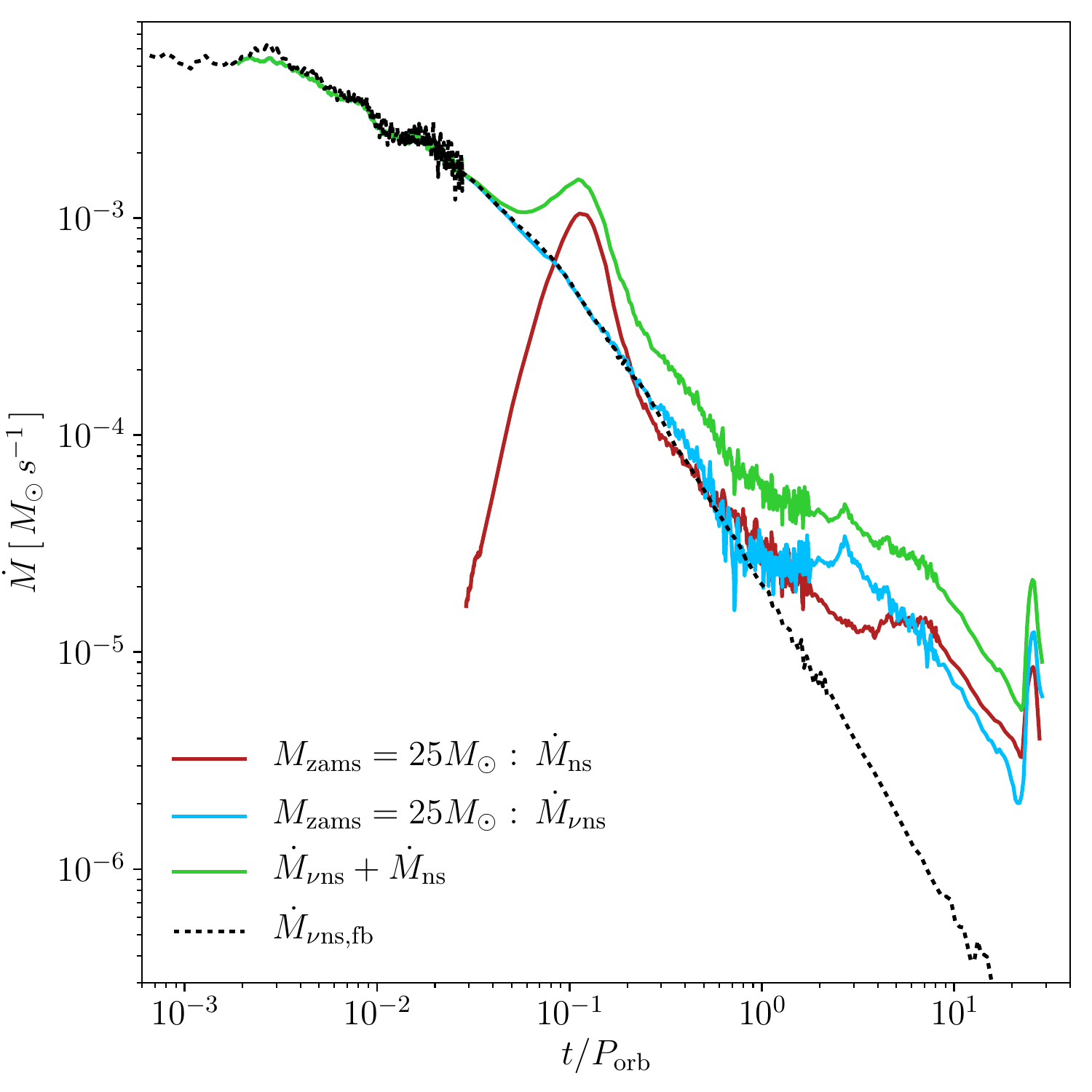}
  \caption{Mass accretion rate onto the NS (red line) and the $\nu$NS (blue line) during the SPH simulation of the expansion of the SN ejecta. The green line shows the sum of both accretion rates. The initial binary system is formed by the CO$_{\rm core}$ of a  $M_{\rm zams}=25\,M_\odot$ progenitor and a NS of $2\,M_\odot$ with an initial binary period of approximately $5$~min. The dotted black line corresponds to the fallback mass-accretion onto the $\nu$NS when the CO$_{\rm core}$ collapses in an single-star configuration, i.e, without the presence of the NS companion.}
  \label{fig:25Mzams_MdotNS}
\end{figure}

In Table~\ref{tab:inconmodels2} we summarize the properties of the SN and the parameters that characterize the state of the initial binary systems with the different CO$_{\rm core}$ obtained with the progenitors of table~\ref{tab:ProgSN}. {The models are labeled as ``$x_1$m$x_2$p$x_3$e'', where $x_1$ is the $M_{\rm zams}$ of the CO$_{\rm core}$ progenitor star, $x_2$ is the fraction between the initial orbital period and the minimum orbital period of the system to have no Roche-lobe overflow; and $x_3$ is the value of the $\eta$ factor by which the SPH particle velocities and internal energy are scaled.} For each model {in Table~\ref{tab:inconmodels2}} we specify the sum of the ejecta kinetic and internal energy, $E_k+U_i$, the initial orbital period, $P_{\rm orb,i}$ as well as the initial binary separation, $a_{\rm orb,i}$.
\begin{table*}
\centering
\caption{SPH Simulations}
\setlength{\tabcolsep}{1.45pt}
\renewcommand{\arraystretch}{1.15}
\footnotesize
\begin{tabular}{c|ccc|cc|cccccccc}
  \hline
  \hline
  Model  & $E_{\rm k}+U_i$&$P_\mathrm{orb,i} $ & $a_\mathrm{orb,i} $ & $M_\mathrm{\nu ns,fb}$ & $V_{\rm kick}$ &$M_\mathrm{\nu ns}$ & $M_{\rm ns}$ &  $V_{\rm CM}$ &  $P_\mathrm{orb,f} $ &  $a_\mathrm{orb,f}$& $e$ & $m_{\rm bound}$& bound   \\
  &     $(\, 10^{51} \,{\rm erg}\,)$ &$(\, \mathrm{min}\,)$ & $(\, 10^{10}\, \mathrm{cm}\,)$ & $(\, m_\odot \,)$&$(\,10^4\, \mathrm{cm/s}\,)$& $ (\, m_\odot\,) $ & $(\, m_\odot$\,) & $(\,10^7\,\mathrm{cm/s}\,)$ & $(\, \mathrm{min}\,)$ & $(\, 10^{10}\, \mathrm{cm}\, )$ & & $(\, m_\odot\,)$  &  \\
   \hline
   \hline
   \multicolumn{12}{c}{ $M_{\rm zams}=15\, M_\odot$ Progenitor} \\ 
   \hline
   \hline 
 15m1p07e   & $1.395$ & $6.58$ & $1.361$ & $1.302$ &  $3.99$ & $1.302$ & $2.003$ & $4.05$ & $19.0$ & $2.437$ & $0.443$ & $2.7\times 10^{-6}$ & yes\\
 15m1p05e  & $1.101$ & $6.58$ & $1.361$ & $1.303$ &  $4.83$ & $1.303$ & $2.006$ & $4.02$ &$18.6$ & $2.393$ & $0.433$ & $4.8\times 10^{-5}$ & yes\\
 15m1p03e   & $0.607$ & $6.58$ & $1.361$ & $1.304$ & $18.93$ & $1.315$ & $2.023$ & $3.91$ & $15.9$ & $2.182$ & $0.398$ & $6.9\times 10^{-4}$ & yes\\
 15m1p01e   & $0.213$ & $6.58$ & $1.361$ & $2.478$ &  $6.62$ & $1.916$ & $2.199$ & $0.39$ & $3.04$ & $0.773$ & $0.233$ & $0.098$ & yes\\
 15m1p005e  & $0.135$ & $6.58$ & $1.361$ & $2.731$ &  $1.73$ & $2.649$ & $2.034$ & $0.08$ & $5.92$ & $1.257$  & $0.036$ & $0.029$ & yes\\
 15m2p03e   & $0.607$ & $11.7$ & $2.000$ & $1.304$ & $18.93$ & $1.304$ & $2.007$ & $0.39$ &   $-$   &    $-$   & $1.192$ & $5.8\times 10^{-3}$ &no \\
 15m2p01e   & $0.213$ & $11.7$ & $2.000$ & $2.478$ &  $6.62$ & $2.238$ & $2.101$ & $0.08$ &$12.9$ & $2.068$ & $0.419$ & $0.0759$  & yes \\
 15m3p01e   & $0.213$ & $17.6$ & $2.000$ & $2.478$ &  $6.62$ & $2.405$ & $2.057$ & $0.32$ &$14.5$ & $2.249$ & $0.103$ & $0.0313$  & yes \\
\hline
\hline
   \multicolumn{12}{c}{ $M_{\rm zams}=25\, M_\odot$ Progenitor} \\ 
   \hline
   \hline
  25m1p1e   & $3.14$ & $4.81$ & $1.352$ & $1.924$ &  $1.35$ & $1.963$ & $2.085$ & $7.49$ & $116.9$ & $8.747$ & $0.866$ & $0.081$ & yes\\
  25m1p09e   & $2.84$ & $4.81$ & $1.352$ & $1.935$ &  $2.97$ & $2.013$ & $2.162$ & $7.17$ & $38.29$ & $4.199$ & $0.744$ & $0.043$ & yes\\
  25m1p08e   & $2.53$ & $4.81$ & $1.352$ & $1.953$ &  $3.57$ & $2.081$ & $2.441$ & $6.09$ & $16.5$ & $2.454$ & $0.600$ & $0.075$ & yes\\
  25m1p07e   & $2.22$ & $4.81$ & $1.352$ & $2.172$ & $41.30$ & $2.371$ & $2.621$ & $3.77$ & $ 4.33 $ & $1.043$ & $0.381$ & $0.160$ & yes\\
  25m2p1e   & $3.14$ & $8.56$ & $1.988$ & $1.924$ &  $1.35$ & $1.929$ & $2.029$ & $6.46$ &     $-$   &   $-$   & $1.005$ & $0.073$ &  no\\
  25m3p1e    & $3.14$ & $11.8$ & $2.605$ & $1.924$ &  $1.35$ & $1.924$ & $2.024$ & $5.76$ &     $-$   &   $-$   & $1.086$ & $0.025$ &  no\\
  25m4p1e     & $3.14$ & $15.9$ & $2.984$ & $1.924$ &  $1.35$ & $1.916$ & $2.014$ & $5.34$ &     $-$   &   $-$   & $1.096$ & $0.013$ &  no\\
  25m2p07e  & $2.22$ & $8.56$ & $1.988$ & $2.172$ & $41.30$ & $2.352$ & $2.522$ & $3.43$ & $18.3$  & $2.699$ & $0.461$ & $0.185$ & yes\\
  25m3p07e   & $2.22$ & $11.8$ & $2.605$ & $2.172$ & $41.30$ & $2.301$ & $2.523$ & $3.25$ & $23.5$ & $3.11$  & $0.477$ & $0.138$ & yes\\
  25m5p07e   & $2.22$ & $19.8$  & $3.463$ & $2.172$ & $41.30$ & $2.166$ & $2.401$ & $3.24$ & $27.8$ & $3.49$  & $0.526$ & $0.0051$& yes\\
  25m15p07e  & $2.22$ & $59.4$  & $7.203$ & $2.172$ & $41.30$ & $2.088$ & $2.208$ & $2.72$ & $ 321.5 $ & $17.41$ & $0.639$ & $3.4\times 10^{-3} $ & yes \\
\hline
\hline
   \multicolumn{12}{c}{ $M_{\rm zams}=30\, M_\odot$ Progenitor\, -\, exp 1} \\ 
\hline
\hline
  30m1p1ea  & $8.43$ & $5.82$ & $1.667$ & $1.756$ & $40.80$ & $1.757$ & $2.007$ & $9.67$ & $-$ & $-$ & $1.701$ &$0.0$ & no\\
  30m1p07ea  & $5.95$ & $5.82$ & $1.667$ & $1.755 $ & $39.36$ & $1.758$ & $2.015$ & $9.58$ & $-$ & $-$ & $1.647$ &$9.6\times 10^{-4}$ & no\\
  30m1p05ea & $4.26$ & $5.82$ & $1.667$ & $1.758$ & $96.20$ & $1.764$ & $2.031$ & $9.46$ & $ - $  & $ - $ & $1.501$ & $0.012$ & no\\
  30m1p03ea & $2.59$ & $5.82$ & $1.667$ & $2.178$ & $3.93\times 10^{3}$ & $1.869$  & $2.455$ & $7.79$ & $ 101.5 $  & $ 8.137 $ & $0.852$  & $0.168$  &yes \\
  30m2p03ea & $2.59$ & $10.8$ & $2.449$ & $2.178$ & $3.93\times 10^{3}$ & $1.837$ & $2.192$  & $7.21$ & $-$ & $-$ & $1.095$ & $0.0854$ & no\\
\hline
\hline
   \multicolumn{12}{c}{ $M_{\rm zams}=30\, M_\odot$ Progenitor\, -\, exp 2} \\ 
\hline
\hline 
30m1p1eb   & $3.26$ & $5.82$ & $1.667$ & $4.184$ & $141.30$ & $3.675$ & $2.382$ & $3.59$ & $12.1$ & $2.209$ &  $0.569$ & $0.331$ &yes\\
30m1p12eb  & $3.91$ & $5.82$ & $1.667$ & $2.462$ & $147.79$ & $2.515$ & $2.376$ & $5.86$ & $46.1$ & $5.009$  & $0.733$ & $0.133$ &yes\\
30m2p12eb & $3.91$ & $10.8$ & $2.410$ & $2.462$ & $147.79$ & $2.621$ & $2.228$ & $4.85$ & $157.1$ & $11.302$ & $0.848$  & $0.029$ &  yes \\
30m1p2eb  & $6.45$ & $5.82$ & $1.667$ & $1.771$  & $13.89$ & $1.783$ & $2.077$ & $9.50$ & $-$ & $-$ & $1.447$  & $5.7\times 10^{-3}$ &no\\
30m1p31eb  & $10.02$ & $5.82$ & $1.667$ & $1.766$ & $5.21$ & $1.768$ & $2.017$ & $9.95$ & $-$ & $-$ & $1.712$  & $6.5\times 10^{-4}$ &no\\
\hline
\hline
   \multicolumn{12}{c}{ $M_{\rm zams}=40\, M_\odot$ Progenitor} \\ 
  \hline
  \hline
 40m1p1e    & $10.723$ & $3.49$ & $1.295$ & $1.871$ & $176.43$ & $1.874$ & $2.119$ & $13.68$  & $-$    &   $-$   & $1.845$ & $0.038$ &  no\\
  40m1p09e  &  $9.670$ & $3.49$ & $1.295$ & $1.872$ & $141.38$ & $1.881$ & $2.274$ & $13.35$  & $-$    &   $-$   & $1.538$ & $0.027$ &  no\\
  40m1p08e  &  $8.618$ & $3.49$ & $1.295$ & $1.873$ & $242.93$ & $1.886$ & $2.545$ & $12.52$  & $-$    &   $-$   & $1.276$ & $0.016$ &  no\\
  40m1p07e  &  $7.506$ & $3.49$ & $1.295$ & $1.879$ & $464.82$ & $2.095$ & $3.033$ & $10.12$ & $85.9$ & $7.712$ & $0.881$ & $0.051$ & yes\\
    40m1p06e  &  $6.513$ & $3.49$ & $1.295$ & $6.568$ & $69.73$  & $3.784$ & $3.209$ & $4.57$ & $2.42$ & $0.792$ & $0.612$ & $1.053$ & yes\\
  40m1p05e  &  $5.145$ & $3.49$ & $-$ &  $-$       &   $0.91$       & $5.430$ & $3.642$ & $-$ & $0.06$ &  $0.339$ &$0.243$ & $1.250$ & yes\\
  40m2p1e   & $10.723$ & $6.22$ & $1.904$ & $1.871$ & $176.43$ & $1.873$ & $2.046$ & $10.79$&$-$ & $-$ & $2.194$ &$6.13\times 10^{-3}$& no \\
  40m2p07e  &  $7.506$ & $6.22$ & $1.904$ & $1.879$ & $464.82$ & $2.064$ & $2.755$ & $8.757$ & $4408$ & $104.26$ & $0.984$ & $0.078$ & yes\\
  40m4p07e  &  $7.506$ & $12.45$ & $3.022$ & $1.879$ & $464.82$ & $1.959$ & $2.507$ & $7.799$ & $-$ & $-$  & $1.180$ & $0.0744$ & no\\
  40m2p06e &  $6.513$ & $6.22$ & $1.904$ & $6.568$ & $69.73$ & $5.581$ & $2.961$ & $2.58$ & $5.79$ & $1.523$ & $0.648$ & $0.509$ &yes\\
\hline 
\end{tabular}
\label{tab:inconmodels2}
\end{table*}

For each model, we first run a simulation of the SN expansion assuming that the CO$_{\rm core}$ collapses and explodes as an isolate star, i.e. without the NS companion. In {Table~\ref{tab:inconmodels2}} we summarize the final mass of the $\nu$NS, indicated as $M_{\nu{\rm ns,fb}}$, and the magnitude of the $\nu$NS kick velocity, $V_{\rm kick}$. {The latter is the $\nu$NS velocity due to the linear momentum accreted by the star from the fallback particles.}

Then, we run simulations of the SN expansion with the CO$_{\rm core}$ as part of a binary system with a $2\,M_\odot$ NS as companion. {We expect a massive initial NS companion because the binary system evolutionary path leading to these systems have at least one common-envelope episode \citep[see e.g.][and section~\ref{sec:6}]{1999ApJ...526..152F,2015ApJ...812..100B}. However, we have also performed simulations with a NS companion with an initial mass $1.4\,M_\odot$ and $1.6\,M_\odot$}. In Table~\ref{tab:inconmodels2} we summarize the parameters that characterize the final outcome of these simulations: the $\nu$NS final mass, $M_{\nu{\rm ns}}$, the NS final mass, $M_{\rm NS}$, the velocity of the final binary system center-of-mass, $V_{\rm CM}$, the final orbital period, $P_{\rm orb,f}$, the major semi-axis and eccentricity of the orbit of the final system, $a_{\rm orb,f}$ and $e$, respectively, and the amount of mass that is still bounded to the binary stars at the moment when the simulation is stopped, $m _{\rm bound}$. This bound material circularizes around the stars and at some moment might be accreted by any of the them. In the final column of Table~\ref{tab:inconmodels2}, we have specified if the system remains bound as a new binary system or if it is disrupted in the explosion.

\subsection{Fiducial model: $25\,M_\odot$ progenitor}

We are going to take the $25\,M_\odot$ progenitor star for the CO$_{\rm core}$ and a $2\,M_\odot$ NS as our fiducial initial binary system in order to describe in detail the main features of the simulations while the SN expands in presence of the NS companion (model 25m1p1e of table~\ref{tab:inconmodels2}). This {fiducial} binary system has the minimum orbital period allowed for the system to have no Roche-lobe overflow, $4.86$ minutes, which corresponds to a binary separation of $1.34\times 10^{10}$~cm. {Thus, this model presents most favorable conditions for the occurrence of the induced collapse: a short period for the binary system and a massive initial NS companion}. Later, we will change one by one the initial conditions and compare the outcomes with the one of this fiducial system.

Figure~\ref{fig:25Mzams2Mns} shows snapshots of the mass density in the x-y plane, the binary equatorial plane (upper panel) and x-z plane (lower panel) at different simulation times. In the plot, the reference system has been rotated and translated in a way that the x-axis is along the line that joins the stars of the binary system and the origin is at the NS companion. In general, when the SN starts to expand, the faster outermost particles of the SN  will  pass almost without being disturbed by the NS gravitational field. The slower-moving material is gravitationally captured by the NS, initially forming a tail and ultimately forming a thick disk around it. In addition, there are particles from the innermost layers of the SN-ejecta that do not have enough kinetic energy to scape, leading to fallback accretion onto the $\nu$NS. Then, at some point, the material that have been capture by the NS companion  start to be also attracted by the $\nu$NS and being accreted by it.

To confirm the formation of the disk around the NS companion, in Figure~\ref{fig:Omega25Mzams} has been calculated  the angular velocity profile with respect the NS companion position at different times and for two different directions: the line that joins the binary stars with the $\nu$NS in the $-x$ direction labeled as $\theta=0$; and the line perpendicular to it on the orbital binary plane, labeled as $\theta=\pi/2$. The angular velocity of the particles closed to the NS companion ($r/a_0<0.25$) superpose the Keplerian angular velocity profile. This confirms the estimates from analytical approximations made in \citet{2015ApJ...812..100B}, where it was shown that the SN ejecta have enough angular momentum to circularize around the NS before being accreted.

Figure~\ref{fig:25Mzams_MdotNS} shows the mass accretion rate as a function of simulation time onto the binary system stars: the NS companion, the $\nu$NS and the sum of both. Either the fallback accretion rate or the NS accretion rate are much greater that the Eddington limit. The NS is allowed to accrete at this high rate by the emission of neutrinos at its surface via $e^{+}e^{-}$ pair annihilation that is the most efficient neutrino emission process at these density and temperature conditions \citep[see][for details]{2016ApJ...833..107B}. This allows the matter to cool fast enough to be incorporated onto the star and we can add the mass of the particles that fulfill the accretion conditions (see section~\ref{sec:2.1} and Equations~\ref{eq:updateStar}). As we have shown the SN ejecta might transport a high amount of angular momentum and form a thick disk around the NS before the accretion take place. There, the densities and temperatures are not high enough to cool the matter by neutrino emission and outflows might occur \citep[see, e.g.,]{1999MNRAS.303L...1B,2005ApJ...629..341K,2013ApJ...772...30D}. Up to $25\, \%$ of the infalling matter can be ejected in strong outflows removing much of the system angular momentum \citep{2006ApJ...646L.131F,2009ApJ...699..409F}. This means that the mass-accretion rate calculated here might overestimate the actual accretion rate onto the star but up to a factor of order unity. It is also important to note that the accretion rate directly depends on the value adopted for the $\xi$ parameter in Equation~(\ref{eq:CaptureRadius}). For this simulation we adopted $\xi=0.1$. In {Section~\ref{sec:6.1}}, we are going to vary this parameter and establish the influence of it on the system final fate.

As we anticipated, we have also run the simulation of the SN ejecta expansion without the NS companion, in order to calculate the fallback accretion rate (black dotted line in Figure~\ref{fig:25Mzams_MdotNS}) and compare it with the accretion rate onto the $\nu$NS in the binary simulation. At the beginning of the simulation,  there is no difference between both accretion rates: an almost flatter high accretion phase at early time and then a decline as $t^{-5/3}$ \citep{1989ApJ...346..847C,2008ApJ...679..639Z,2009ApJ...699..409F,2013ApJ...772...30D,2014arXiv1401.3032W}. However, at around $t/P_{\rm orb} \lesssim 1.0$, there is a jump in the fallback accretion rate of the binary simulation, that can be associated with the time at which the $\nu$NS starts to accrete the material decelerated by the NS companion. The high early time accretion rate calculated here, is due to  the fallback of those particles that did not have enough kinetic energy to escape from the $\nu$NS gravitational field. This can occur, either because after the forward shock is launched the proto-NS cools and contracts sending a rarefaction wave to the ejecta that decelerates it \citep{1971ApJ...163..221C}, or because the SN shock is smoothly decelerated when it goes outward pushing the star material out \citep{1999ApJ...522..413F,1995ApJS..101..181W}.

\subsection{SN explosion energy}

\begin{figure*}
	\centering
    \subfigure[NS star mass accretion rate]{\includegraphics[scale=0.5]{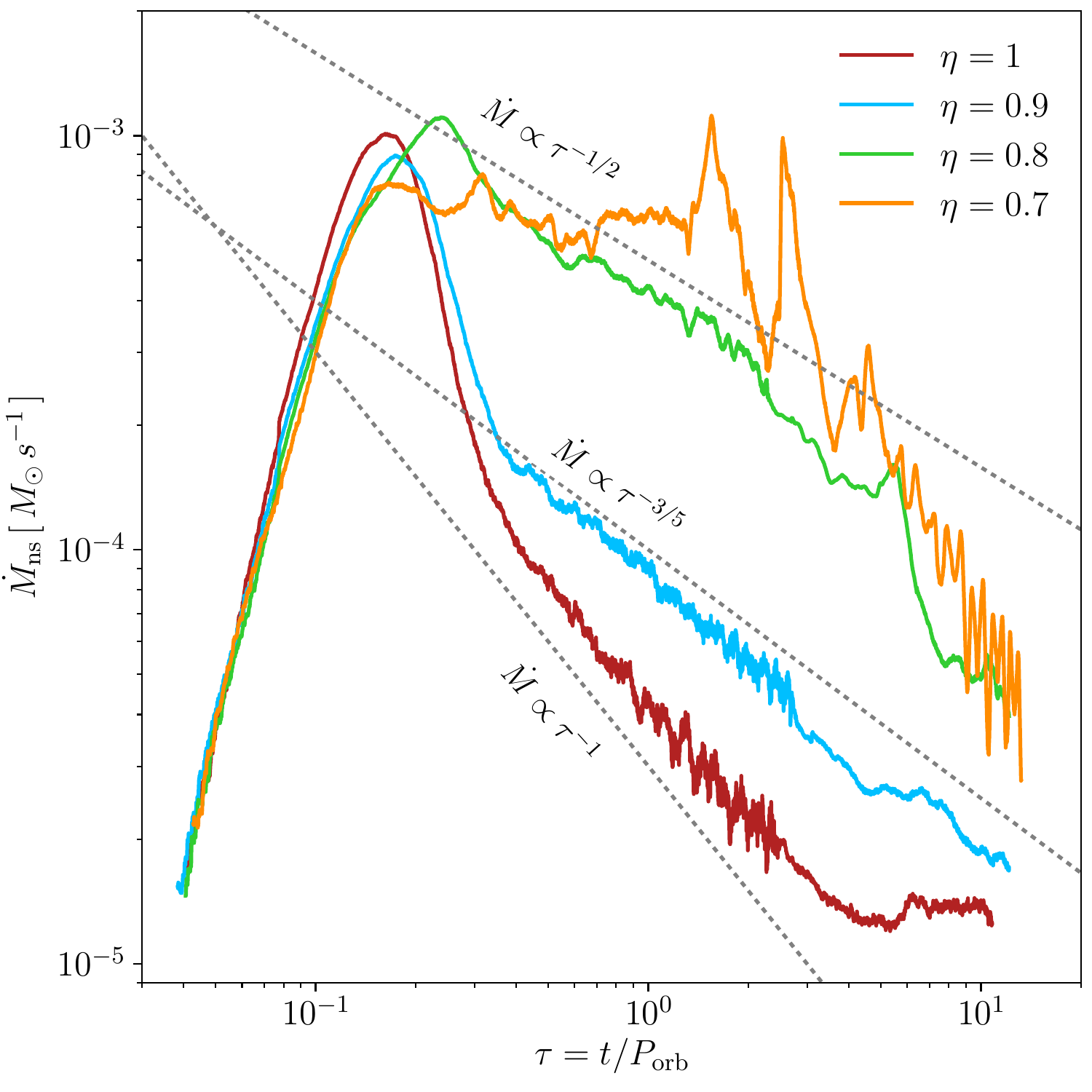}}
    \subfigure[$\nu$NS mass accretion rate]{\includegraphics[scale=0.5]{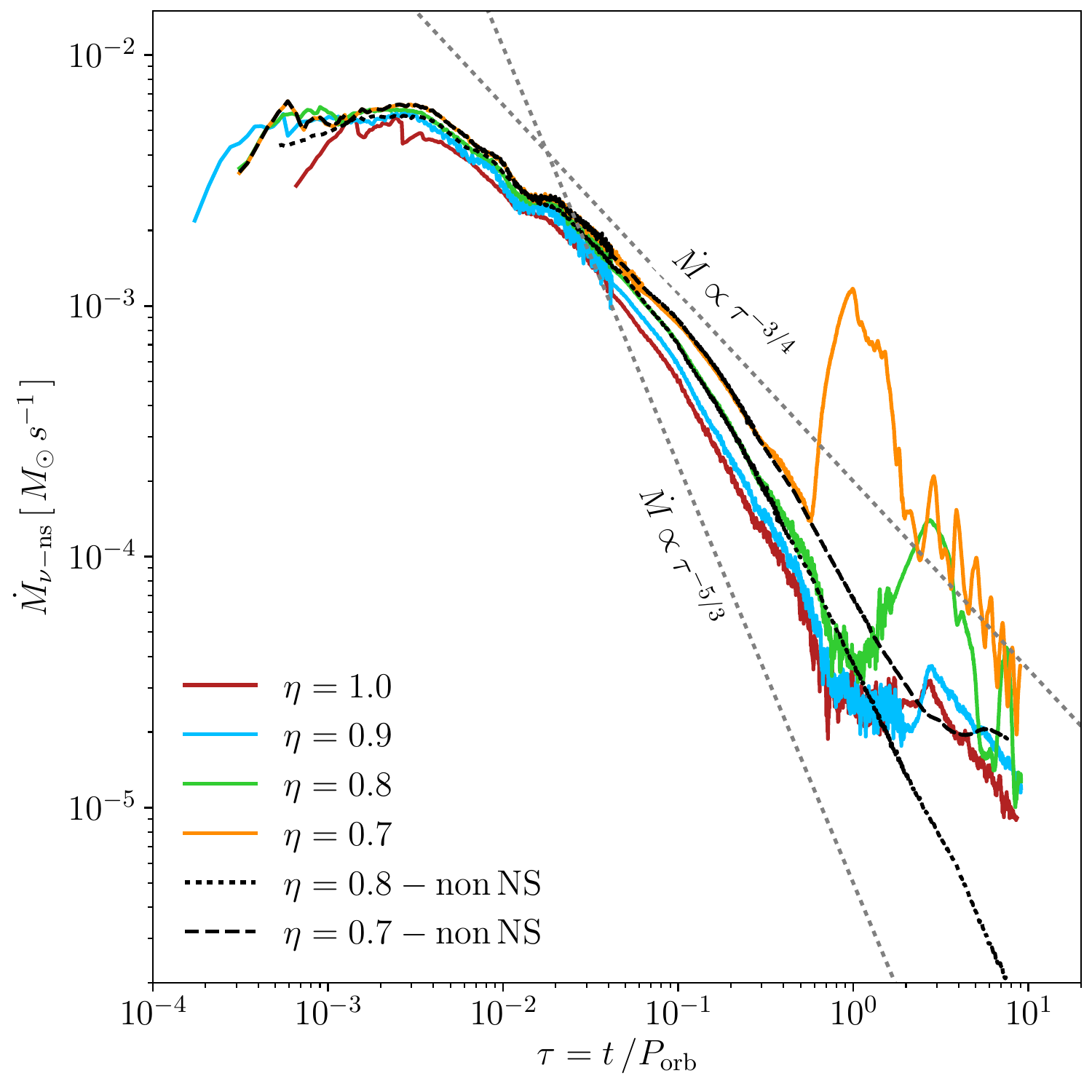}}
    \caption{Mass-accretion rate on the NS (left panel) and the $\nu$NS (right panel) in the IGC scenario. The initial binary system is the same as the one of Figure~\ref{fig:25Mzams_MdotNS} but the SN explosion energy has been varied scaling the kinetic and internal energy of the SPH particles by the factor $\eta<1$. The accretion rate onto the NS presents a peak for the more energetic SN, while in the weaker ones this peak is flattened, i.e. the accretion happens at a nearly constant rate for a longer time, making the star to increase its mass faster. At early times, the fallback accretion rate onto the $\nu$NS is nearly independent on the SN energy, although the late bump induced by the accretion of the matter gravitational capture by the NS companion is stronger for the weakest explosions.}
	\label{fig:25Mzams_Mnsdot_Energy}
    \vspace{0.5\floatsep}
	\centering
    \subfigure[Surface density]{\includegraphics[scale=0.42]{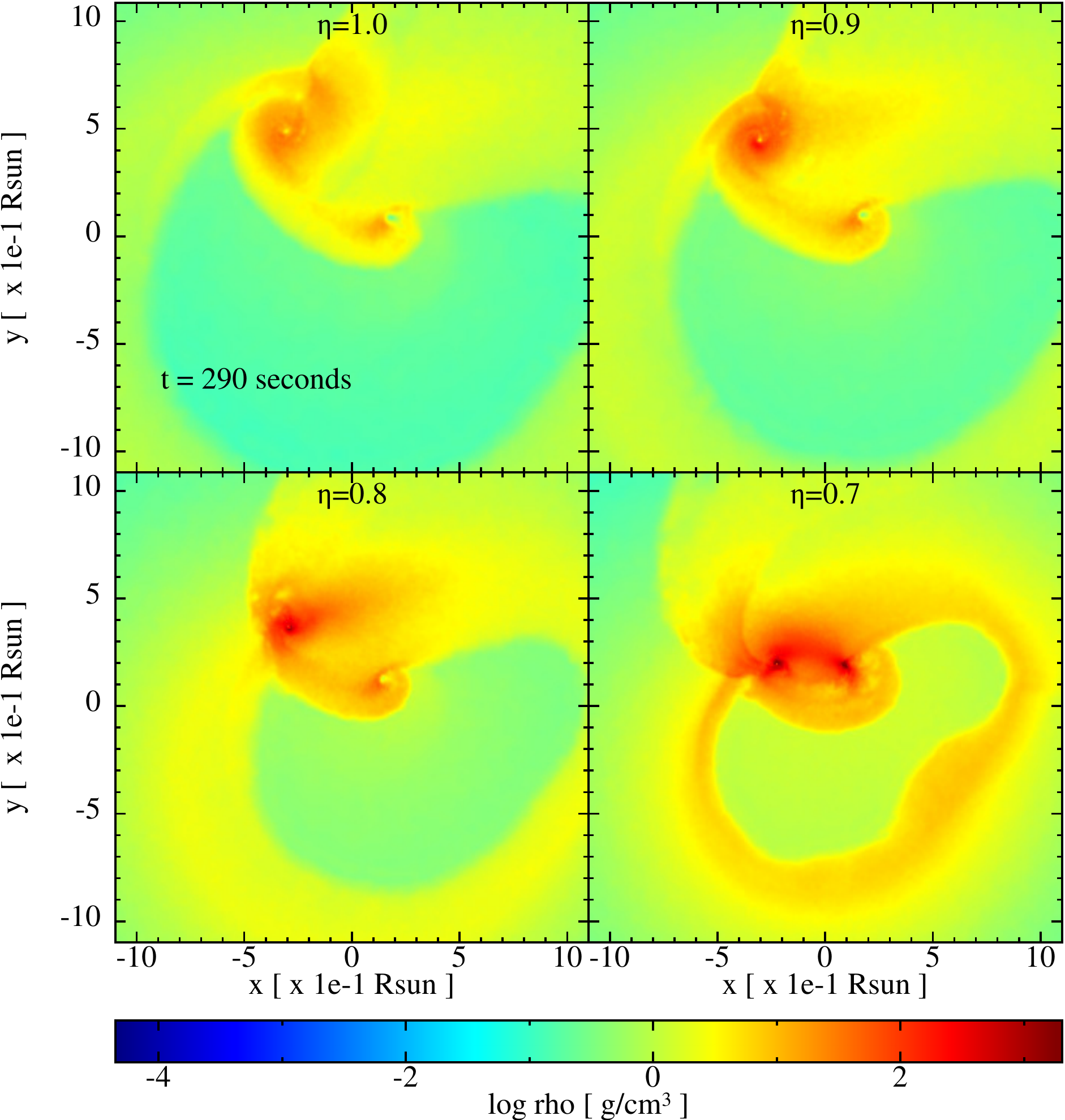}}\hspace{0.3cm}
	\subfigure[Surface specific internal energy]{\includegraphics[scale=0.42]{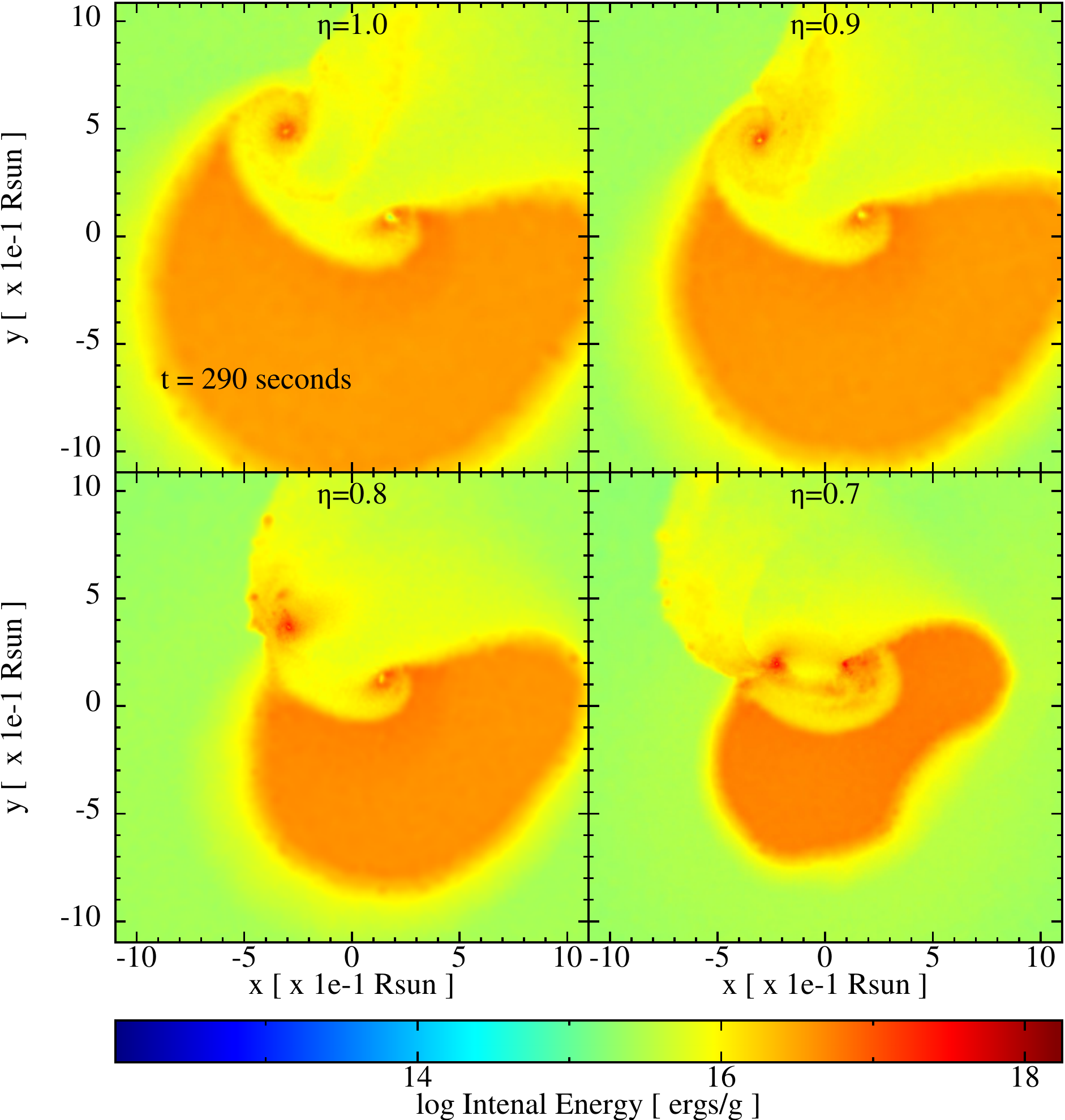}}
  \caption{Snapshots of the mass density (left panel) and the specific internal energy (right panel) on the equatorial plane after $290.0$~s from the beginning of the SPH simulation (around one orbital period of the initial binary system). The initial binary system parameters are the same as the one represented in Figure~\ref{fig:25Mzams2Mns} but the SN explosion energy has been scaled by a factor $\eta$ shown at the upper side of each panel (these simulations correspond to models 25M1p1e with $\eta=1$, 25M1p09e with $\eta=0.9$, 25M1p08e with $\eta=0.8$ and 25M1p07e with $\eta=0.7$ of table~\ref{tab:inconmodels2}). The accretion onto the $\nu$NS as the NS companion is higher for the weaker SN explosions, then the star masses increases faster and the final orbital period of the system shortens. In these explosions also the amount of mass accreted by the $\nu$NS is larger.}
	\label{fig:25Mzams_Energies}
\end{figure*}

In the following, we start to change systematically  the initial parameters that will affect the fate of the final configuration. We will do it one by one, in order to determinate the most favorable conditions that increase the accretion rate onto the NS companion. Figure~\ref{fig:25Mzams_Mnsdot_Energy} shows the mass accretion rate onto the NS and onto the $\nu$NS for different energies of the SN explosion, with the same progenitor star: the $25\, M_\odot$ of Table~\ref{tab:ProgSN}. As we explained in Section~\ref{sec:2}, in these simulations we scale the kinetic energy and the internal energy by a factor $\eta$ (i.e. the velocities of the particles by $\sqrt{\eta}$) once we map the 1D exploded configuration to the 3D one. As expected, the total mass accreted by the  NS companion is larger for low energetic SN than for high energetic ones, and then  more favorable to the collapse of the NS (see models from 25m1p1e to 25m1p07e in Table~\ref{tab:inconmodels2}). On the other hand, the energy of the SN explosion needs to be enough high, otherwise  a considerable part of the ejected mass causes fallback and can instead induce the collapse of the $\nu$NS to a BH. This is a novel and alternative possibility not considered in the original version of the IGC scenario \citep[see, e.g.,][]{2012ApJ...758L...7R,2014ApJ...793L..36F}. Additionally, the energy of the SN explosion do not have a big influence on the magnitude of the peak of the mass-accretion rate but do have on its shape. The NS companion accretes more mass in the weakest SN explosion ($\eta=0.7$) because the accretion rate is maintained almost constant for a longer time than in the strongest explosion ($\eta=1$) where a clear  peak appears. {We can see that the late decay of the accretion rate depends on the SN explosion energy.}

For these simulations, Figure~\ref{fig:25Mzams_Energies} shows snapshots of the mass density and the specific internal energy on the equatorial plane after about one orbital period of the initial configuration ($\sim 5 $~min). Each panel corresponds to a different value of the $\eta$ parameter: $\eta=1.0$ and $\eta=0.9$ for the left and right upper panels, $\eta=0.8$ and $\eta=0.7$ for the left and right bottom panels. The asymmetries of the interior ejecta layer are more pronounced for the less energetic explosion. The orbital period of the final configuration shortens with the decreases of the SN energy, i.e. with the accretion of mass by the binaries stars. For example, the accretion onto the $\nu$NS and onto the NS companion is around $20\%$ and $16\%$, respectively, more efficient for the weakest explosion ($\eta=0.7$) with respect to the strongest one ($\eta=1$), and the final orbital period is almost $90\%$ shorter than the one of the final system from the most energetic explosion (see table~\ref{tab:inconmodels2}).

\begin{figure*}
  \centering
  \subfigure[Single star ( $\eta=1.0$ )]{ \includegraphics[width=0.48\hsize]{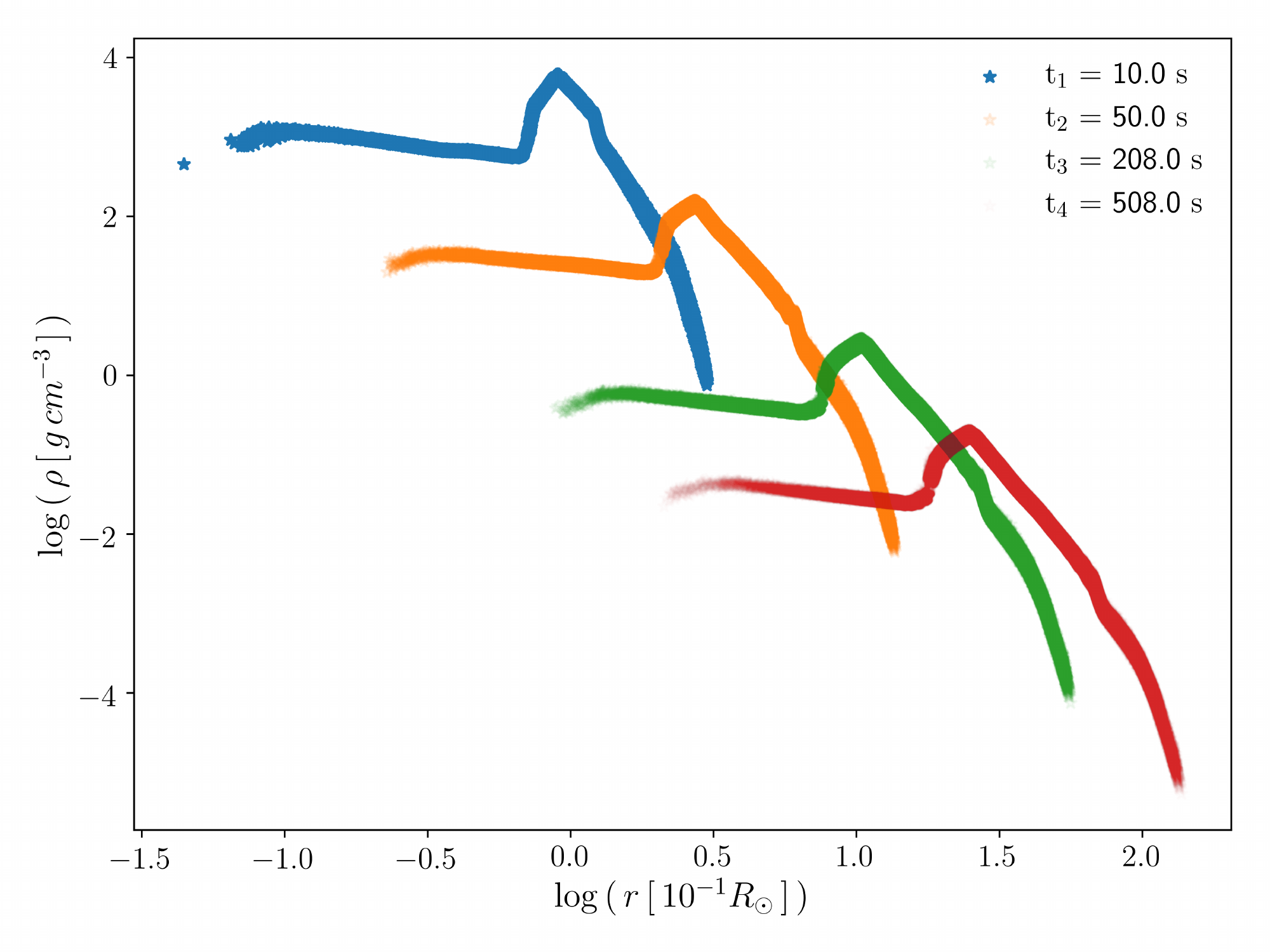}}
  \subfigure[Binary system ( $\eta=1.0$ )]{ \includegraphics[width=0.48\hsize]{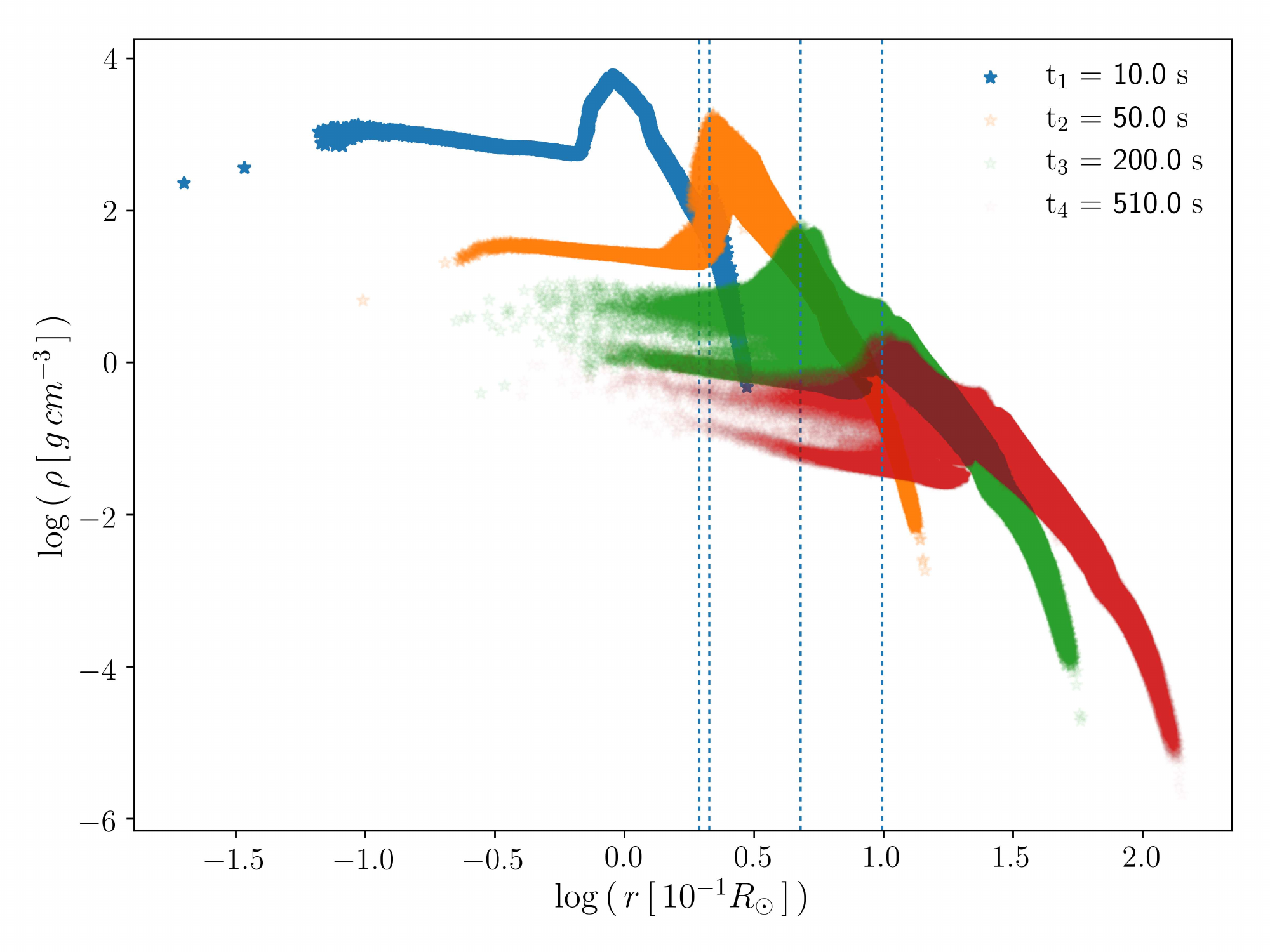}}
  \subfigure[Single star ( $\eta=0.7$ )]{ \includegraphics[width=0.48\hsize]{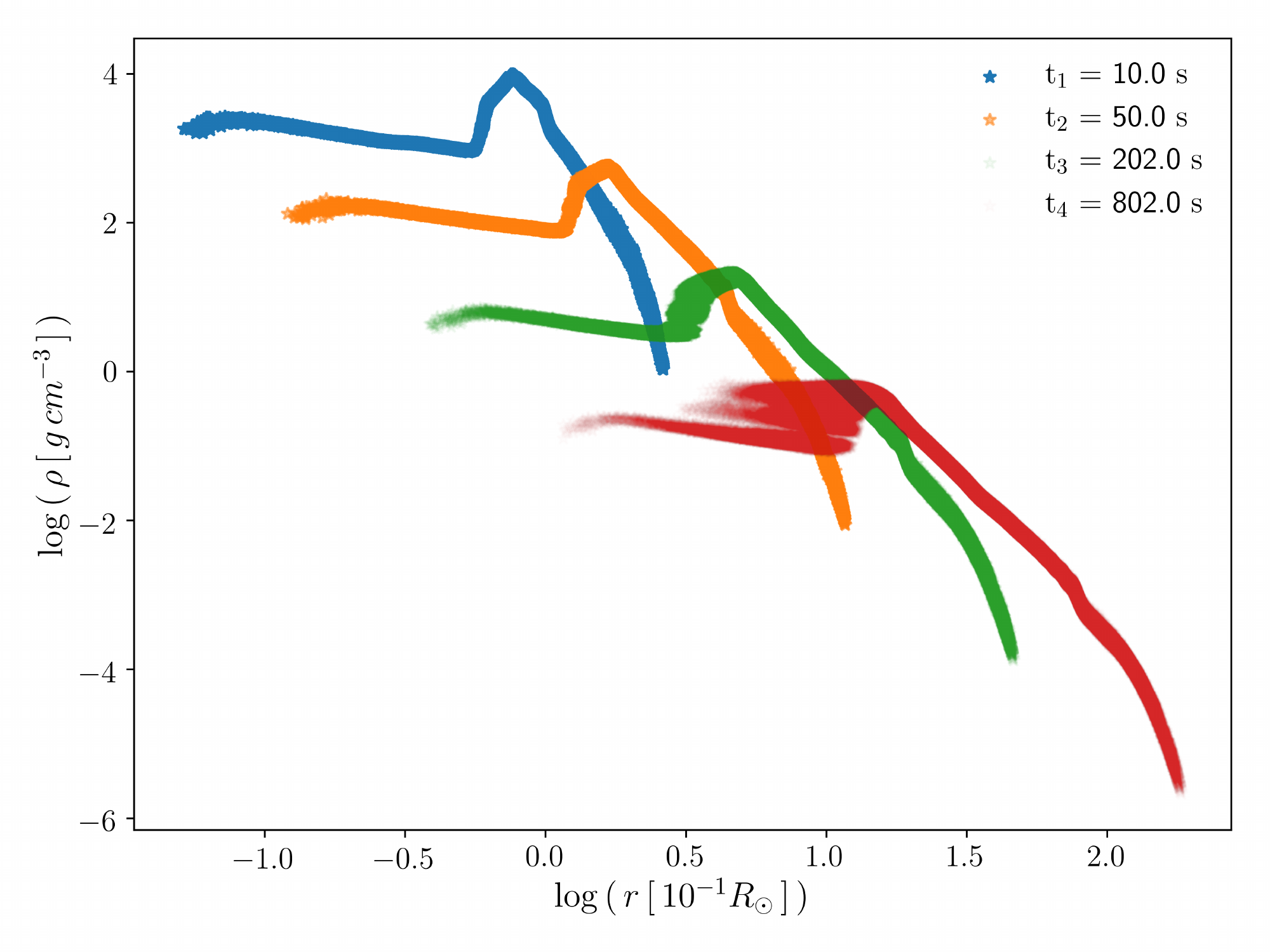}}
  \subfigure[Binary system ($\eta=0.7$ )]{ \includegraphics[width=0.48\hsize]{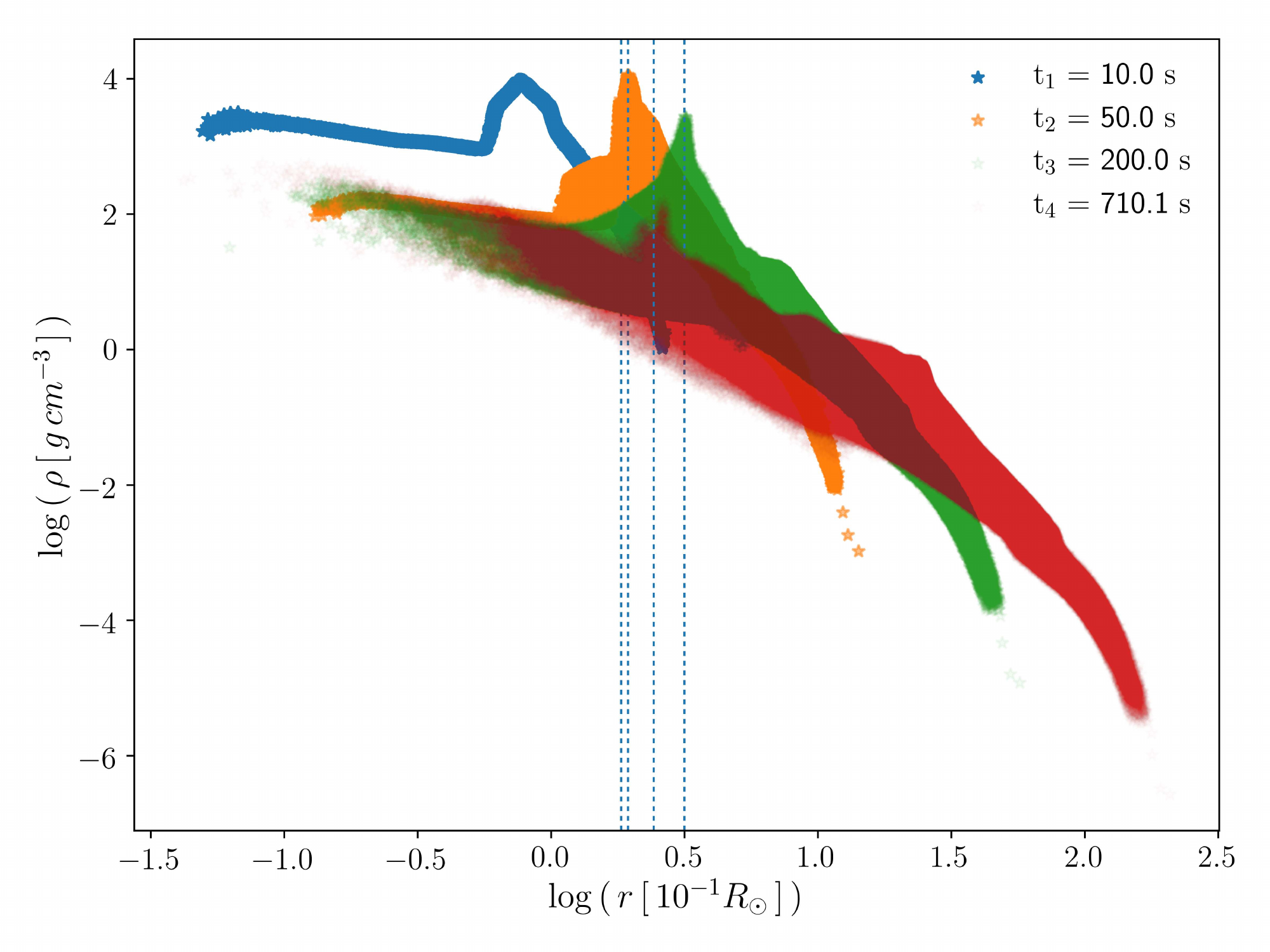}}
  \caption{
  Density profile evolution of the SN ejecta after the core-collapse of the CO$_{\rm core}$ of a $25\,M_\odot$ progenitor. The $r$ coordinate is measured from the $\nu$NS position. The plots at the left panel correspond to the evolution of the SN in a single-star system while, in the ones at the right panel, the CO$_{\rm core}$ belongs to a binary system with a NS  companion of $2\, M_\odot$ and an initial binary separation of $1.36\times 10^{10}$~cm. The blue-dotted lines indicate the position of the NS companion. The SN energy of the upper plots is $1.56\times 10^{51}$~erg (Model 25M1p1e of table~\ref{tab:inconmodels2}) and for the lower plots the SN energy is $6.4\times 10^{50}$~erg (Model 25M1p07e of Table~\ref{tab:inconmodels2}). For isolated SN explosions (or for very wide binaries), the density of the SN ejecta would approximately follow the homologous evolution as is seen in the left panel plots. For explosions occurring in close binaries with compact companions (as it is the case of the IGC progenitors), the SN ejecta is subjected to a strong gravitational field which produces an accretion process onto the NS companion and a deformation of the SN fronts closer to the accreting NS companion, as seen in the right panel plots.}
  \label{fig:25Mzams_rhoProfile}
\end{figure*}

As we did before, we calculate the fallback accretion rate for these explosions onto the $\nu$NS for the isolated progenitors and compare and contrast it with its binary system counterpart. The right panel of Figure~\ref{fig:25Mzams_Mnsdot_Energy} shows the evolution of the mass accretion rate onto the $\nu$NS. The black lines correspond to the single progenitors simulations. The accretion rate peaks at an early time and then decays as {$\dot{M}\propto t^{-5/3}$} \citep{1989ApJ...346..847C}. For the binary simulations (colors lines) a late peak on the fallback accretion rate is produced from the accretion of the material captured by the gravitational field of the NS companion. This is higher and even early for low energetic supernova (around one order of magnitude for the less energetic explosion).

In order to compare the SN evolution in the single and binary simulations, in Figure~\ref{fig:25Mzams_rhoProfile} we show the SN density profile as seen from the $\nu$NS at different times and for two different SN energies (models 25M1p1e and 25M1p07e of table~\ref{tab:inconmodels2}). The left-side plots correspond to the explosion of the CO$_{\rm core}$ of the $25\,M_\odot$ progenitor of table~\ref{tab:ProgSN}, while the right-side plots show its binary counterpart with a $2\, M_\odot$ NS companion. For the isolated SN explosions, the SN ejecta density profiles evolve approximately following an homologous evolution keeping its spherical symmetry around the explosion center. For the explosions occurring in the close binaries, the NS companion gravitational field induced asymmetries in the SN fronts closer to it, that will be more pronounced for the low energetic explosions (see also Figure~\ref{fig:25Mzams_Energies}). We have shown in \citet{2016ApJ...833..107B} that these asymmetries lead to observational effects both in the supernova optical emission and in the GRB X-ray afterglow.

\subsection{Initial binary period}
%
\begin{figure*}
  \centering
  \includegraphics[scale=0.45]{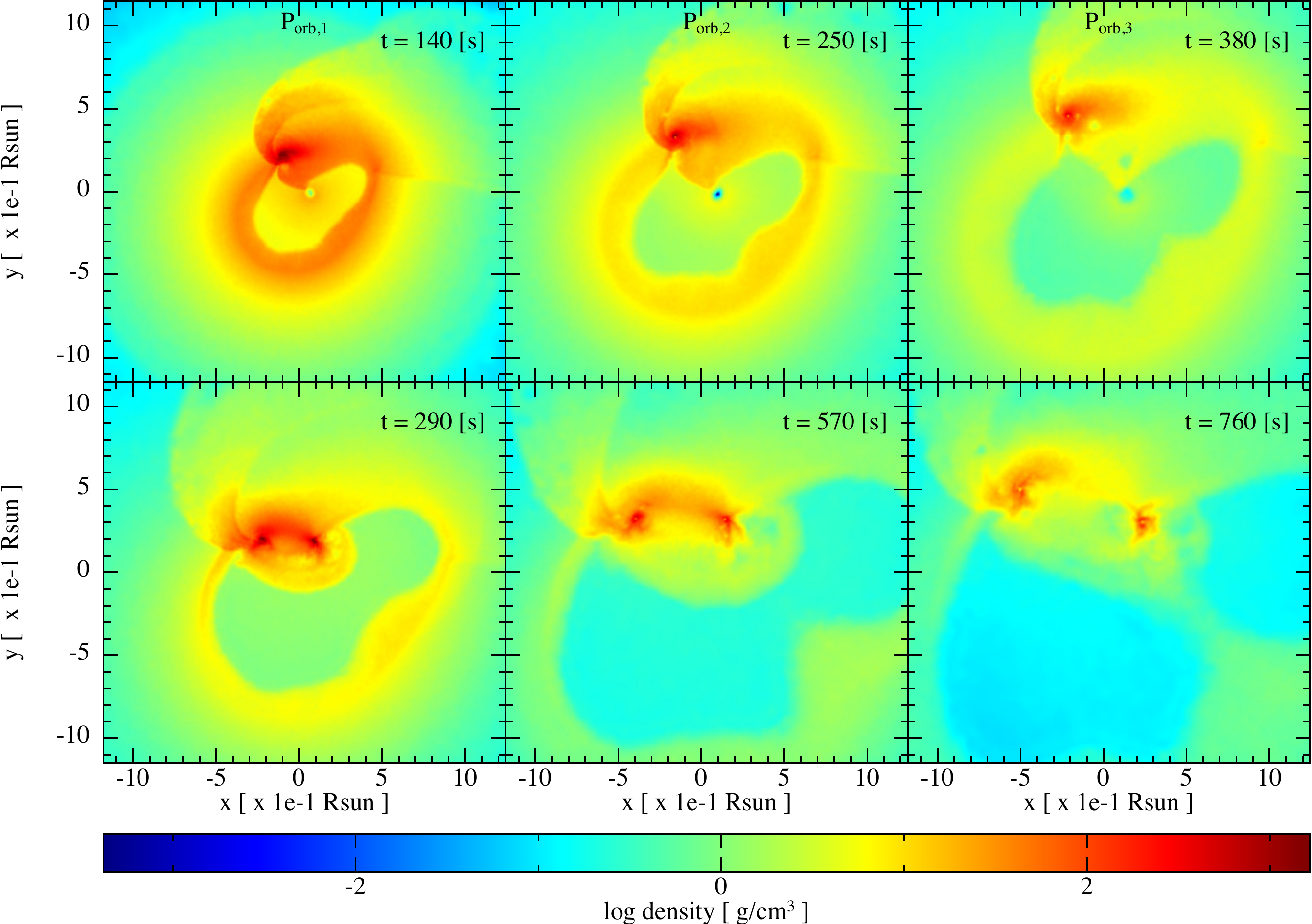}
  \caption{Snapshots of the surface density on the equatorial plane for systems with three different initial binary periods. The initial binary system is formed by the CO$_{\rm core}$ of the $M_{\rm zams}=25\, M_\odot$ progenitor (see table~\ref{tab:ProgSN}) and a $2\, M_\odot$ NS. The SN energy has been reduced to $6.5\times 10^{50}$~erg, scaling the particles velocity and the internal energy by a factor $\eta=0.7$. The periods of the labels are $P_{\rm orb,1}=4.8$~min, $P_{\rm orb,1}=8.1$~min and $P_{\rm orb,1}=11.8$~min that correspond to Models 25m1p07e, 25m3p07e and 25m2p07e of table~\ref{tab:inconmodels2}, respectively.}
  \label{fig:25M580Periods07E}
\end{figure*}

\begin{figure}
  \centering
  \includegraphics[width=0.95\hsize]{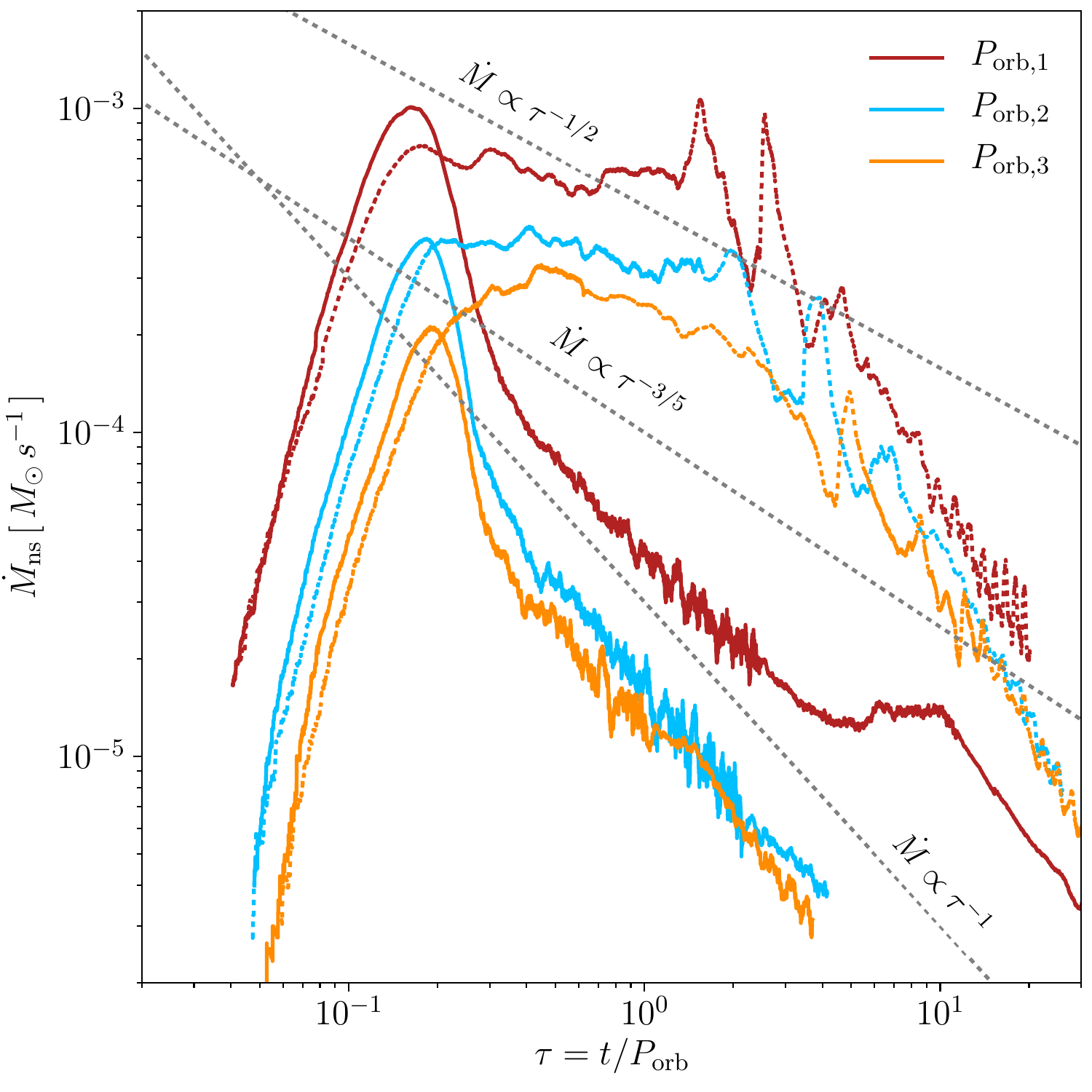}
  \caption{Mass-accretion rate onto the NS companion in the IGC scenario. Different colors correspond to different initial orbital periods: $P_{\rm orb,1}=4.8$~min (red line), $P_{\rm orb,1}=8.1$~min (blue line), $P_{\rm orb,1}=11.8$~min (orange line). The other parameters that characterize the initial binary system are the same as in Figure~\ref{fig:25Mzams_MdotNS}. The solid lines correspond to a SN energy of $1.57\times 10^{51}$~erg, while the dotted ones correspond to a lower SN energy of $6.5 \times 10^{50}$~erg. It can be seen that the mass-accretion rate scales with the binary orbital period.}
  \label{fig:25Mzmans_Period}
\end{figure}

We continue the exploration of the parameter space of the initial binary configuration by running simulations with different values of the initial orbital period. Figure~\ref{fig:25Mzmans_Period} shows the mass-accretion rate onto the NS companion for three different initial orbital periods: $4.8$~min, $8.1$~min and $11.8$~min, and for two different SN energies: the fiducial explosion (with $\eta=1$) and the $\eta=0.7$ modified explosion. For the two explosion energies, the accretion rate seems to scale with the initial binary period of the configuration and follow the same power law at the late times of the accretion process. {On the other hand, for longer binary periods, as expected, the accretion onto the $\nu$NS tends to equal the fallback accretion when the CO$_{\rm core}$ explodes as an isolated star. This can be seen by comparing the final mass of the $\nu$NS in both scenarios (columns $M_{\nu{\rm ns,fb}}$ and $M_{\nu{\rm ns}}$ in Table~\ref{tab:inconmodels2}).}

Figure~\ref{fig:25M580Periods07E} shows snapshots, at two different times, of the surface density on the orbital plane for the same initial binary periods of Figure~\ref{fig:25Mzmans_Period} and the modified explosion with $\eta=0.7$. The system appears to evolve self-similarly with the increase of the binary period.

%
\subsection{Initial mass of the NS}
%
%
{
Up to now we have considered a massive initial NS companion of $2.0\, M_\odot$, since we expect the progenitor of the CO$_{\rm core}$ losses its hydrogen and helium layers interacting with its companion (through common envelope and Roche-lobe overflow episodes \citep{2014NuPhA.928..296L}). Observationally, the measured NS masses in double NS binaries are lighter than $1.5\,M_\odot$ \citep{2007PhR...442..109L} and massive, $\sim 2~M_\odot$ NSs have been measured in binaries with white dwarf companions, i.e.~PSR J1614--2230 \citep{2010Natur.467.1081D} and PSR J0348+0432 \citep{2013Sci...340..448A}.
}

{
In order to increase the parameter space, we have also run simulations with different initial masses for the NS companion: $1.4\, M_\odot$ and $1.6\, M_\odot$. In Table~\ref{tab:NSmass} we summarize the results of these simulations. In these cases, the models are labeled as ``$x_1$Mns$x_2$p$x_3$e", where $x_1$ is the initial mass of the NS companion, $x_2$ and $x_3$ have the same meaning as the model labels of Table~\ref{tab:inconmodels2}. We report also the same columns as in Table~\ref{tab:inconmodels2}. For the progenitor of the CO$_{\rm core}$ we used $M_{\rm zams}=25\,M_\odot$.
}

\begin{figure}
  \centering
  \includegraphics[width=0.95\hsize]{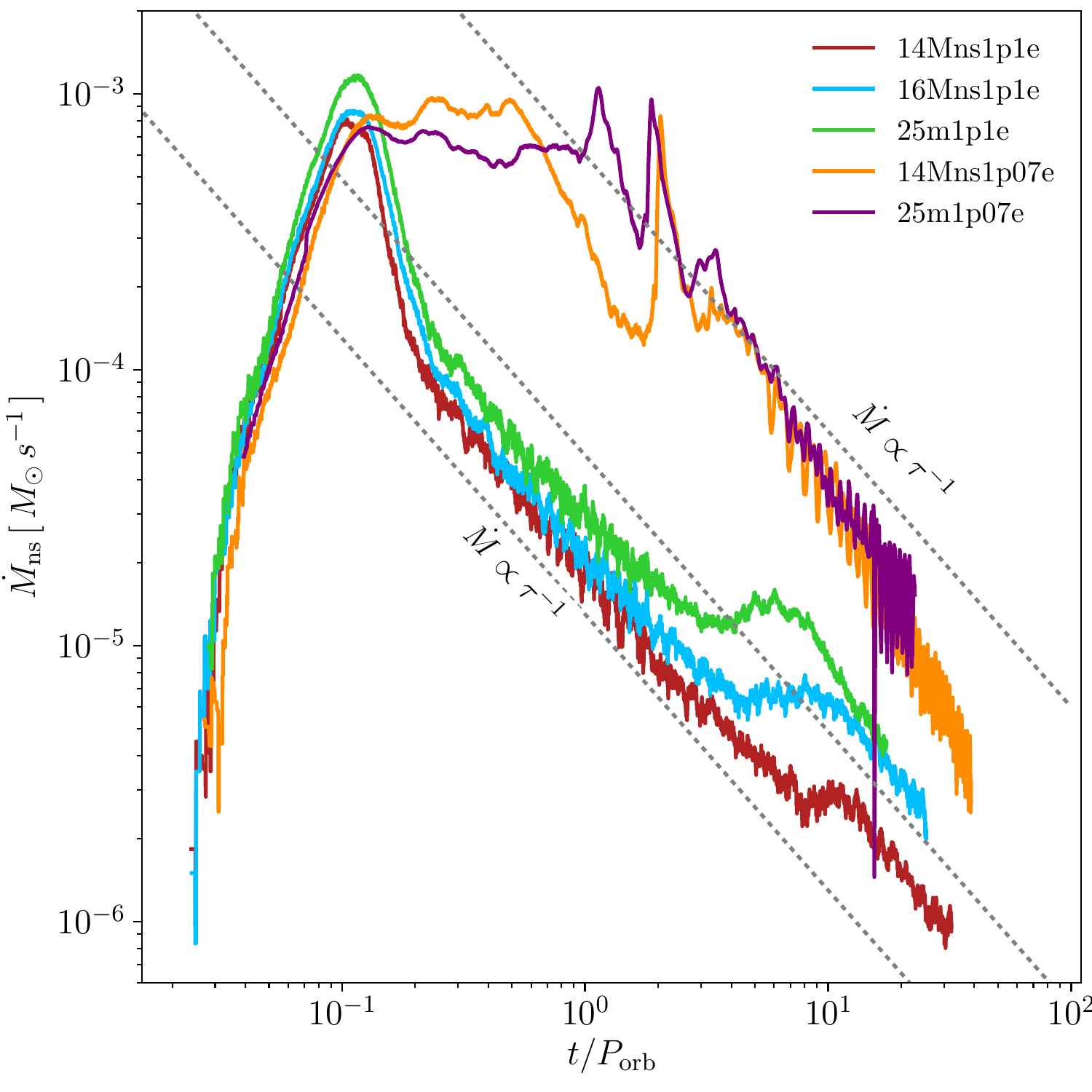}
  \caption{Mass-accretion rate on the NS companion for different initial masses. The  initial orbital period is close to the minimum period that the system can have in order that there is no Roche Overflow before the collapse of the CO$_{\rm core}$.}
  \label{fig:Mdot_initialNSmass}
\end{figure}

{
Figure~\ref{fig:Mdot_initialNSmass} shows the evolution of the mass accretion rate onto the NS companion as a function of the time normalized to the initial orbital period. As expected, for an initial binary system with a massive NS companion, the mass accretion rate onto it increases during the SN expansion, i.e.~the gravitational force due to a massive NS is stronger. For example, the NS companion gained around $4.2$\%, $3.9$\% and $3.1$\% of its initial mass for the 25Mns1p1e, 16Mns1p1e and 14Mns1p1e models, respectively and $36.3$\% and $31.1$\% for the 14Mns1p07e and 25Mns1p07e, respectively. On the other hand, the evolution of the mass accretion rate seems to not be influenced by the NS companion initial mass. For the most energetic explosion ($\eta=1.0$), there is an early peak, follow by a power-law decay and a late bump. This bump is early and bigger for a more massive NS companion. For the less energetic explosion ($\eta=0.7$), there is a nearly constant mass accretion phase, followed by some oscillations and a power-law decay. Finally, the total accreted mass on the $\nu$NS from the fallback seems not to be affected by the initial mass of the NS companion. In the case of the most energetic explosion, the $\nu$NS final mass, in these simulations, is around $1.9\,M_\odot$ and for the less energetic one is around $2.4\,M_\odot$. The differences between the simulations of equal SN energy could be due to numerical errors.
}
\begin{table*}
\centering
\caption{{SPH Simulations with different mass for the NS companion.}}
\setlength{\tabcolsep}{1.55pt}
\renewcommand{\arraystretch}{1.25}
\footnotesize
\begin{tabular}{cccccccccccc}
  \hline
  \hline
  Model   & $M_{\rm ns,0}$ & $P_{\rm orb,i}$ & $a_{\rm orb,i}$  & $M_\mathrm{\nu ns}$ & $M_{\rm ns}$ &  $V_{\rm CM}$ &  $P_\mathrm{orb,f} $ &  $a_\mathrm{orb,f}$& $e$ & $m_{\rm bound}$& bound   \\
     & ($\, m_\odot$\,) & $(\, \mathrm{min}\,)$ & $(\, 10^{10}\, \mathrm{cm}\, )$ &  ($\, m_\odot\,) $ & $(\, m_\odot$\,) & $(\,10^7\,\mathrm{cm/s}\,)$ & $(\, \mathrm{min}\,)$ & $(\, 10^{10}\, \mathrm{cm}\, )$ & & $(\, M_\odot\,)$  &  \\
   \hline
   \hline
  14Mns1p1e & $1.4$ & $4.29$ &  $1.216$ & $1.971$  & $1.443$  & $6.47$ & $293$  & $15.28$ & $0.929$ & $4.2\times 10^{-3}$& yes\\
  16Mns1p1e & $1.6$ &$4.35$ &  $1.244$ & $1.956$  & $1.664$ & $7.19$ &  $1872$ & $53.566$  & $0.979$ &  $5.1\times 10^{-2}$ &yes \\
  14Mns1p07e & $1.4$ & $4.29$& $1.216$  & $2.457$   & $1.909$ & $3.33$ & $4.20$ &  $0.977$&  $0.375$ &$9.8\times 10^{-2}$ &yes \\
 \hline
 \hline
\end{tabular}
\tablecomments{For the progenitor of CO$_{\rm core}$ we used the one with $M_{\rm zams}=25\, M_\odot$.}
\label{tab:NSmass}
\end{table*}
%

\subsection{Asymmetric SN expansion}

\begin{table*}
\centering
\caption{{SPH Simulations with an asymmetric blastwave.}}
\setlength{\tabcolsep}{1.55pt}
\renewcommand{\arraystretch}{1.25}
\footnotesize
\begin{tabular}{ccccccccccc}
  \hline
  \hline
  Model   & $M_\mathrm{\nu ns,fb}$ & $V_{\rm kick}$ &$M_\mathrm{\nu ns}$ & $M_{\rm ns}$ &  $V_{\rm CM}$ &  $P_\mathrm{orb,f} $ &  $a_\mathrm{orb,f}$& $e$ & $m_{\rm bound}$& bound   \\
   & $(\, m_\odot \,)$&$(\,10^4\, \mathrm{cm/s}\,)$& $ (\, m_\odot\,) $ & $(\, m_\odot$\,) & $(\,10^7\,\mathrm{cm/s}\,)$ & $(\, \mathrm{s}\,)$ & $(\, 10^{10}\, \mathrm{cm}\, )$ & & $(\, M_\odot\,)$  &  \\
   \hline
   \hline
  2f20$\Theta:$ z    & $1.931$ & $1.18\times 10^3$ & $2.009$ & $2.161$ & $7.31$ & $2852.15$ & $4.849$ & $0.774$ & $0.065$ & yes\\
  2f20$\Theta:$ x    & $1.931$ & $1.18\times 10^3$ & $1.959$ & $2.142$ & $7.13$ & $8225.79$ & $9.772$ & $0.892$ & $0.056$ & yes\\
  2f20$\Theta:$ -x   & $1.931$ & $1.18\times 10^3$ & $2.002$ & $2.182$ & $7.82$ & $3170.17$ & $5.209$ & $0.801$ & $0.054$ & yes\\
  4f20$\Theta:$ z   & $2.826$ & $5.08\times 10^3$ & $2.382$ & $2.424$ & $5.16$ & $456.51$ & $1.496$ & $0.438$ & $0.189$ & yes\\
  2f40$\Theta:$ z   & $2.364$ & $5.38\times 10^3$ & $2.316$ & $2.395$ & $5.84$ & $529.37$ & $1.643$ & $0.589$ & $0.105$ & yes\\
\hline
\hline
\end{tabular}
\tablecomments{The initial binary system is formed by the CO$_{\rm core}$ of the $M_{\rm zams}=25\,M_\odot$ progenitor and a NS of $2\,M_\odot$. The initial binary parameters are the same of the model 25m1p1e of table~\ref{tab:inconmodels2} but the SN velocity profile was modified with equations~(\ref{eq:AssyVel_in}) and (\ref{eq:AssyVel_out}).}
\label{tab:assy_table}
\end{table*}

\begin{figure}
  \centering
  \includegraphics[width=0.9\hsize]{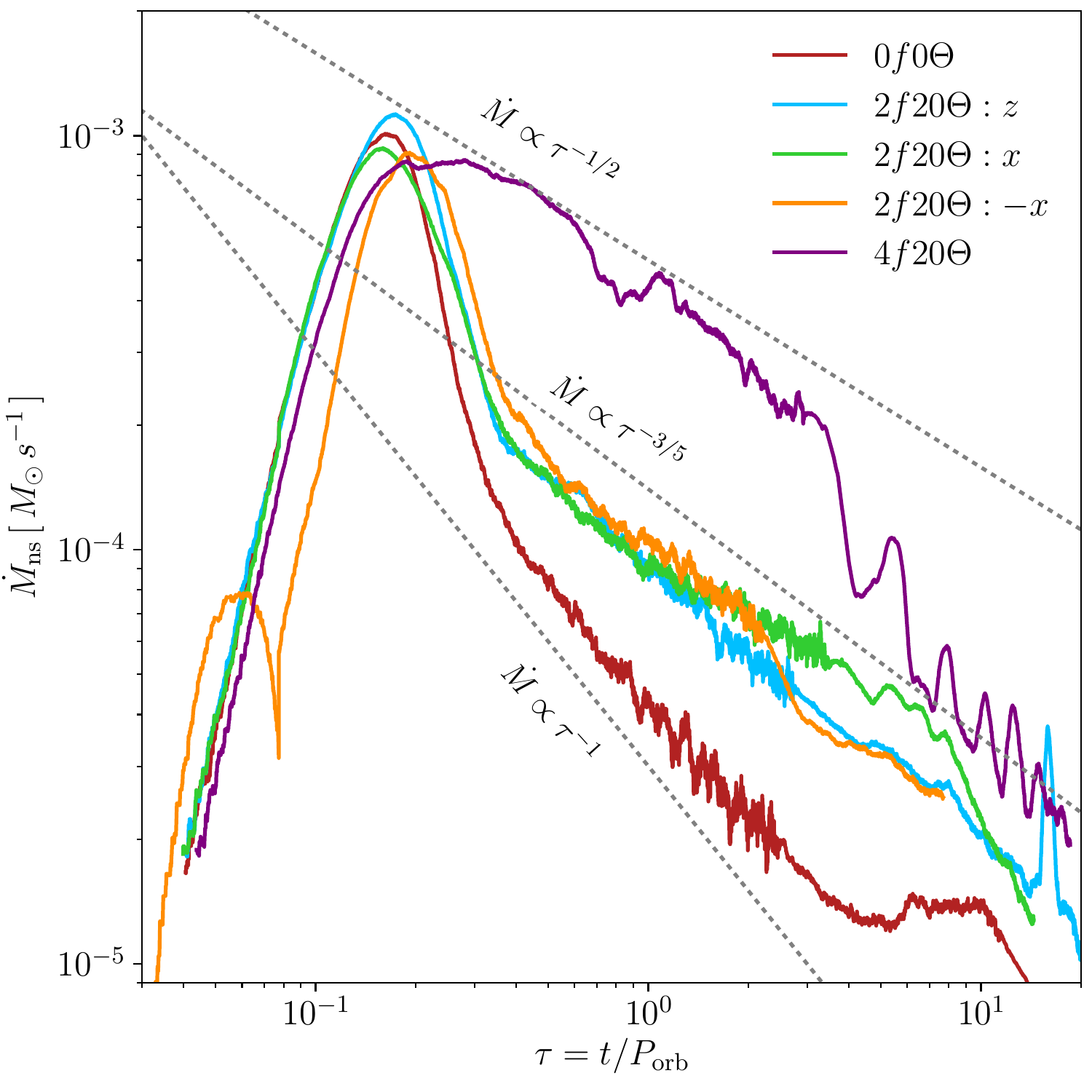}
  \caption{Mass-accretion rate onto the NS companion introducing a conical geometry for the SN velocity profile of the $M_{\rm zams}=25\,M_\odot$ progenitor (see Table~\ref{tab:ProgSN}). The parameter of the initial binary system are the same as the one for Figure~\ref{fig:25Mzams_MdotNS}. The cone was opened along the $z$-axis (perpendicular to the orbital plane-blue and purple lines) and along the $x$-axis (on the orbital plane-green and orange lines). Since the SN energy is conserved, the introduction of the asymmetry reduces the particles velocity and increases the accretion rate.}
  \label{fig:25Mzmans_Assymetries}
\end{figure}

We now explore how a SN explosion with an asymmetry blastwave affects the evolution of the system. {In Table~\ref{tab:assy_table} we summarize the results of these simulations. In these cases, the models are labeled as ``$x_1$f$x_2\Theta$:$x_3$", where $x_1$ and $x_2$ are the values of the parameter $f$ and $\Theta$, respectively, of Equations~(\ref{eq:AssyVel_in}) and (\ref{eq:AssyVel_out}), and $x_3$ is the direction of the lobe}. We started the simulations with the parameters $f=2.0$ (in-cone to out-cone velocity ratio) and $\Theta=20.0$ (cone amplitude), with the direction of the lobe in the z-axis (perpendicular to the equatorial binary plane) and in the +x-axis (directed to the NS companion) and -x-axis (opposed to the NS companion). Additionally, we explore the dependence with the opening angle and the velocities ratio, running simulations with $f=4.0$ and with $\Theta=40$. Figure~\ref{fig:25Mzmans_Assymetries} shows the accretion rate onto the NS companion for these simulations. We can see that the introduction of asymmetries in the SN expansion velocity increases the accretion rate onto the NS companion as well as the fallback accretion rate onto the $\nu$NS. This is expected  because, in order to conserved the SN energy explosion, we increased the velocities of the particles inside the cone while we decreased the velocities of the particles out-side it, then these slower particles are more probably captured by the stars. The direction on the lobe does not introduce great changes in the evolution of the accretion rate or the final star mass of the configurations but it does when the parameters $f$ or $\Theta$ are increased.

\begin{figure*}
  \centering
  \subfigure[]{\includegraphics[scale=0.42]{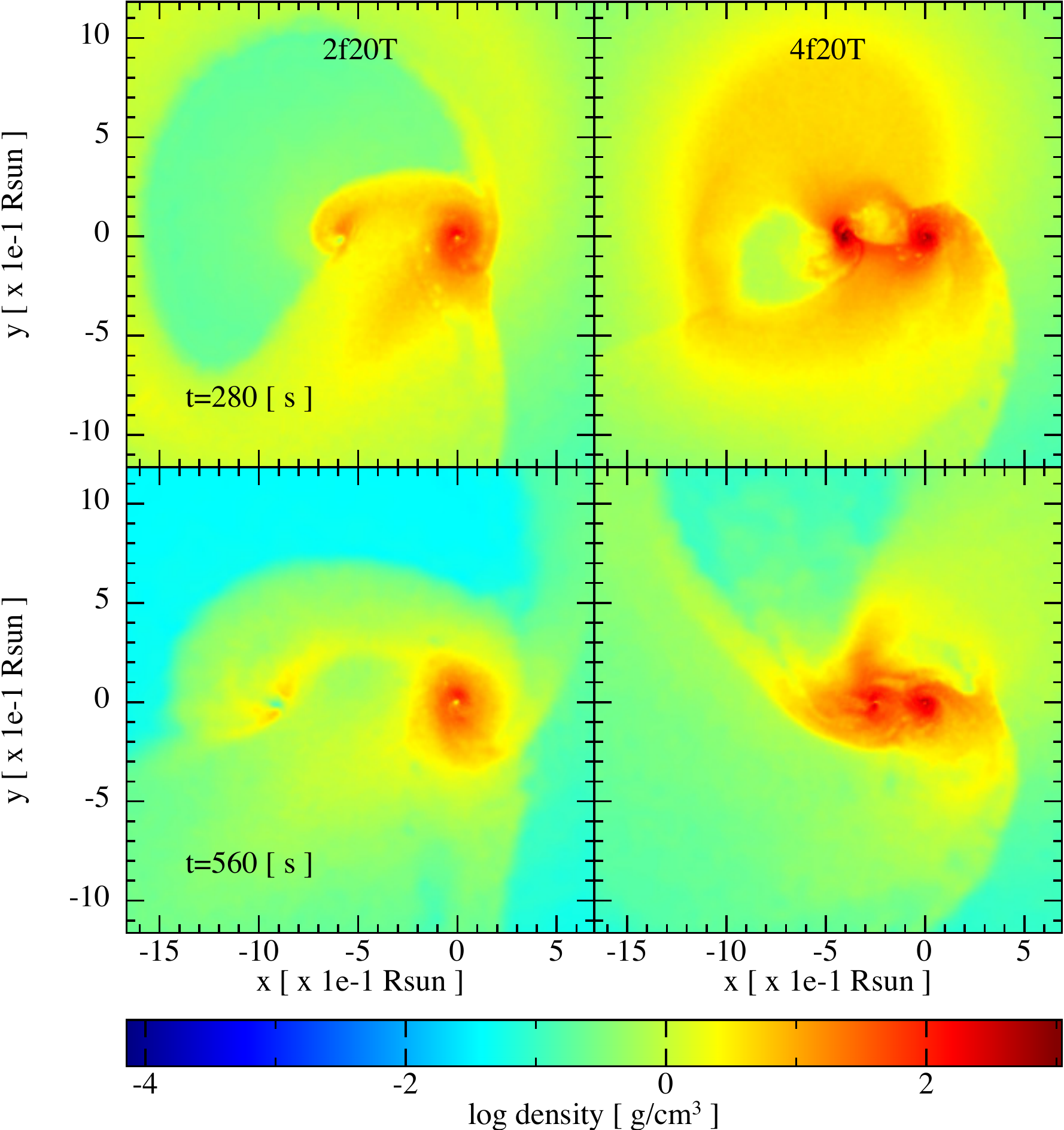}}\hspace{0.2cm}
  \subfigure[]{\includegraphics[scale=0.42]{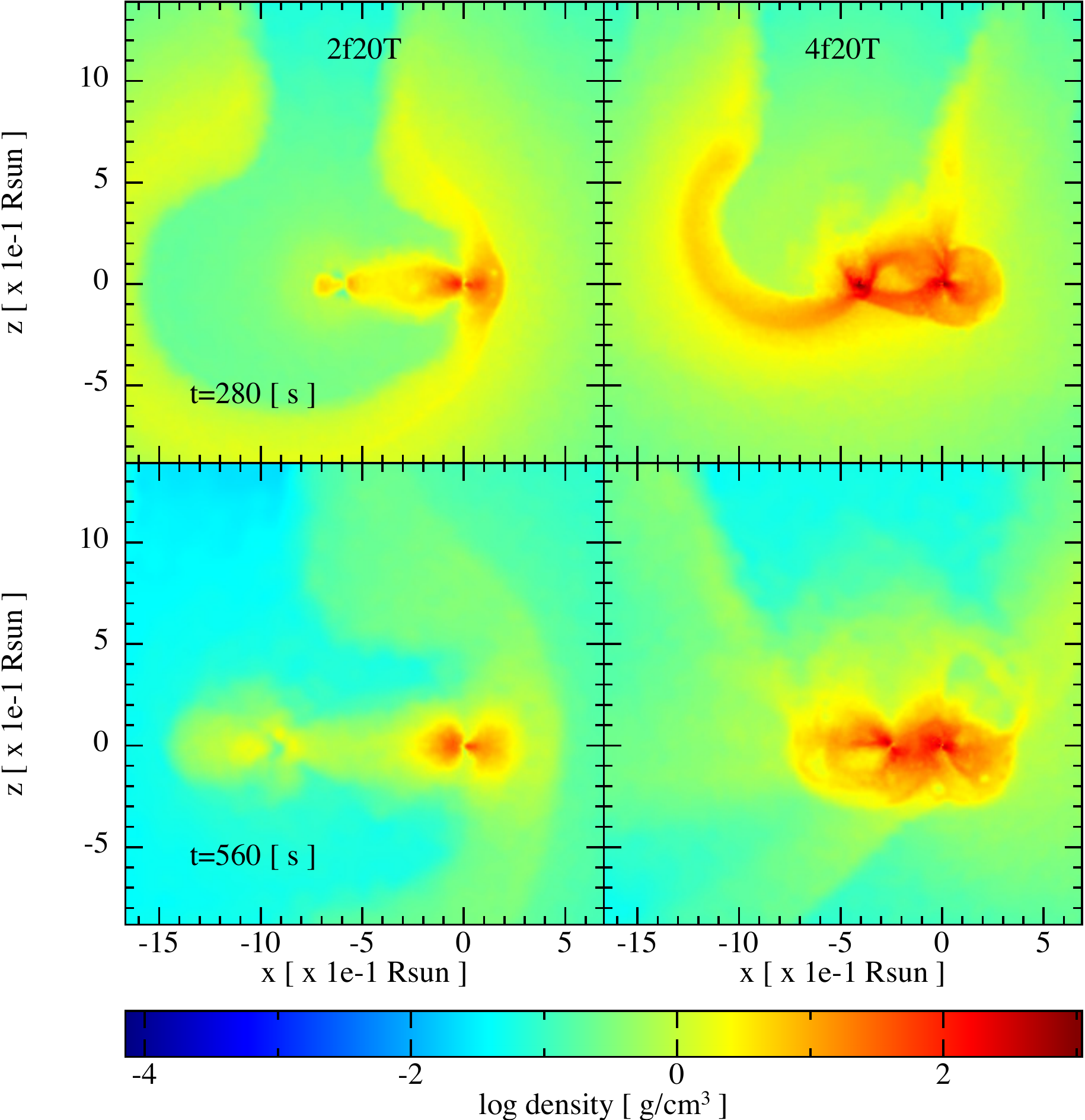}}
  \caption{Snapshots of the surface density on the binary equatorial plane (left panel) and the plane orthogonal to it (right panel). The reference system have been rotated and translated in a way that the $x$-axis becomes directed in the line that joins the binary stars and its origin is at the  NS companion companion. The initial binary system is the same as the one represented in Figure~\ref{fig:25Mzams2Mns} but the SN velocity profile has been modified to a conical geometry following Equations~(\ref{eq:AssyVel_in}) and (\ref{eq:AssyVel_out}). In all the cases, the cone opens along the $z$-axis. The left frames of each plot have parameters $f=2.0$ and $\Theta=20.0$ (Model 25m2f20tz of table~\ref{tab:inconmodels2}) while the right ones have $f=4.0$ and $\Theta=20.0$  (Model 25m4f20tz of table~\ref{tab:inconmodels2}). If the lobe of the explosion is directed outside of the initial orbital plane, as in here, the orbital plane of the final configuration changes. For the case $f=2.0$, the final orbital plane makes an angle of $2.55^\circ$ with respect to the initial orbital plane and for $f=4.0$, this angle grows up to $11.5^\circ$.}
  \label{fig:25M580Assy}
\end{figure*}

Figure~\ref{fig:25M580Assy} shows snapshots of the surface density for the simulations of the asymmetric SN expansion. As it was previously done, the reference system was rotated and translated in the way that its center  corresponds to the NS companion position and the x-axis joins the binary stars. Contrary to the symmetry cases, the binary orbital plane change after the SN explosion if the lobe of the explosion is outside the equatorial plane of the initial binary. For example, for $f=2.0$, the final orbital plane makes an angle of $2.55^\circ$ with respect to the initial orbital plane and for $f=4.0$, this angle grows up to $11.5^\circ$ . However, either for the symmetry or the asymmetry explosion, the magnitude of the velocity of the center-of-mass of the final binary system remains around  $100$--$800$~km~s$^{-1}$.
The kick velocities given in Table~\ref{tab:inconmodels2} are due to the accretion of linear momentum from the accreted particles and the gravitational attraction that the ejecta material do on the $\nu$NS \citep{1994A&A...290..496J}. Another source for the NS kick velocity can be due to the anisotropic emission of the neutrino during the star collapse \citep{1987IAUS..125..255W,1993A&AT....3..287B,2006ApJS..163..335F}.
\begin{figure}
  \centering
  \includegraphics[width=0.9\hsize]{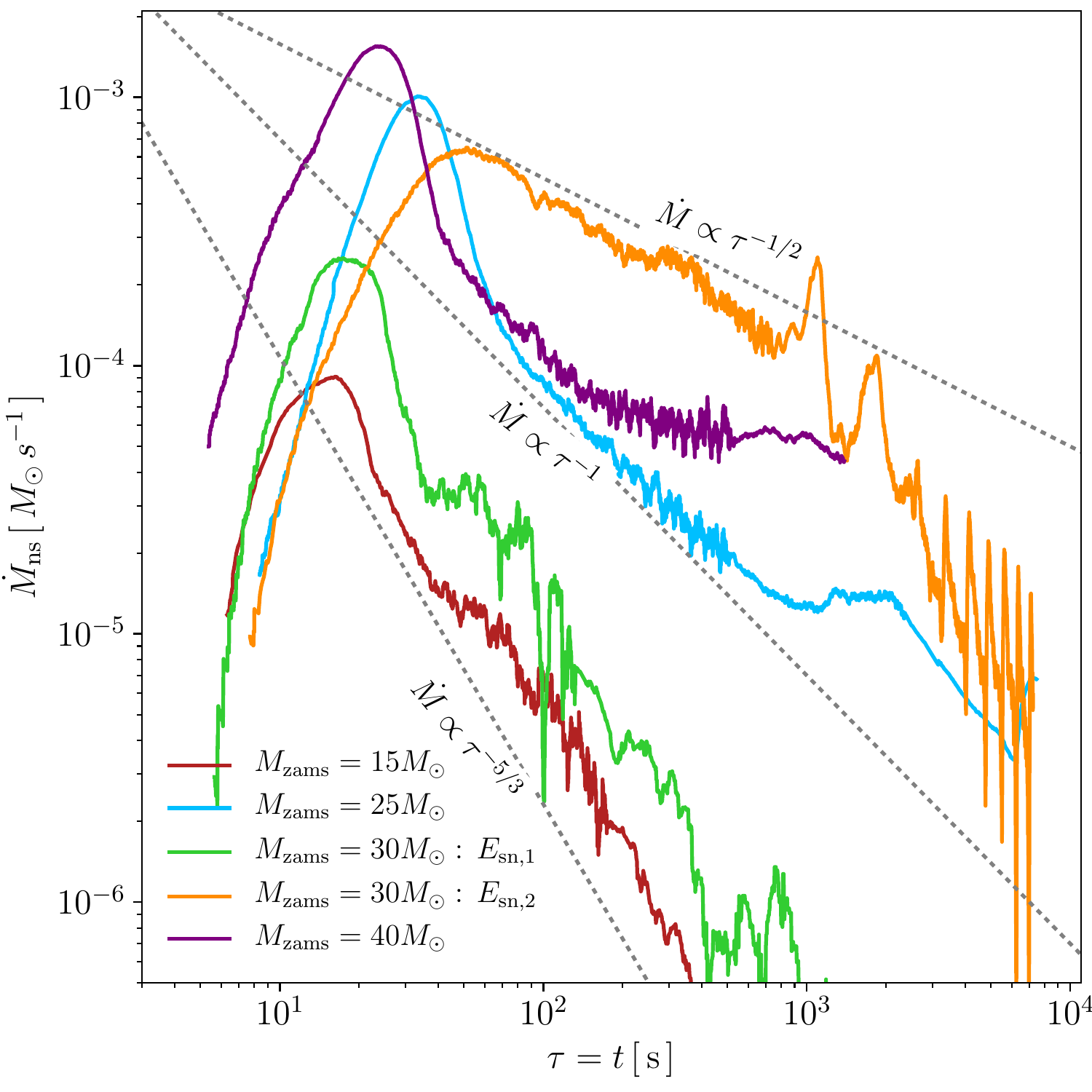}
  \caption{Mass-accretion rate on the NS companion using the explosion of all the CO$_{\rm core}$ progenitors summarized in Table~\ref{tab:ProgSN}. The NS companion has an initial mass of $2\,M_\odot$ and the orbital period is close to the minimum period that the system can have in order that there is no Roche-lobe overflow before the collapse of the CO$_{\rm core}$: $6.5$~min, $4.8$~min, $6.0$~min and $4.4$~min for the $M_{\rm zams}=15 M_\odot$, $25 M_\odot$, $30M_\odot$ and $40M_\odot$ progenitors, respectively.}
  \label{fig:Mdot_Progenitor}
\end{figure}
%

\subsection{CO$_{\rm core}$ progenitor mass}

Finally, we have varied the progenitor of the CO$_{\rm core}$. Figure~\ref{fig:Mdot_Progenitor} shows the mass-accretion rate onto the NS companion for all the progenitors listed in Table~\ref{tab:ProgSN}. In Table~\ref{tab:inconmodels2} we summarize the results for these simulations, and additionally we have run more simulations with each of these progenitors changing the SN energy and the initial binary separation. We now present the salient properties of these simulations.

The $15\,M_\odot$ ejects just around $1.6\,M_\odot$, then the energy of the SN explosion needs to be low (on the order of $10^{50}$~erg) and the binary system enough compact, in order to have a significant accretion onto the NS companion. However, if the SN energy is considerably reduced, most of the SN material will do fallback onto the $\nu$NS. For example, scaling the SN kinetic energy and internal energy by $\eta=0.05$, almost $80\%$ of the SN ejecta is accreted by the $\nu$NS, while scaling the parameter to $\eta=0.1$, the accreted material via fallback is reduced to the $30\%$.  It is important to point out that, if the initial orbital period is increased by a factor of $1.7$, the amount of ejected mass that can not escape the $\nu$NS gravitational field grows to the $55\%$. Namely, the presence of a close NS companion could avoid the collapse of the $\nu$NS in the weak explosion cases.

For the $30\,M_\odot$ progenitor, we worked with two simulated explosions with different energies, one almost one order of magnitude stronger than the other. For the lowest energetic explosion, $E_{\rm sn,1}$, a significant amount of mass is making fallback, then the collapse due to the hypercritical accretion onto the $\nu$NS is more probable than the one of the NS companion. {The mass accretion rate on the NS companion flattens for this SN explosion (see Figure~\ref{fig:Mdot_Progenitor})}.  On the other hand, the velocities of the stronger energetic explosion, $E_{\rm sn,2}$, are so high that almost all the SN ejecta surpass the NS companion without being captured by it. For these explosions, the ratio between the total SN energy and the kinetic energy is $0.45$ and $0.81$, respectively.
\begin{figure*}
  \centering
  \subfigure[$E_{\rm sn}=1.09\times 10^{52}\, {\rm erg}$]{\includegraphics[scale=0.45]{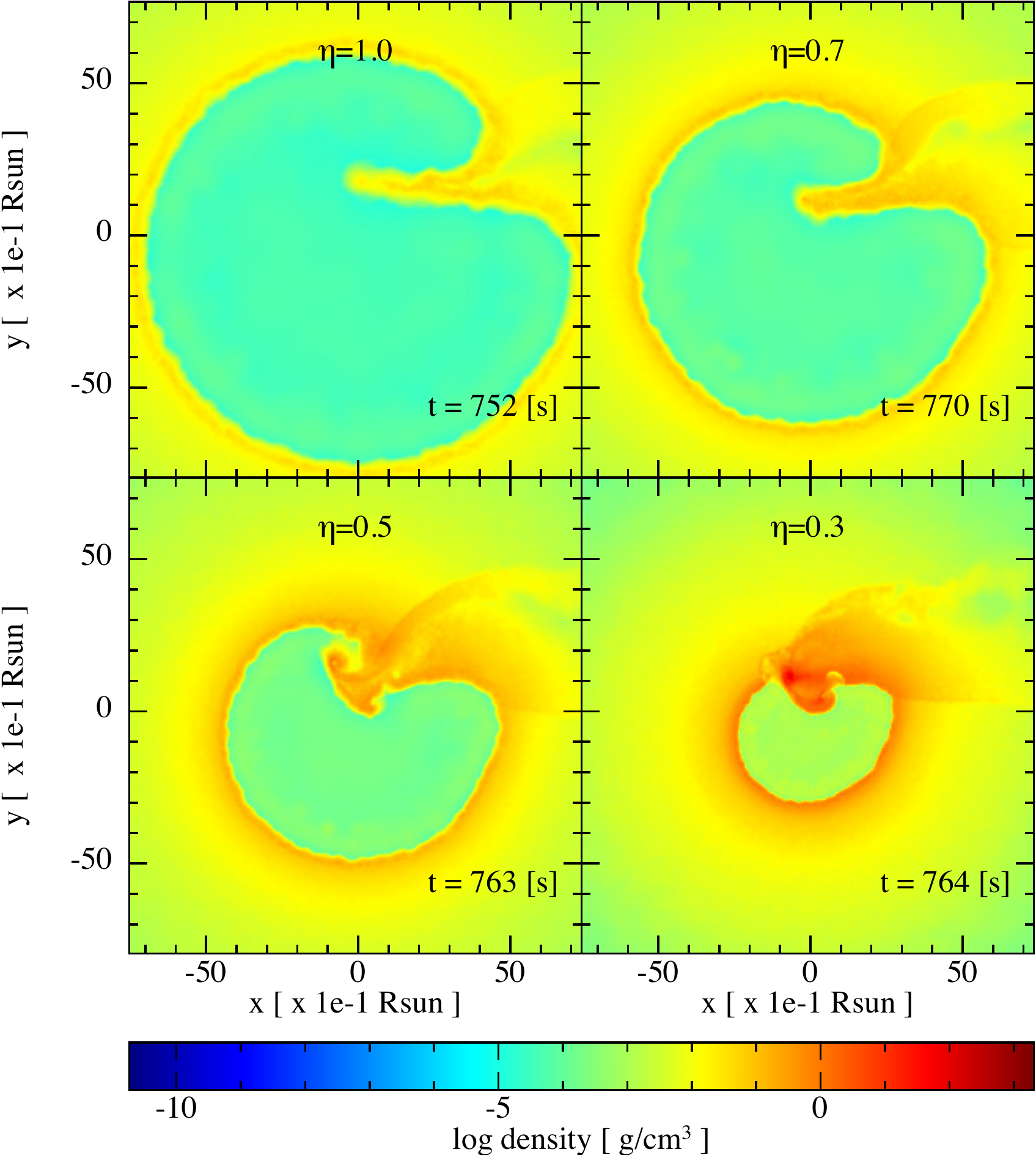}}\hspace{0.3cm}
  \subfigure[$E_{\rm sn}=2.19\times 10^{51}\, {\rm erg}$]{\includegraphics[scale=0.45]{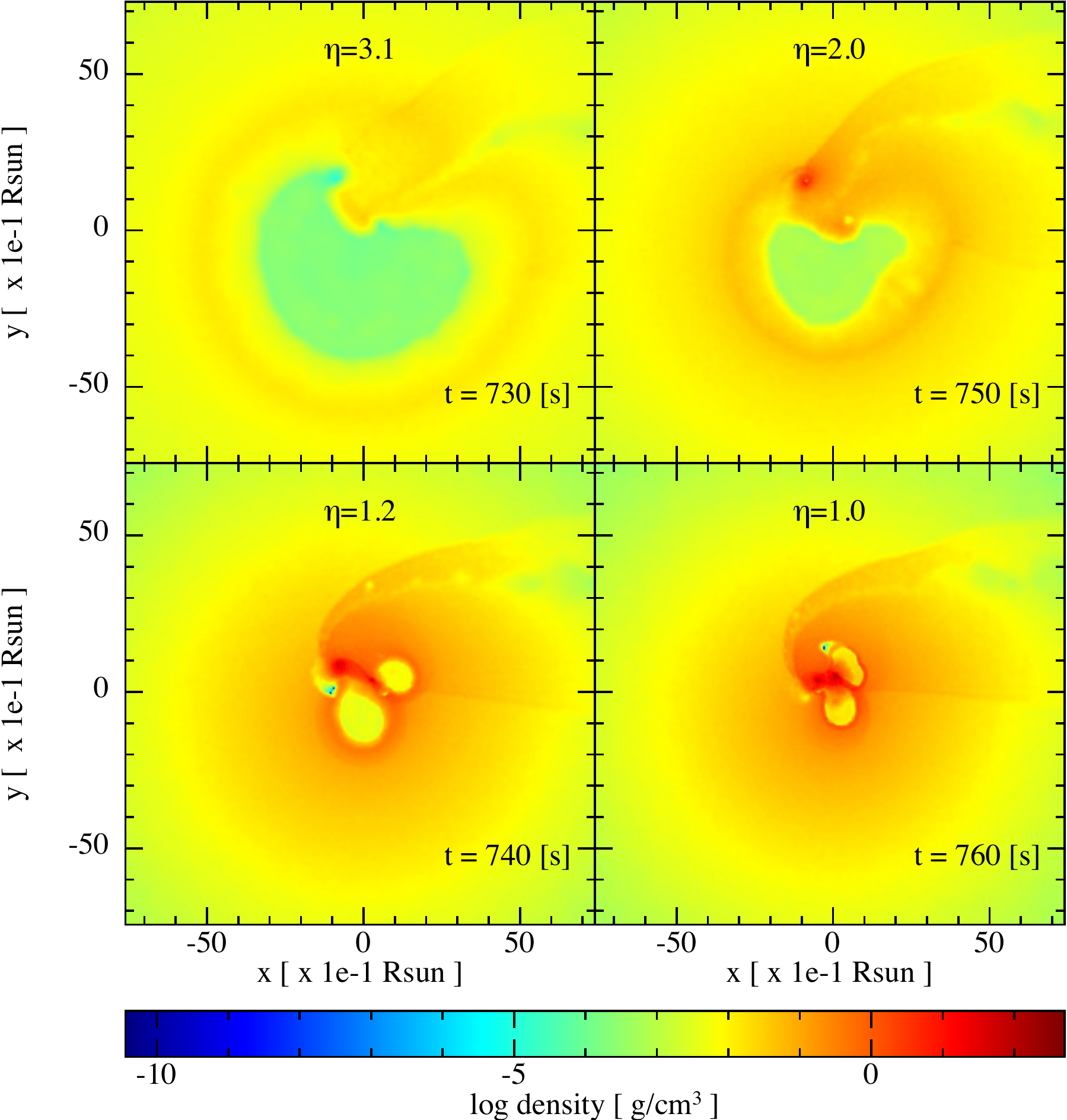}}
  \caption{Snapshots of the surface density on the binary equatorial plane. The initial binary system is formed by the CO$_{\rm core}$ of the $M_{\rm zams}=30\, M_\odot$ progenitor and a $2\,M_\odot$ NS with a orbital period of around $6$~min. We have simulated the collapse and bounce of the $30\, M_\odot$ progenitor with two different SN energies as is specified in the left and right panel labels. We show the simulations scaling the SN energy of these two explosions by a factor $\eta$ (specified in the upper part of each frame). In general, the internal radius of the low energetic explosion is about two times smaller that the one of the high energetic explosion. This increases the fallback accretion onto the $\nu$NS and the region near to the binary system becomes denser.}
  \label{fig:30Mzams_density}
\end{figure*}

We have performed more simulations scaling the energy of these two explosions, and summarize their results in Table~\ref{tab:inconmodels2}. In these cases, we can evaluate how accurate is our alternative path of changing the explosion energy by scaling the velocities and the internal energy of the SPH particles, instead of re-run new simulations of the CO$_{\rm core}$ collapse and bounce of the shock. Figure~\ref{fig:30Mzams_density} shows the density profile at around the same time for the two explosions of the $30\, M_\odot$ progenitor and their respective simulations with the scaled SN energy. In general, the internal radius of the low energetic explosion is about two times smaller that the one of the high energetic explosion (see Figure~\ref{fig:Progrho}).  These will increase the  material that will make fallback in the single star as well as in the binary system simulations, even when the scaled energy of the explosion becomes comparable.

Finally, we use the $40\,M_\odot$ progenitor of the CO$_{\rm core}$. Since for this progenitor the ejected mass in the SN is around $11\,M_\odot$, the energy of the explosion needs to be low to allow the configuration to remain bound and also for the NS companion to be able to accrete enough mass to collapse. If we use a factor $\eta=0.7$ to reduce the SPH particles velocity and internal energy, we see that the amount of mass accreted by the $\nu$NS is low but the mass accreted by the NS-companion could make it to induced its collapse. Instead, for $\eta=0.5$, most of the ejecta make fallback accretion onto the $\nu$NS.
 \begin{table*}
  \centering
  \caption{{Supernova initial and final mean bulk velocity }} 
  \setlength{\tabcolsep}{5pt}
  \begin{tabular}{cccc|cccc}
  \hline \hline
    Progenitor & $\eta$  & $\langle v\rangle_{\rm initial} $ & $\langle v \rangle_{\rm final}$ &Progenitor & $\eta$  & $\langle v\rangle_{\rm initial} $ & $\langle v \rangle_{\rm final}$ \\  
    $M_{\rm zams}$ & &  $10^8$~cm~s$^{-1}$ & $10^8$~cm~s$^{-1}$ &  $M_{\rm zams}$ & &  $10^8$~cm~s$^{-1}$ & $10^8$~cm~s$^{-1}$\\
    \hline \hline
 \multirow{3}{*}{15 $M_\odot$} & 0.7 & $7.316$ & $8.763$ &   \multirow{2}{*}{30 $M_\odot$ $^a$} & 0.7 &  & $7.931$ \\ 
                  & 0.5 & $6.183$ & $ 7.103$ &                   & 0.7 & $4.912$ & $6.227 $\\ 
                   \cline{5-8}
                  & 0.3 & $4.789$ & $5.066$ &  \multirow{2}{*}{30 $M_\odot$ $^b$} & 1.0 & $3.097$ & $4.845$  \\ 
                  &     &          &        &                          & 1.2 & $3.393 $ & $5.538$ \\
      \hline
 \multirow{4}{*}{25 $M_\odot$ } & 1.0  & $4.186$  & $5.805$ & 
 \multirow{4}{*}{40 $M_\odot$} & 1.0 & $4.949$ & $7.232$\\ 
     & 0.9 & $3.889$ & $5.079$ & & 0.9 & $4.667$ & $6.624$   \\
     & 0.8 & $3.560$ & $4.341$ & & 0.8 & $4.405$ & $5.976$ \\
     & 0.7 & $3.217$ & $3.945$ & & 0.7 & $4.114$ & $5.288$\\
   \hline \hline
      \end{tabular}
  \label{tab:SN_v}
\end{table*}
{In table~\ref{tab:SN_v}, we estimate the initial and final mean bulk SN velocity in the simulations, $\langle v\rangle_{\rm inital}$ and $\langle v\rangle_{\rm inital}$, computed as:}
 \begin{equation}
 {\langle v \rangle=\sqrt{\frac{2E_k}{M_{\rm ej}}}\, ,}
\end{equation}
{where $E_{k}$ is the total kinetic energy of the SPH-particles and $M_{\rm ej}$ is the total mass of the expanding material particles. We compute this for all the CO$_{\rm core}$ progenitors and for different values of the $\eta$ parameter. We do not account here for the acceleration of the SN ejecta arising from the energy injection in the accretion process and in the GRB emission via the impact onto the SN by the $e^+e^-$ plasma. As it has been shown in \citet{2018ApJ...852...53R}, the high velocities that characterize the HNe associated with GRBs are explained by this mechanism.}

\citet{2014ApJ...793L..36F} performed a 1D numerical simulations of the CO$_{\rm core}$ collapse, bounce and explosion and it was estimated the accretion rate onto the NS companion using the Bondy-Hoyle formalism \citep{1939PCPS...35..405H,1944MNRAS.104..273B,1952MNRAS.112..195B}. In these simulations, at the beginning of the accretion process, there is a burst in the accretion rate, growing up to the $10^{-1}~M_\odot$~s$^{-1}$, that is two order of magnitude greater than the accretion rate that we are obtaining with the SPH simulation. However, the total time of the accretion process is much shorter in those simulations that the one presented here, making the amount of mass accreted by the NS companion comparable between the simulations. This discrepancy is a direct consequence of the increase of dimension of the simulation. While the 1D simulation is stopped when the SN innermost layer reaches the NS, the 3D simulation can continue because there are particles that remain bound to the NS companion in a kind of disk structure, and the star continues the accretion process.

\section{Bound and unbound systems}\label{sec:5}

We have also studied the evolution of the binary parameters during the SN expansion.

If we were to assume that the ejected mass in the SN explosion leaves the system instantaneously, the semi-major axis of the post-explosion system, $a$, is given by $a/a_0=(M_0-\Delta M)/(M_0-2a_0\Delta M/r)$, where $M_0$ and $a_0$ is the total mass and the semi-major axis of the initial binary system, $\Delta M$ is the mass ejected in the SN and $r$ is the star separation at the moment of the explosion \citep{1983ApJ...267..322H}. Then, binaries with circular initial orbits become unbound after a SN event if more than half of its total initial mass is lost. However, as it was shown in \citet{2015PhRvL.115w1102F}, in the IGC scenario the mass loss can not be considered instantaneous because the binary initial periods are of the same order than the time it takes for the slowest SN layer to reach the NS companion. For example, the innermost layers of the $25\,M_\odot$ progenitor have a velocity of the order of $10^8$~cm~s$^{-1}$ so they reach the NS initial position in a time $\sim 100$~s, nearly $2/5$ of the initial binary period. Moreover, it has to be considered that either the $\nu$NS or the NS companion will accrete mass and momentum from the SN ejecta, and this will reduce the system mass loss.

\begin{figure}
  \centering
  \includegraphics[width=0.95\hsize]{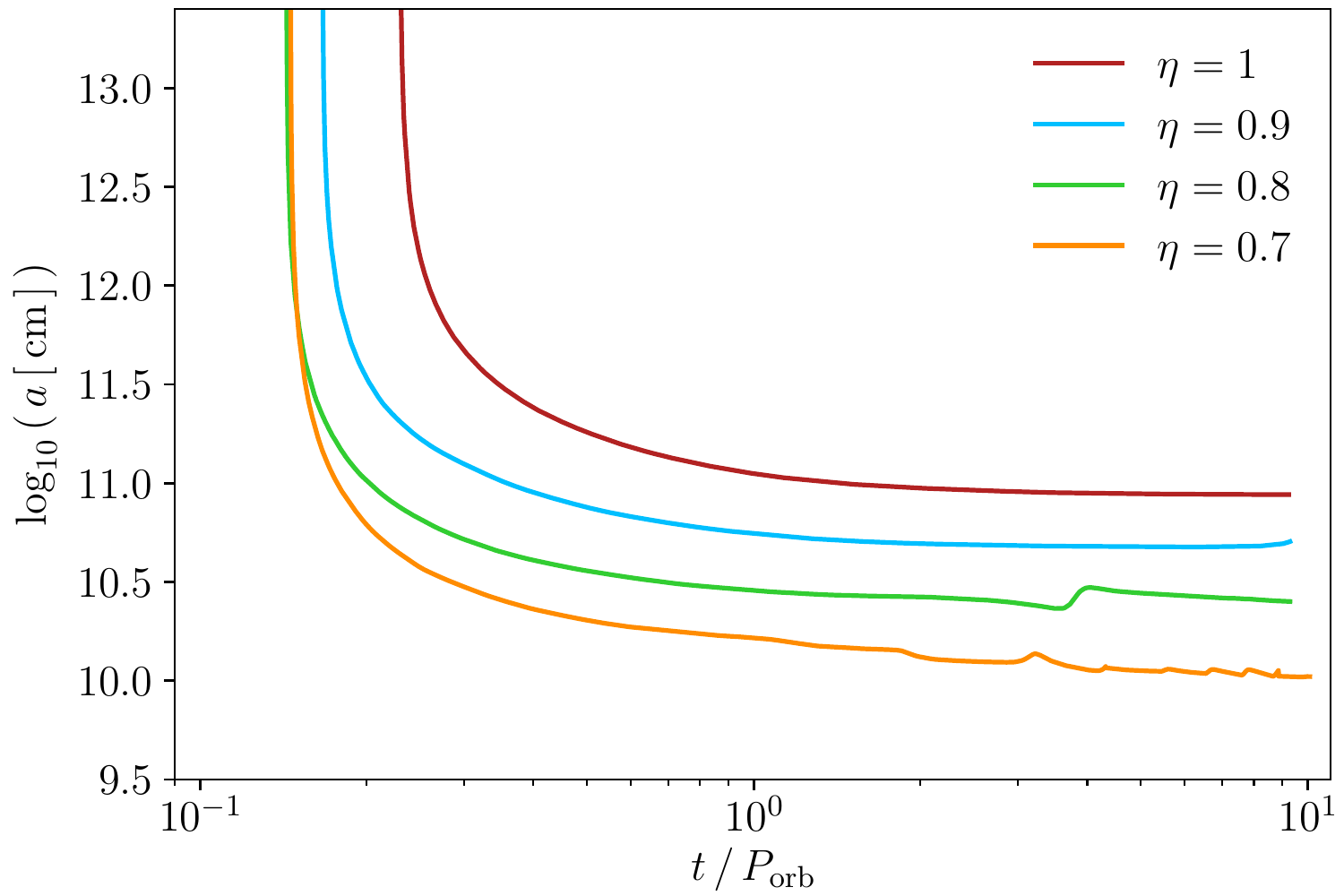}
  \caption{Evolution of the semi-major axis of the $\nu$NS-NS binary system. The initial configuration is a binary system formed by the CO$_{\rm core}$ of a $M_{\rm zams}=25\, M_\odot$ progenitor and a NS of $2\,M_\odot$ with an initial binary period of approximately $5$~min. The CO$_{\rm core}$ collapses and undergoes a SN explosion ejecting around $5\,M_\odot$ and leaving a proto-NS of $1.85\,M_\odot$. Not all the material ejected in the SN leaves the binary system. Some part of this material fallback and is accreted by the remnant star from the collapse of the CO$_{\rm core}$ while, some other part, is accreted by the NS companion. The final binary configuration can remain bound even when the system lost most of the half of its initial total mass.}
  \label{fig:25MzmansEnergie_aorb}
\end{figure}

Figure~\ref{fig:25MzmansEnergie_aorb} shows the evolution of the binary semi-major axis with time for the same cases of Figure~\ref{fig:25Mzams_Mnsdot_Energy}. In these simulations, that correspond to the minimum orbital period of the initial binary system ($\sim 5$~min), independently on the SN energy, the post-explosion system remains bound. This occurs even if the system loses more than the half of its total initial mass. For example, for the $\eta=1$ case, the mass loss in the system is around $4.82\,M_\odot$, namely about $54.7\,\% $ of its total initial mass.

For the fiducial SN model, an increase by a factor $1.7$ of the orbital period changes the fate of the post-explosion binary system, leading to an unbound final configuration. While, for the less energetic explosion, the system remains bounded in all scenarios in which we have  increased the initial binary period (up to $\approx 60$~min, i.e. $\approx 12$ times the minimum orbital period). {The determination of the maximum orbital period that the system can have in oder to remain bound after the SN explosion is left for a future work.}

\section{Mass-accretion rate, NS critical mass and gravitational collapse}
\label{sec:6}
%
%

We turn now to evaluating whether the NS companion collapses or not to a BH due to the accretion of part of the SN ejecta. For this, we need both to study how the NS gravitational mass and angular momentum evolve with time, and to set a value to the NS critical mass. As a first approximation, we assume that the NS evolves from an equilibrium configuration to the next, using the uniformly rotating NS equilibrium configurations in \citet{2015PhRvD..92b3007C,PhysRevD.96.024046}. These configurations were constructed using the public code RNS \citep{1995ApJ...444..306S} to solve the axisymmetric Einstein equations for three selected relativistic mean-field nuclear matter equations of state (EOS) models: NL3, GM1 and TM1. {The mass-radius relations obtained for these EOS satisfy current observational constraints (see figure~5 in \citealp{2015PhRvD..92b3007C}). The most stringent constraint is that the critical mass for gravitational collapse of a non-rotating NS must be larger than the mass of the most massive NS observed: the pulsar PSR J0348+0432 with $2.01\pm 0.04\,M_\odot$ \citep{2013Sci...340..448A}\footnote{{We recall that this pulsar constrains the non-rotating NS mass-radius relation because the structure of a NS rotating at the measured rate of PSR J0348+0432, $f \approx 26$~Hz \citep{2013Sci...340..448A}, in practice ``overlaps'' with the one of a non-rotating NS \citep[see e.g.][]{2015PhRvD..92b3007C}.}}. Recently, it was claimed the measurement of a more massive NS, PSR J2215+5135 with $2.27^{+0.17}_{-0.15}\,M_\odot$ \citep{2018ApJ...859...54L} and $f \approx 383$~Hz \citep{2013ApJ...769..108B}. This measurement is still under debate and, for this paper, we use constraint imposed by the mass of PSR J0348+0432. The current measurements of NS radii e.g. from X-ray observations, unfortunately, poorly constrain the mass-radius relation and so the nuclear EOS \citep[see e.g.][for details]{2015PhRvD..92b3007C}}


In general, for uniformly rotating NS configurations, when the star accretes an amount of baryonic mass, $M_{\rm b}$, and angular momentum, $J_{\rm ns}$, its  gravitational mass, $M_{\rm ns}$, evolves as:
\begin{equation}
  \dot{M}_{\rm ns}=\left( \frac{\partial M_{\rm ns}}{\partial M_b} \right)_{J_{\rm ns}}\, \dot{M}_b + \left( \frac{\partial M_{\rm ns}} {\partial J_{\rm ns}}\right)_{M_b}\, \dot{J}_{\rm ns}\,.
  \label{eq:Mgrav_Evol}
\end{equation}
From \citet{2015PhRvD..92b3007C}, we know the relation $M_{\rm ns}(M_b,J_{\rm ns})$:
\begin{equation}
  \frac{M_b}{M_\odot}=\frac{M_{\rm ns}}{M_\odot}+\frac{13}{200}\left( \frac{M_{\rm ns}}{M_\odot} \right)^2\left(1-\frac{1}{130}j_{\rm ns}^{1.7} \right) \, ,
  \label{eq:MbMns}
\end{equation}
where $j_{\rm ns}\equiv cJ_{\rm ns}/(GM_\odot^2)$. This relation is common to all the NS matter EOS studied in \citet{2015PhRvD..92b3007C} within an error of 2\%. The mass and angular momentum of the accreted particle add to the baryonic mass and the angular momentum of NS, thus we can integrate Equation~(\ref{eq:Mgrav_Evol}) and follow the evolution of the gravitational mass of both the $\nu$NS and the NS companion. The NS accretes mass until it reaches an instability point: the mass shedding limit or the secular axisymmetric instability. The first limit is met if the star angular velocity has growth enough that the gravitational force at the stellar equator equals the centrifugal force, then a faster rotation will induce the ejection of matter. {Although the dimensionless angular momentum parameter has a nearly EOS-independent value along the mass-shedding sequence, $cJ_{\rm ns}/(GM^2_{\rm ns}) \approx 0.7$ \citep{2015PhRvD..92b3007C}, the numerical value of $J_{\rm ns}$ at the shedding point depends on the specific EOS since $M_{\rm ns}$ is different.} On the order hand, the secular axisymmetric instability which defines the critical mass for gravitational collapse arises because the star became unstable to axisymmetric perturbations. From \citet{2015PhRvD..92b3007C}, we have that, within at maximum error of $0.45\%$, the critical mass is given by:
\begin{equation}
  M_{\rm max}^{J_{\rm ns}\neq 0}=M_{\rm max}^{J_{\rm ns}=0}\left( 1+kj_{\rm ns}^l \right)\, ,
  \label{eq:Mcrit}
\end{equation}
where $M_{\rm max}^{J_{\rm ns}=0}=[2.81,2.39,2.20]$, $l=[1.68,1.69,1.61]$ and $k=[0.006,0.011,0.017]$ for the NL3, GM1 and TM1 EOS.  
\begin{table}
\centering
\caption{{Properties of the NS critical mass for NL3, TM1 and GM1 EOS.}}
\setlength{\tabcolsep}{5pt}
\begin{tabular}{c|ccc}
  \hline
  \hline
  EOS & $M_\mathrm{max}^{J_{\rm ns}=0}$ & $M_\mathrm{max}^{J_{\rm ns}\neq 0}$ & $f_{\rm max}$\\       
  &  $(\,M_\odot\,)$  & $(\,M_{\odot}\,)$ & $(\,$kHz $\,)$ \\
\hline
NL3 & $2.81$ & $3.38$ & $1.40$\\
GM1 & $2.39$ & $2.84$ & $1.49$ \\
TM1 & $2.20$ & $2.62$ &  $1.34$ \\
\hline
\hline
\end{tabular}
\tablecomments{{$M_\mathrm{max}^{J_{\rm ns}=0}$: critical mass for non-rotating case. $M_\mathrm{max}^{J_{\rm ns}\neq 0}$: maximum critical mass value, i.e. the mass of the configuration along the secular instability line and rotating at the Keplerian value. $f_{\rm max}$: rotation frequency of the NS with $M_\mathrm{max}^{J_{\rm ns}\neq 0}$. Table taken from \cite{2015PhRvD..92b3007C}.}}
\label{tab:NS_EOS}
\end{table}

{
Table~\ref{tab:NS_EOS} lists some properties of the NS critical mass configurations for the NL3, GM1 and TM1 EOS. In particular, we show the minimum and maximum values of the critical mass, $M_\mathrm{max}^{J_{\rm ns}=0}$ and $M_\mathrm{max}^{J_{\rm ns}\neq 0}$, respectively: $M_\mathrm{max}^{J_{\rm ns}=0}$ corresponds to the configuration along the secular axisymmetric instability line which is the non-rotating and $M_\mathrm{max}^{J_{\rm ns}\neq 0}$ to the maximally rotating one which is the configuration that intersects the Keplerian, mass-shedding sequence. We also list the rotation frequency of the maximally rotating critical configuration, $f_{\rm max}$.
}

\begin{figure}
  \centering
  \includegraphics[width=0.95\hsize]{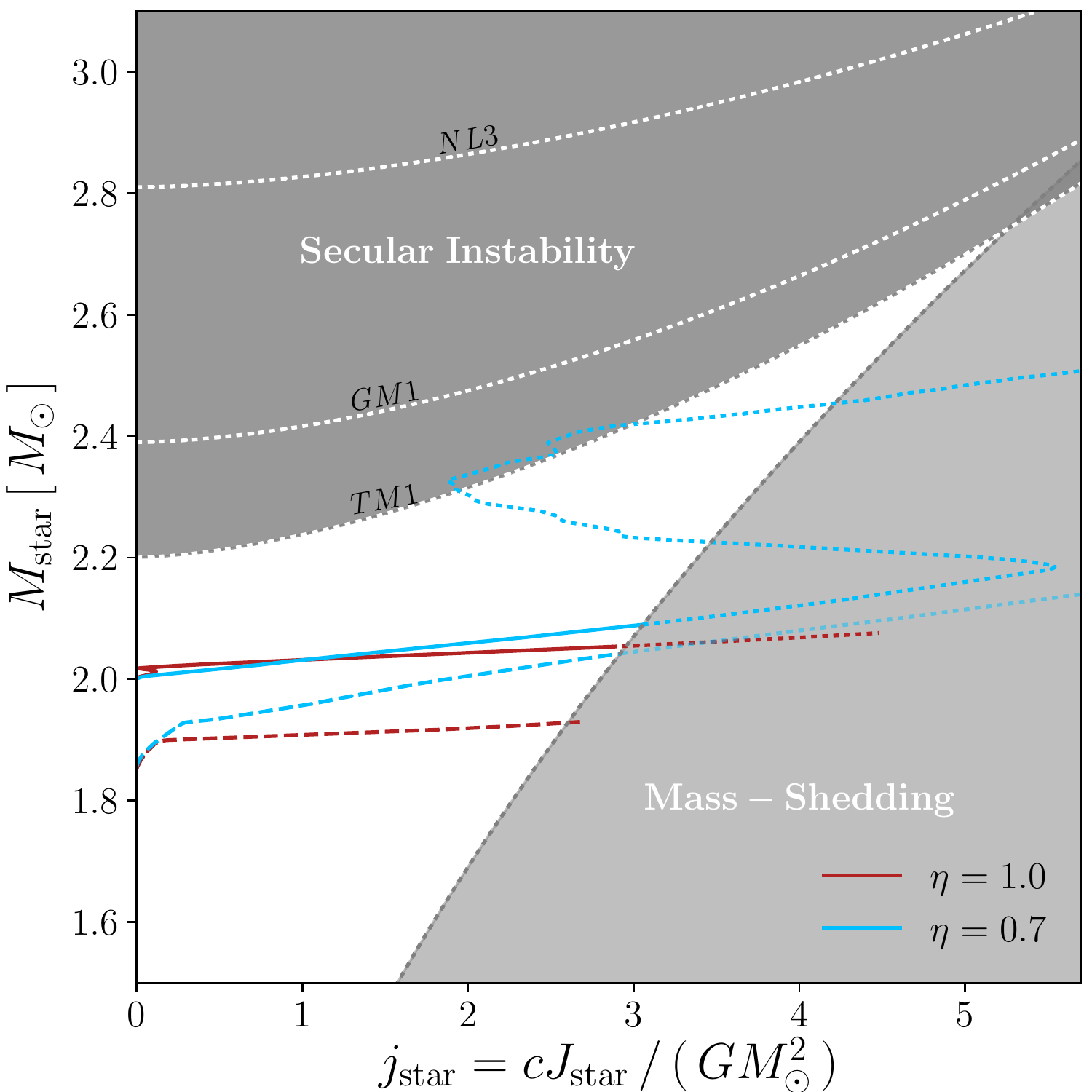}
  \includegraphics[width=0.95\hsize]{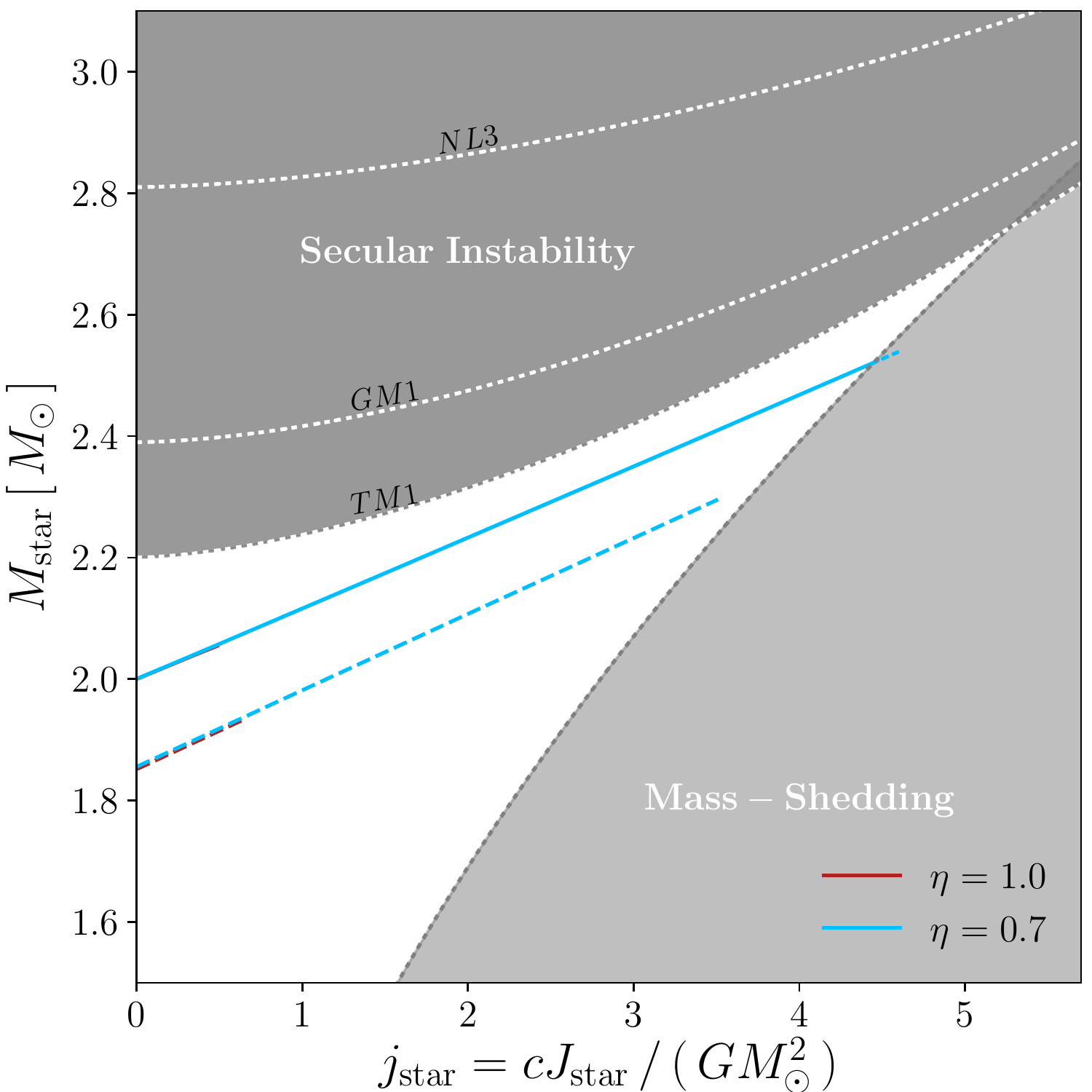}
  \caption{Evolution of the $\nu$NS (dashed line) and the NS companion (solid line) in the mass-dimensionless angular momentum ($M_{\rm star}$-$j_{\rm star}$) plane. The mass of the particles accreted contributes to the NS baryonic mass. In the upper plot we adopt that the star accretes all the particles angular momentum. In the lower plot we adopt that the star accretes from a disk-like structure, namely that the angular momentum evolution is dictated by the disk-accretion torque (see text for details). In this example the initial binary system is formed by the CO$_{\rm core}$ of the $M_{\rm zams}=25\,M_\odot$ progenitor and a $2\, M_\odot$ NS with an orbital period of about $5$~min. The red lines correspond to a SN explosion of $\times 10^{51}$~erg, while  for the blue line the explosion energy has been scaled by a factor $\eta=0.7$, leading to $\times 10^{51}$~erg.}
  \label{fig:25Mzamns_Mjplane}
\end{figure}

In the upper panel of Figure~\ref{fig:25Mzamns_Mjplane} we show the track followed by the NS companion (solid line) and the $\nu$NS (dashed line) in the $M_{\rm star}-j_{\rm star}$ plane for the  $25\,M_\odot$ progenitor of the CO$_{\rm core}$, for two different SN explosion energies (models 25M1p1e and 25M1p07e of table~\ref{tab:inconmodels2}). For the system with the stronger SN explosion, the $\nu$NS and the NS companion reach the mass-shedding limit at $t = 21.66$~min with $1.93\, M_\odot$ and $t=20.01$~min with $2.055\, M_\odot$, respectively. For the less energetic SN explosion this occurs early, at $t=5.51$~min with $2.04\,M_\odot$ for the $\nu$NS, and at $t=2.91$~min with $2.09\, M_\odot$ for the NS companion. The dotted line shows the continuation of the integration of Equation~(\ref{eq:MbMns}) for all the simulation time. For the NS companion, there is a decreasing of angular momentum. This occurs because there is a change in the direction of rotation of the accreted particles with respect to the one of the accreting NS.

Since we are assuming that the angular momentum of the accreted particles is totally transferred to the NS, even the accretion of a little amount of mass might bring it soon to the mass-shedding limit (see Figure~\ref{fig:25Mzamns_Mjplane}). However, we see in the simulations that a kind of disk is formed around the NS, then before they being actually accreted, circularize and loose angular momentum owing to the friction force developed in the disk. In this picture, we need to integrate Equation~(\ref{eq:Mgrav_Evol}) assuming that the star angular momentum evolution is given by the disk accretion torque, i.e.:
\begin{equation}
  \dot{J}_{\rm ns}=l(R_d)\dot{M}_b\,,
  \label{eq:jtorque}
\end{equation}
where $l(R_d)$ is the specific angular momentum of the particles at the disk interior radius, $R_d$. We adopt now that the disk interior radius is given by the radius of the last circular orbit (LCO) of a test particle around the NS. From \citet{PhysRevD.96.024046}, we have that the specific angular momentum of the LCO, independently on the studied EOS, is given by:
\begin{equation}
  l(R_d) = l_{\rm LCO} = \frac{G M_{\rm ns}}{c}\left[2\sqrt{3}\mp 0.37\left( \frac{j_{\rm ns}}{M_{\rm ns}/M_\odot} \right)^{0.85}\right]\,,
  \label{eq:lISO}
\end{equation}
where the upper sign corresponds to the corotating particles and the lower sign to the counter-rotating particles. This fitting formula is accurate with a maximum error of $0.3\%$.

In the bottom panel of Figure~\ref{fig:25Mzamns_Mjplane} we show the evolution of the $\nu$NS and the NS companion in the $M_{\rm star}-j_{\rm star}$ plane, for the same models of the upper panel. In this case, we have integrated Equations~(\ref{eq:Mgrav_Evol}) and (\ref{eq:lISO}) assuming, again, that the particles mass sums to the star baryonic mass and the disk viscous timescale is smaller that the accretion timescale, i.e. we used the mass-accretion rate obtained from the SPH simulations. In this case only the NS companion for the less energetic SN simulation reaches the mass-shedding limit.

\begin{figure}
\centering
  \includegraphics[width=0.95\hsize]{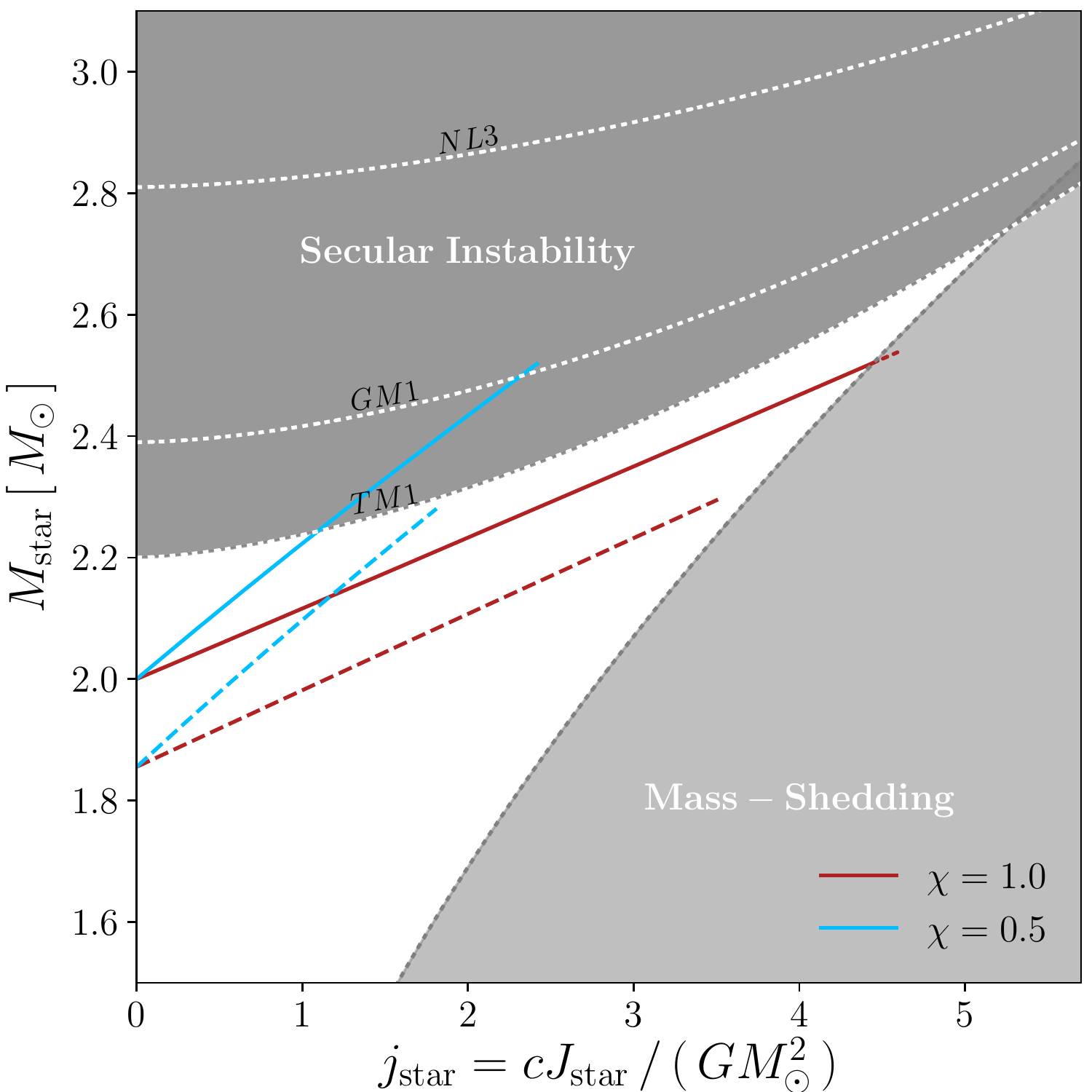}
  \caption{Evolution of the $\nu$NS (dashed line) and the NS companion (solid line) in the mass-dimensionless angular momentum ($M_{\rm star}$-$j_{\rm star}$) plane. The angular momentum evolution is dictated by the disk-accretion torque given by Equation~(\ref{eq:chi}), where we have introduced the efficiency parameter, $\chi\leq 1$, that accounts for angular momentum losses between the disk and the stellar surface. The initial binary system is formed by the CO$_{\rm core}$ of the $M_{\rm zams}=25\,M_\odot$ progenitor and a $2\,M_\odot$ NS with an orbital period of about $5$~min. The SN explosion energy has been scaled by $\eta=0.7$ (Model 25M1p07e of Table~\ref{tab:inconmodels2}).}
  \label{fig:eff_torque}
\end{figure}

{The evolution equations (\ref{eq:Mgrav_Evol}) and (\ref{eq:jtorque}) are general while the binding energy and LCO angular momentum equations are, in general, EOS-dependent relations. Fortunately, it has been shown in \citet{PhysRevD.96.024046} that equations (\ref{eq:MbMns}) and (\ref{eq:lISO}) remain valid for the set of EOS used in this work (and for a wider variety of EOS) with high accuracy, namely they are nearly EOS-independent. This implies that the evolution track followed by the NS, given the initial mass, is the same for all these EOS. However, This does not imply that the fate of the NS is the same since the instability boundaries depend on the EOS, namely the numerical values of the mass and angular momentum of the NS at the mass-shedding and at the secular instability depend on the EOS, as it can be seen from figures \ref{fig:25Mzamns_Mjplane} and \ref{fig:eff_torque}.}

\begin{table*}
  \centering
  \caption{Final state of the $\nu$NS and the NS companion}
  \setlength{\tabcolsep}{1.5pt}
\renewcommand{\arraystretch}{1.45}
\footnotesize
  \begin{tabular}{cc|c|ccc|ccc|c|ccc|ccc}
    &  & \multicolumn{7}{|c}{$\nu$NS} & \multicolumn{7}{|c}{NS}\\
    \hline
    \hline
    &  &  & \multicolumn{3}{|c}{$\chi=0.5$} &\multicolumn{3}{|c|}{$\chi=1.0$} &  & \multicolumn{3}{|c}{$\chi=0.5$} &\multicolumn{3}{|c}{$\chi=1.0$} \\ \hline
   CO$_{\rm core}$ & Model  & $L_{\rm tot}$ & $M_{\nu{\rm ns}}$ & $j_{\nu{\rm ns}}$ &Fate& $M_{\nu{\rm ns}}$ & $j_{\nu{\rm ns}}$ &Fate& $L_{\rm tot}$ & $M_{\rm ns}$ & $j_{\rm ns }$ &Fate& $M_{\rm ns}$ & $j_{\rm ns}$ &Fate \\
 $M_{\rm zams}$& & $c/(GM_\odot^2)$ & $M_\odot$ & $c/(GM_\odot^2)$ & & $M_\odot$ & $c/(GM_\odot^2)$ & & $c/(GM_\odot^2)$ & $M_\odot$ & $c/(GM_\odot^2)$ & & $M_\odot$ & $c/(GM_\odot^2)$ & \\
    \hline\hline
 \multirow{6}{*}{$15\,M_\odot$} &15m1p07e   & $0.027$  & $1.302$ & $0.007$ & Stb & $1.302$ & $0.008$ & Stb &  $0.085$ & $2.004$ & $0.009$& Stb & $2.002$ & $0.018$ & Stb  \\
     \cline{2-16}
               & 15m1p05e  & $0.069$   & $1.303 $ & $0.009 $& Stb & $1.303$ & $0.016$ & Stb       & $0.323$  & $2.004$ & $0.019$& Stb & $2.004$ & $0.037$ & Stb  \\
     \cline{2-16}
     \cline{2-16}
  & 15m1p03e  & $0.019$ & $1.315$ &  $0.041$ & Stb& $1.315$ & $0.077$ & Stb   & $0.362$ & $2.023$ & $0.101$ & Stb  &  $2.023$ & $0.204$ & Stb  \\
      &  15m2p03e  & $0.091$ &  $1.303$ & $0.08$ & Stb & $1.303$ & $0.017$ & Stb  & $0.579$ &  $2.006$ & $0.026$ & Stb & $2.006$ & $0.056$ & Stb \\
     \cline{2-16}
 &15m1p01e  & $13.63$ & $1.815$ & $1.571$ & Stb & $1.636$ & $1.874$ & {\bf M-sh} & $19.373$  & $2.157$ & $0.701$ & Stb & $2.159$ & $1.377$ & Stb  \\
    & 15m2p01e   & $38.38$ &  $2.077$ & $2.534$ & Stb & $1.639$ & $1.892$ & {\bf M-sh} & $12.533$ & $2.080$ & $0.35$ & Stb & $2.080$ & $0.693$ & Stb  \\
    & 15m3p01e    &  $30.95$ & $1.759$ & $1.377$ &  Stb & $1.862$ & $3.253$ & {\bf M-sh} & $2.004$ & $2.045$ & $1.197$ &Stb &  $2.045$ & $0.388$ & Stb  \\
    \hline\hline
 \multirow{8}{*}{ $25M_\odot$} & 25m1p1e  & $3.469$  & $1.931$ & $0.321$  & Stb & $1.933$ & $0.627$ & Stb & $4.746$ & $2.055$ & $0.2467$ & Stb & $2.056$ & $0.497$ & Stb  \\
  & 25m2p1e &  $1.779$  & $1.912$ & $0.242$ & Stb & $1.914$ & $0.472$ & Stb & $1.927$ & $2.022$ & $0.099$ & Stb & $2.022$ & $0.198$ & Stb  \\
  & 25m3p1e & $1.085$  & $1.912$ & $0.229$ & Stb & $1.912$ & $0.399$ & Stb & $1.944$ &  $2.018$ & $0.0813$ & Stb & $2.019$ & $0.1639$ & Stb  \\
 \cline{2-16}
 & 25m1p09e  & $6.243$ & $1.982$ & $0.513$ & Stb & $1.983$ & $1.010$ & Stb & $6.538$ & $2.127$ & $0.584$ & Stb & $2.129$ & $1.187$ & Stb  \\
  \cline{2-16}
   & 25m1p08e  & $7.331$ &$2.038$ & $1.449$ & Stb & $2.031$ & $1.365$ & Stb & $9.870$ & $2.258$ & $1.242$ & {\bf Sc-in} & $2.348$ & $3.576$ & Stb  \\
  \cline{2-16}
 & 25m1p07e   & $18.146$  & $2.284$ & $1.826$ & Stb & $2.289$ & $3.434$ & Stb & $8.491$ & $2.246$ &  $1.105$ & {\bf Sc-in} & $2.528$ & $4.506$ & {\bf M-sh } \\
   & 25m2p07e  & $19.51$  &$2.250$ & $1.663$ & Stb & $2.265$ & $3.215$ & Stb & $9.908$ & $2.252$ & $1.135$ & {\bf Sc-in} & $2.426$ & $3.648$ & Stb \\
  & 25m3p07e   & $21.34$  & $2.214$ & $1.476$ & Stb & $2.226$ & $2.812$ & Stb & $17.292$ & $2.004$ & $2.246$ & {\bf Sc-in} & $2.425$ & $3.638$ & Stb  \\
  \hline
   & 14M1p1e   &  $6.61$ & $1.945$  & $0.399$  & Stb & $1.951$ & $0.791$& Stb  & $3.933$ & $1.438$ & $0.110$ & Stb&  $1.437$& $0.221$ & Stb \\
   & 16M1p1e   & $5.172$   & $1.935$ & $0.672$ & Stb  & $1.935$ &  $0.672$ & Stb &$5.742$  & $1.652$ & $0.357$ & Stb& $1.652$  &  $0.178$ & Stb \\
   & 14M1p07e   &  $22.87$ & $2.312$  & $1.972$ & {\bf Sc-in} & $2.327$ & $3.788$ & {\bf M-sh} & $3.079$  & $1.837$& $1.509$ &Stb &  $1.711$ & $2.05$ & {\bf M-sh}  \\
  \hline\hline
      \multirow{5}{*}{$30\, M_\odot$\footnote{$E_{\rm sn}=1.09\times 10^{52}\, {\rm erg}$}} & 30m1p1ea  & $0.077$  & $1.756$ & $0.021$ & Stb & $1.756$ & $0.044$ & Stb  & $0.059$ &  $2.006$ & $0.026$ & Stb & $2.006$ & $0.052$ & Stb  \\
    \cline{2-16}
 & 30m1p07ea &  $0.954$  & $1.758$ & $0.032$ & Stb & $1.758$ & $0.062$ & Stb & $0.366$ & $2.012$ & $0.053$& Stb &  $2.012$ & $0.106$ & Stb  \\
     \cline{2-16}
 & 30m1p05ea  & $1.828$  & $1.764$ & $0.053$ & Stb & $1.764$ & $0.107$ & Stb & $4.073$ & $2.028$ & $0.125$ & Stb & $2.028$ & $0.251$ & Stb  \\
     \cline{2-16}
& 30m1p03ea & $3.560$ & $1.842$ & $0.3494$ & Stb  & $1.843$ & $0.692$ & Stb &$33.083$ & $2.246$ &$2.358$ & {\bf Sc-in}& $2.356$ &$3.532$  & Stb  \\
  &  30m2p03ea  & $4.266$  &$1.821$ & $0.267$ & Stb & $1.821$ & $0.522$  & Stb & $36.161$& $2.151$ & $0.7006$ &Stb & $2.154$ & $1.426$  & Stb \\
 \hline \hline
  \multirow{5}{*}{$30\, M_\odot\footnote{$E_{\rm sn}=2.19\times 10^{51}\, {\rm erg}$} $} &  30m1p1eb  & $64.935$ &$2.379$ & $2.614$ & {\bf Sc-in} & $2.215$ & $3.507$ & {\bf M-sh} & $19.995$ & $2.244$ & $1.099$& {\bf Sc-in} & $2.307$ & $2.634$ & Stb  \\
    \cline{2-16}
    &  30m1p12eb   & $28.432$ & $2.362$ & $2.541$& Stb & $2.200$ & $3.392$ & {\bf M-sh} & $33.681$ & $2.244$  & $1.100$ & {\bf Sc-in}&  $2.304$  & $2.606$ & Stb  \\
     &  30m2p12eb  &  $26.508$ & $2.397$ & $2.807$  & {\bf Sc-in} & $2.162$ & $3.297$  & {\bf M-sh} & $23.922$ & $2.1801$ &  $0.802$ & Stb & $2.1827$ &  $1.572$ &  Stb    \\
    \cline{2-16}
      &  30m1p2eb  & $2.819$ & $1.777$ & $0.106$ & Stb  & $1.777$ & $0.196$ & Stb & $7.846$ & $2.061$ & $0.271$ & Stb & $2.061$ & $0.546$ & Stb  \\
     \cline{2-16}
   &  30m1p31eb  & $0.721$  &$1.766$ & $0.0611$ & Stb &  $1.766$ & $0.105$ & Stb & $1.6715$ & $2.014$ & $0.062$ & Stb &  $2.014$ & $0.122$ & Stb  \\
      \hline\hline
  \multirow{6}{*}{ $40\,M_\odot$} & 40m1p1e  & $1.125$ & $1.874$ & $0.081$ & Stb & $1.873$ & $0.132$ & Stb & $11.504$ & $2.056$ & $0.2453$ &Stb  & $2.056$ & $0.4918$ & Stb  \\
    \cline{2-16}
                  & 40m1p09e  & $2.189$ & $1.875$ & $0.087$ & Stb  & $1.875$ & $0.164$ & Stb & $25.669$ & $2.236$ & $1.097$ & Stb & $2.236$ & $2.247$ & Stb  \\
     \cline{2-16}
                   & 40m1p08e  & $3.468$ & $1.879$ & $0.105$ & Stb  & $1.879$ & $0.199$ & Stb & $30.146$ & $2.254$ & $1.22$ & {\bf Sc-in} &  $2.405$ & $4.060$ & {\bf M-sh}  \\
    \cline{2-16}
                   & 40m1p07e  &  $9.963$ & $2.042$& $0.7631$ & Stb & $2.042$& $1.486$ & Stb & $34.074$ & $2.243$  & $1.09$ & {\bf Sc-in} & $2.526$  & $4.491$ & {\bf M-sh}  \\
                 & 40m2p07e  & $9.6264$  & $2.023$ & $0.689$ & Stb & $2.024$ & $1.343$ & Stb & $42.97$ & $2.245$ & $1.100$ & {\bf Sc-in} & $2.522$ & $4.458$ & {\bf M-sh}  \\
       & 40m4p07e  & $10.333$ & $1.942$ & $0.342$ & Stb & $1.942$ & $0.652$ & Stb &$27.474$ & $2.246$   & $1.129$ &  {\bf Sc-in} & $2.4135$   & $3.6104$ &  Stb \\
     \cline{2-16}
                  & 40m1p06e  & $171.063$  & $2.310$ &$1.948$ & {\bf Sc-in} & $2.338$ &$3.857$ & {\bf M-sh} & $63.31$ & $2.244$  & $1.098$ & {\bf Sc-in} & $2.529$  & $4.518$ & {\bf M-sh}  \\
          & 40m2p06e & $121.594$ & $2.318$ &$1.987$ & {\bf Sc-in}   & $2.331$ &$3.793$ & {\bf M-sh} & $45.944$ & $2.244$  & $1.096$ & {\bf Sc-in} & $2.527$  & $4.506$ & {\bf M-sh} \\
     \cline{2-16}
                   & 40m1p05e & $11.593$ & $2.316$ & $1.197$& {\bf Sc-in}   & $2.338$ & $3.861$& {\bf M-sh} & $5.434$ & $2.133$ & $0.588$  & Stb  & $2.134$ & $1.158$  & Stb  \\
  \hline\hline   
\end{tabular}
\tablecomments{Stb: stable configuration; M-sh: mass-shedding limit; Sc-in: secular axisymmetric instability. {In this table, we report the results for the TM1 EOS and the gravitational mass value takes into account the angular momentum transfer by accretion. These values could differ from those reported in table~\ref{tab:inconmodels2} where the angular momentum transfer is not considered.}}
  \label{tab:AngularMomentum}
\end{table*}

We have assumed until now a totally efficient angular momentum transfer of the particles from the inner disk to the NS surface. However, additional angular momentum losses should be taken into account. We model these losses introducing a parameter for the efficiency of the angular momentum transfer, $\chi\leq 1$ \citep[see e.g.][]{2016ApJ...833..107B}, defined as:
\begin{equation}\label{eq:chi}
  \dot{J}_{\rm ns}=\chi l(R_d)\dot{M}_b.
\end{equation}
{Therefore, $\chi=1.0$ implies that the particles lose angular momentum only in their downward motion within the disk, while $\chi<1$ introduces the possible losses in their final infall to the NS, e.g. from accretion outflows and/or from the deceleration of the matter from the inner disk radius to its final incorporation to the NS surface \citep[see e.g.][and references therein]{1988AdSpR...8..135S,1999AstL...25..269I,2008MNRAS.386.1038B,2010AstL...36..848I,2016ApJ...817...62P}.} In Figure~\ref{fig:eff_torque} we compare the evolutionary path on the mass-dimensionless angular momentum plane for two values of the efficiency parameter, $\chi=0.5$ and $\chi=1.0$. It can be seen that angular momentum losses make the star to reach the secular instability limit {of the TM1 and the GM1 EOS}, namely the critical mass to collapse to a BH, instead of the mass-shedding limit. This result is in agreement with previous results presented in \citet{2015ApJ...812..100B,2016ApJ...833..107B}.

Table~\ref{tab:AngularMomentum} lists the total angular momentum of the particles accreted by the stars when it crosses an instability limit (if it does) or when the simulation was stopped. {In these cases we worked with the TM1 EOS.} We show the results for two selected values of the angular momentum transfer efficiency parameter, $\chi = 1.0$ and $0.5$. For low energetic SN explosions, it is more probable that the $\nu$NS arrives to the mass-shedding limit {when $\chi=1.0$ or the secular instability limit when $\chi=0.5$}. This is the case for the less energetic explosion of the $15\,M_\odot$ progenitor, the $30\,M_\odot$ and the $40\,M_\odot$ ones. On the other hand, there are few cases in which the NS companion arrive first to the mass-shedding limit. The case for the $25 M_\odot$ progenitor with SN energy scaled by $\eta=0.7$ and the minimum orbital period, and for the $40\, M_\odot$ progenitor with SN energy scaled by $\eta=0.8$, $\eta=0.7$ and $\eta=0.6$. Notice that in the case with $\eta=0.8$ and $\chi = 1.0$ the NS companion arrives to the mass-shedding limit, but the final system is unbound after the SN ejecta leave the system.

\subsection{Accuracy of the mass accretion rate}\label{sec:6.1}

We turn now to analyze the relevance of the $\xi$-parameter of Equation~(\ref{eq:CaptureRadius}) on the mass-accretion rate onto the star. Until now, we have assumed a value $\xi=0.1$ for this parameter. A larger value of $\xi$ results in higher accretion rates: it allows a bigger gravitational capture radius and then more particles can be accreted by the star. In this way, a $\xi=1.0$, the maximum value that $\xi$ can have, will establish an upper limit for the mass accretion rate onto the star. We can also establish a lower limit for the accretion rate, if we allow the star to just accrete the particle which angular momentum is equals or smaller that the angular momentum that a particle orbiting the star would have at the LCO. An equivalent condition is to adopt a varying $\xi$ parameter that equals the capture radius to the radius of the last circular orbit, $r_{\rm LCO}$. But, since the $r_{\rm LCO}$ for the NS is of the order of the NS radius, i.e.~$\sim 10$~km, we will need to increase considerably the number of particles to be able to resolve the NS surface.
 \begin{figure}
   \centering
   \includegraphics[width=0.95\hsize]{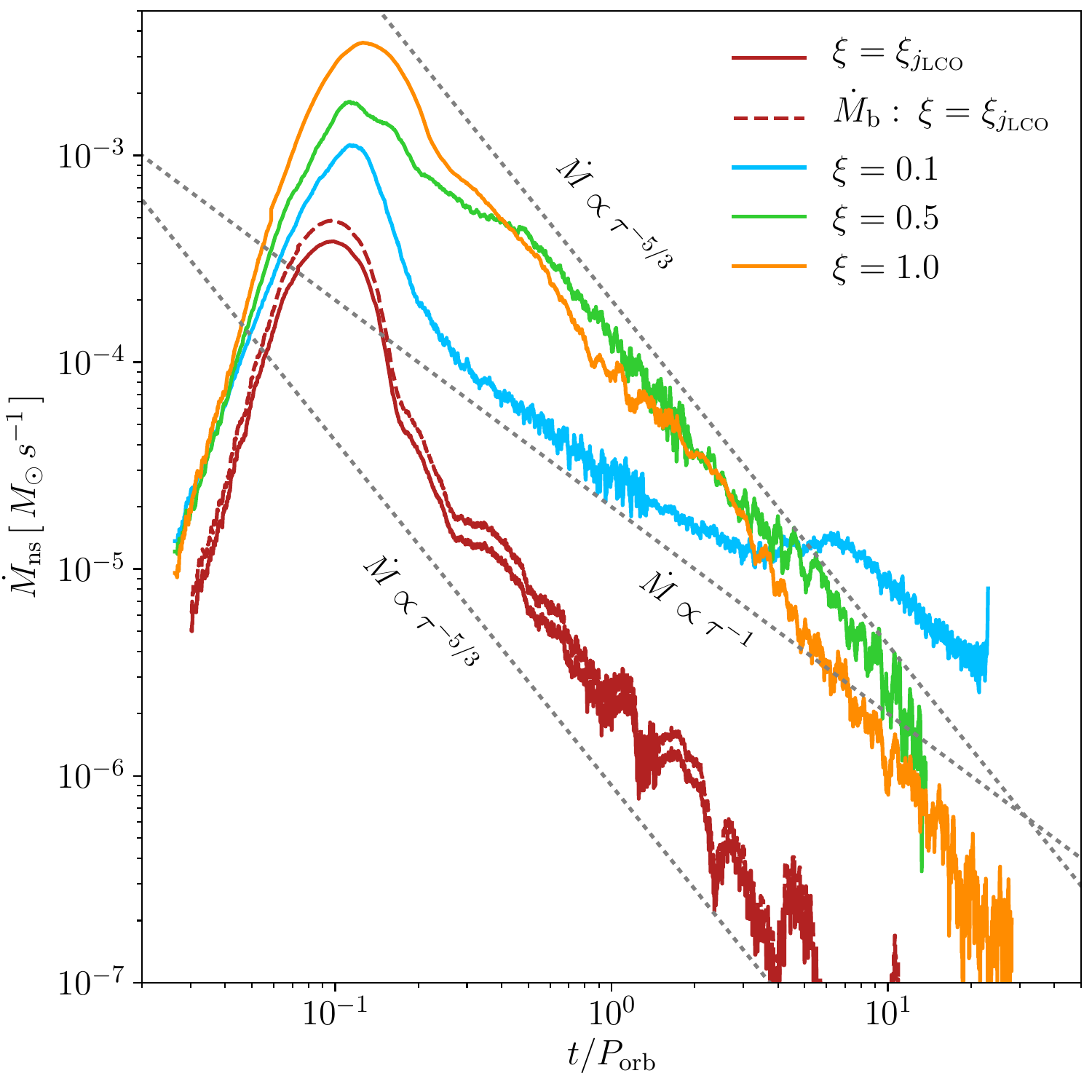}
   \caption{Mass-accretion rate onto the NS companion adopting different values for the $\xi$ parameter in Equation~(\ref{eq:CaptureRadius}). This parameter controls the size of the NS capture radius. A value $\xi=1.0$ establishes an upper limit on the mass-accretion rate. We also defined a lower limit on the accretion (red line), allowing to the star just accreted those particles that have an angular momentum smaller than the one of the LCO around the NS. The initial binary configuration is the one of Figure~\ref{fig:25Mzams_MdotNS}.}
   \label{fig:Mdot_xi}
 \end{figure}

\begin{figure*}
  \centering
  \includegraphics[width=0.9\hsize]{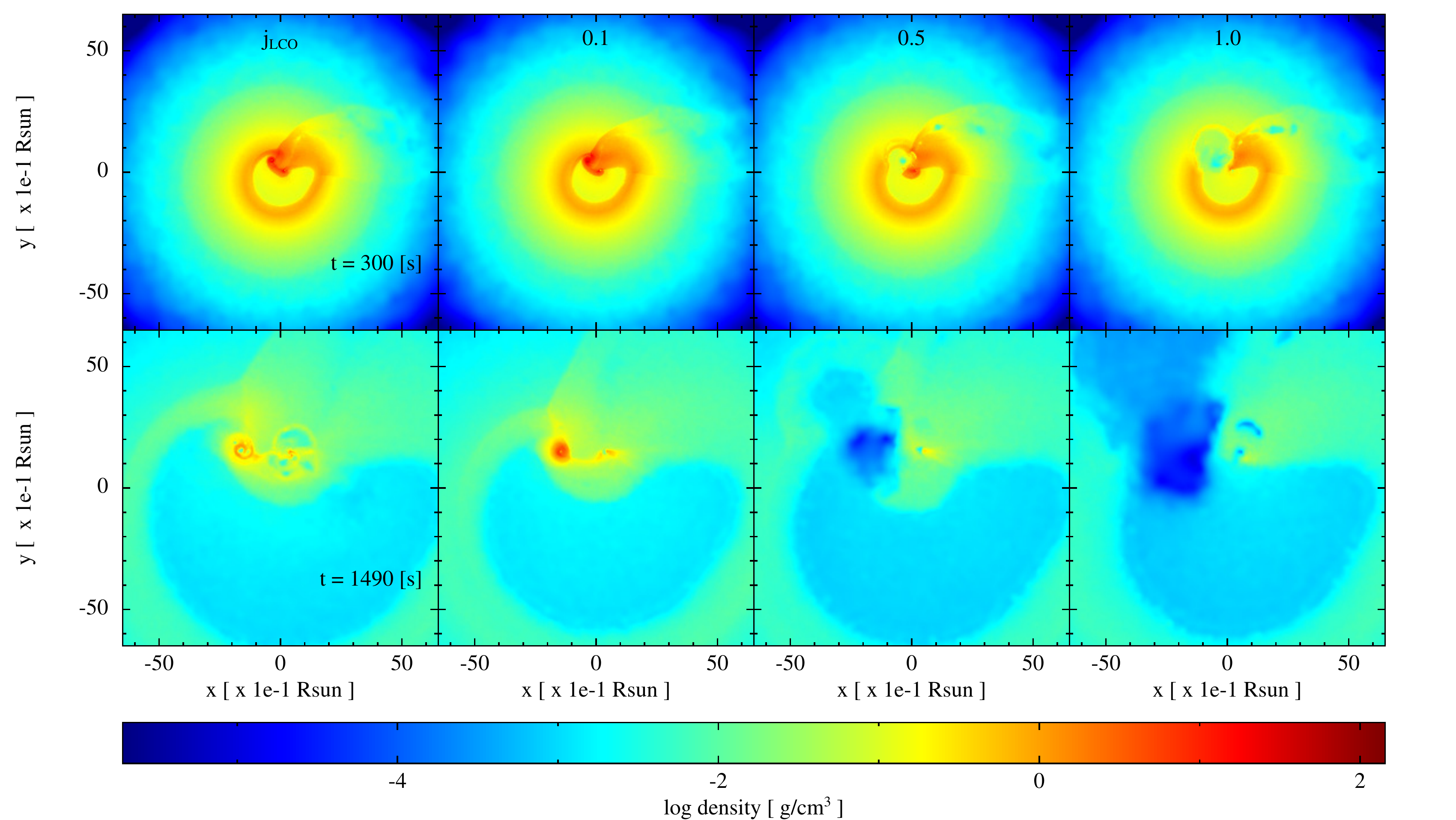}
  \caption{Snapshots of the surface mass density. The initial binary system is formed by the CO$_{\rm core}$ of a $25 M_\odot$ progenitor and a $2M_\odot$ NS with an initial orbital period of around $5$~min. For the plots in the upper panel the system has evolved a time close to one initial orbital period, $300$~s, and for the plots of the bottom panel, the time corresponds to about $24$~min from the beginning of the simulation. The vertical panels correspond to different values for the $\xi$ parameter in Equation~(\ref{eq:CaptureRadius}), from the second to the fourth it is: $0.1$, $0.5$ and $1.0$. In the simulation label $j_{\rm LCO}$ the star accretes just the particles with angular momentum lower than the one of the NS LCO.}
  \label{fig:25Mzams_rhoXi}
\end{figure*}

We have re-run the SPH simulations adopting different values of $\xi$, for the binary system formed by the CO$_{\rm core}$ of the $25\,M_\odot$ progenitor and a $2\, M_\odot$ NS companion, and orbital period of about $5$~min. In Figure~\ref{fig:Mdot_xi} we show the accretion rate onto the NS companion for $\xi=1.0$, $0.5$ and $0.1$. The label $\xi=\xi_{j_{\rm LCO}}$ corresponds to the case in which the star just accretes the particles with angular momentum lower than the one of the LCO. The late mass-accretion rate of the simulations with $\xi=\xi_{j_{\rm LCO}}$ and $\xi=0.5$ and $1.0$ fall almost with the same power-law, $\dot{M}\propto t^{-5/3}$. Also, for the same simulation, Figure~\ref{fig:25Mzams_rhoXi} shows snapshots of the density surface at the binary equatorial plane at two different times.
 \begin{table*}
  \centering
  \caption{NS companion final state}
  \setlength{\tabcolsep}{5pt}
  \begin{tabular}{ccccccc}
    Progenitor& $\eta$ & $\xi$ & $\Delta M_{\rm acc}$ & $\Delta l_{\rm acc}$ & $m_{\rm bound}$ & $a_{\rm orb,f}$  \\
    $M_{\rm zams}$ & & & $M_\odot$ & $c/(GM_\odot^2)$ & $M_\odot$ & $10^{10}$~cm \\
   \hline \hline
   $25M_\odot$ & $1.0$ & $j_{\rm LCO}$ & $0.011$ & $0.0097$ &      $0.00238$       & $14.96$       \\
                    &       &       $0.1$    & $0.078$ &  $4.7460$ &        $0.08100$       & $8.110$ \\
                    &       &       $0.5$    & $0.171$ & $31.921$ & $0.00497$ & $9.780$ \\
                    &       &       $1.0$    & $0.211$ & $42.148$ & $0.00131$ & $48.87$ \\
                     & $0.7$ & $j_{\rm LCO}$ & $0.049$ & $0.0870$ &  $0.07700$ &  $1.724$      \\
                      & $0.7$ & $0.1$ & $0.659$ &  $7.7650$ & $0.16030$ &    $1.021$    \\
                     &  & $1.0$ & $0.633$ & $225.64$ & $0.00453$ &    $1.575$    \\
    \hline
   $30M_\odot$ & $2.0$ &       $0.1$    & $0.077$ &  $7.8460$ & $0.00560$  & $-$ \\
                    & $2.0$ &       $1.0$    & $0.172$ & $22.166$ & $0.00053$  & $-$ \\
   \hline
   $40M_\odot$ & $0.8$ &       $0.1$    & $0.457$ & $30.147$ &         $0.01650$     & $-$ \\
                    & $0.8$ &       $1.0$    & $0.545$ & $56.835$ & $0.00703$ & $-$ \\
\hline
  \end{tabular}
  \label{tab:Var_xi}
\end{table*}
The simulation with $\xi=1.0$ and $\xi =0.5$ gives greater peaks for the mass-accretion rate. This is expected since in these cases the NS capture radius is larger, the star cleans up its surrounding and produces, at later times, a quickly dropping accretion rate. Comparing this simulation with the one with $\xi=0.1$, we can deduce that there is a delay time between the particles that are gravitationally captured by the star and the time it is actually accreted by it. We expect that the simulation with $\xi_{j_{\rm LCO}}$ gives a better resolution of the disk around the NS companion. However, the artificial viscosity used in the code was introduced in order to resolve shocks and does not model the disk viscosity. Then, we are seeing that the particles that circularize around the star, at some point scape from the NS gravitational field, making the mass-accretion rate to drop (see Figure~\ref{fig:25Mzams_rhoXi}). In Table~\ref{tab:Var_xi} we summarize the parameters that characterize the final state of the NS companion as well as the final binary system. We have re-run simulation with the $25\,M_\odot$, $30 \,M_\odot$ and $40 \,M_\odot$ progenitors of the CO$_{\rm core}$. {For each model, we summarized the total mass and angular momentum on the accreted particles, $\Delta\, M_{\rm acc}$ and $\Delta l_{\rm acc}$, the mass bound to the system when the simulation is stopped, $m_{\rm bound}$, and the final orbital separation, $a_{\rm orb,f}$. In general, the larger the $\xi$ parameter the larger the $\Delta l_{\rm acc}$ and $\Delta\, M_{\rm acc}$ depends both on $\xi$ and on the SN explosion energy. The accreted mass increases with $\xi$ but the increment decreases with a decrease of the SN energy, e.g.: for $M_{\rm zams}=25\,M_\odot$ and $\eta=1.0$, the accreted mass increases by $\approx 170$\% when going from $\xi=0.1$ to $\xi=1.0$ while, for $\eta=0.7$, the accreted mass is almost the same for both values of $\xi$.
}

\section{Consequences of the simulations on the XRF/BdHN model}\label{sec:7}

\subsection{Parameters leading to a successful IGC}\label{sec:7.1}

{
The accretion rate increases for higher densities and lower velocities, so we expect it increases with time as the inner ejecta layers, which are denser and slower, pass by the NS. Using an homologously expanding SN ejecta density profile, \citet{2016ApJ...833..107B} derived approximate, analytic formulas for the peak accretion rate, $\dot{M}_{\rm peak}$, and the corresponding peak time, $t_{\rm peak}$ (see Equations~(33)--(34) there). From this estimate they show that, at the lowest order in $M_{\rm ns}/M$, the system satisfies:
}
\begin{eqnarray}\label{eq:tpeak_Mpeak}
t_{\rm peak} \propto \frac{M^{1/3} P_{\rm orb,i}^{2/3}}{v_{\rm star,0}},\quad \dot{M}_{\rm peak} \propto \left(\frac{M_{\rm ns}}{M}\right)^{5/2}\frac{1}{P_{\rm orb,i}},\label{eq:Mpeak}
\end{eqnarray}
{
where $M_{\rm ns}$ is the initial mass of the NS companion, $M = M_{\rm ns}+M_{\rm CO}$ is the initial total binary mass, $M_{\rm CO} = M_{\nu \rm ns}+M_{\rm ej}$ is the mass of the CO$_{\rm core}$, $P_{\rm orb,i}$ is the initial orbital period and $v_{\rm star,0}$ is the velocity of the outermost layer of the SN ejecta. The above dependence confirms that the shorter/smaller the orbital period/separation, the higher the peak accretion rate, $\dot{M}_{\rm peak}$ and the shorter the peak time, $t_{\rm peak}$. It is also confirmed that the highest velocity layer have a negligible contribution to the accretion rate but it is important in the determination of the time at which accretion starts and, consequently, of the peak time. These formulas are relevant to have an insight on the properties of the system and are valid to obtain typical and/or order-of-magnitude estimates of the accretion rate. An analysis of the performance of the analytic formulas with respect to the values obtained from a full numerical integration can be found in the appendix A of \citet{2016ApJ...833..107B}.
}

{
Using the above and approximating the accreted mass as $\Delta M_{\rm acc,an} \approx \dot{M}_{\rm peak} t_{\rm peak}$, it can be checked that it satisfies:
}
\begin{equation}\label{eq:Macc}
\Delta M_{\rm acc, an} \propto \frac{M_{\rm ns}^2}{P_{\rm orb,1}^{1/3} v_{\rm star,0}},
\end{equation}
{
for a fixed CO$_{\rm core}$ mass. For some of the models of table~\ref{tab:inconmodels2}, we summarize in table~\ref{tab:Pred_models} their corresponding order-of-magnitude values of $t_{\rm peak}$, $M_{\rm peak}$ and $\Delta M_{\rm acc}$ from equations~(\ref{eq:tpeak_Mpeak}) and (\ref{eq:Macc}). In order to evaluate the accuracy of these analytic estimates, we have calculated the ratio 
\begin{equation}
\eta*\equiv \frac{\Delta\, M_{\rm acc, sim}}{\Delta\, M_{\rm acc,an}},
\end{equation}
between the mass accreted by the NS companion obtained numerically in the simulation and the one predicted by equation~(\ref{eq:Macc}). We confirm the behavior found in \citet{2016ApJ...833..107B} that the accuracy of the analytic estimate increases ($\eta*$ parameter approaches unity) for longer orbital periods (less compact binaries) and we found here that also for more energetic SNe. We also confirm that the analytic formula underestimates ($\eta*$ much larger than unity) the accretion rate for the short orbital period binaries. Indeed, we found that when $\eta*$ is in excess of unity the gravitational collapse of the NS is more probable. It is important to recall that the equations~(\ref{eq:tpeak_Mpeak}) were derived assuming that the binary period is constant while the SN ejecta expand and neglecting the contribution of the $\nu$NS in the evolution of the system. In order to have an idea of the goodness of those assumptions, we also show in table~\ref{tab:Pred_models} the ratio between the final and initial binary separation, and the final and initial mass of the $\nu$NS.
}

{From equation~(\ref{eq:Macc}), we see that not only the larger the $M_{\rm ns}$ the less the mass it needs to accrete to reach the critical mass for gravitational collapse and BH formation,
}
\begin{equation}
\Delta M_{\rm crit} \equiv M_{\rm crit} - M_{\rm ns},
\end{equation}
{
where $M_{\rm crit} = M_{\rm max}^{J_{\rm ns}\neq 0}$ (see Equation~\ref{eq:Mcrit}), but also the larger the $M_{\rm ns}$ the higher the accretion rate and the larger the mass it accretes in a time $t_{\rm peak}$. The above analysis places as main parameters of the system for the occurrence of an XRF or a BdHN are the initial mass of the NS companion, the orbital period and the SN velocity, or equivalently, the SN explosion/kinetic energy.
}

{
We defined in \citet{2016ApJ...833..107B} the orbital period $P_{\rm max}$ separating the XRFs and BdHNe subclasses, namely the maximum orbital period up to which the collapse of the NS companion to a BH, induced by accretion, occurs. Thus, $P_{\rm max}$ is set by the orbital period for which $\Delta M_{\rm acc} = \Delta M_{\rm crit}$, which leads to
}
\begin{equation}\label{eq:Pmax}
P_{\rm max} \propto \frac{M_{\rm ns}^6}{v_{\rm star,0}^3(M_{\rm crit}-M_{\rm ns})^3},
\end{equation}
{
which is a monotonically increasing function of $M_{\rm ns}$, as obtained in \citet{2016ApJ...833..107B} (see Figure~5 there) from the full numerical integration of the NS evolution equations.
}

\begin{table*}
\centering
\caption{{Comparison between the analytic estimate of the accretion process and the full numerical results. In the first two columns is presented the label of the model and the SN velocity of the last layer at the beginning of the simulation. The next three columns correspond to the estimations of $t_{\rm peak}$, $\dot{M}_{\rm peak}$ and $M_{\rm acc}$, calculated from equations (\ref{eq:tpeak_Mpeak}) and (\ref{eq:Macc}). The $\eta*$ parameter from sixth columns corresponds to the ratio between the mass accreted obtained from the simulations and the one calculated with equation (\ref{eq:Macc}). The last two columns are the ratio between the initial and final orbital separation and the initial and final mass of the $\nu$NS, respectively. We reported these two last quantities since in the analytical approximation they were assumed as constant.}}
\setlength{\tabcolsep}{1.9pt}
\fontsize{8}{8}
{\scriptsize
\begin{tabular}{lccccccc|lccccccc}
  \hline
  \hline
  Model &  $v_{\rm star,0}$ & $t_{\rm peak}$  & $\dot{M}_{\rm peak}$ & $\Delta M_{\rm acc,an}$ & $\eta*$ & $\frac{a_{\rm orb,f}}{a_{\rm orb, i}}$ & $\frac{M_{\nu ns}}{M_{\nu ns,0}}$ & Model &  $v_{\rm star,0}$ & $t_{\rm peak}$  & $\dot{M}_{\rm peak}$ & $\Delta M_{\rm acc,an}$ & $\eta*$ & $\frac{a_{\rm orb,f}}{a_{\rm orb, i}}$ & $\frac{M_{\nu ns}}{M_{\nu ns,0}}$  \\
      & $10^{8}$~cm/s & s & $10^{-4} \, M_\odot$/s & $M_\odot$ &  &  &  & & $ 10^{8}$~cm/s & s & $10^{-4} \, M_\odot$/s & $M_\odot$ &  & &  \\
 \hline \hline
   \multicolumn{8}{c|}{ $M_{\rm ZAMS}=25\, M_{\odot}$ Progenitor ( $\rho_{\rm core}R_{\rm star}^{3}=3.237\, M_\odot$ )} & \multicolumn{8}{c}{ $M_{\rm ZAMS}=30\, M_{\odot}$ Progenitor - exp2 ( $\rho_{\rm core}R_{\rm star}^{3}=5.280\, M_\odot$ )}   \\ \hline
    25m1p1e    & $9.5$ & $48.14$ & $5.87$ & $0.028$ & $3.01$ &  $6.47$ & $1.06$ & 30m1p1e   &  $5.21$ & $340.06$ & $0.984$ & $0.033$ & $11.428$ & $ 1.32 $ & $2.10$ \\
  25m2p1e   & $9.5$ & $70.69$ & $3.30 $ & $0.023$ & $1.24$ &  $  -  $ & $1.04$ &30m2p1e   &  $5.21$ & $513.53$ & $0.767$ & $0.039$ & $5.784$ & $ - $ & $1.02$ \\ 
  25m3p1e    & $9.5$ & $87.56 $ & $2.39$ & $0.021$ & $1.14$ &  $  -  $ & $1.04$ & 30m1p12e   & $5.71$ & $310.43$ & $0.984$ & $0.031$ & $11.428$ & $ 3.01 $ & $1.44$ \\
  25m4p1e  & $9.5$ & $106.8$ & $1.77$ & $0.019$ & $0.74$ &  $  -  $ & $1.04$ & 30m1p2e  & $5.21$ & $239.91$ & $0.984$ & $0.023$ & $0.719$ & $ - $ & $1.02$ \\
  25m1p09e   & $9.1$ & $50.74$ & $5.87$ & $0.029$ & $5.43$  &  $3.10 $ & $1.09 $ &30m1p31e   & $9.18$ & $193.15$ & $0.984$ & $0.019$ & $0.895$ & $ - $ & $1.01$  \\ 
 \cline{9-16} \cline{9-16}
  25m1p08e  & $8.5$ & $53.82$  & $5.87$ & $0.032$ & $13.95$ &  $1.82 $ & $ 1.12 $ & \multicolumn{8}{c}{ $M_{\rm ZAMS}=40\, M_{\odot}$ Progenitor ( $\rho_{\rm core}R_{\rm star}^{3}=7.47\, M_\odot$ )}  \\ \cline{9-16}
  25m1p07e   & $7.9$ & $57.54$ & $5.87$ & $0.034$ & $18.37$ &  $0.77$ & $1.28 $ & 40m1p1e  & $6.58$ & $33.74$ & $9.39$ & $0.032$ & $3.74$ & $ - $ & $1.01$ \\
  25m2p07e   & $7.9$ & $84.49$  & $3.30$ & $0.028$ & $18.71$ &  $1.99$ & $1.27 $ & 40m2p1e  & $6.58$ & $49.67$ & $5.58$ & $0.028$ & $1.65$ & $ - $ & $1.01$  \\
  25m3p07e   & $7.9$ & $104.65$ & $2.39$ & $0.025$ & $20.87$ &  $2.30$ & $1.24$  &  40m1p09e  & $6.24$ & $35.62 $ & $9.39$ & $0.033$ & $8.14$ & - $ $ & $ 1.012$ \\
  25m5p07e   & $7.9$ & $147.78$ & $1.43$ & $0.021$ & $19.02$ &  $2.58$ & $1.17 $ &  40m1p08e  & $5.88$ & $37.78$  & $9.39$ & $0.035$ & $15.06$ & $ - $ & $ 1.02 $ \\
  14Mns1p1e   & $9.5$ & $43.58$ & $3.71$ & $0.016$ & $2.65$ &$12.6 $    & $1.07$ & 40m1p07e   & $5.51$ & $40.39$  & $9.39$ & $0.038$ & $27.21$ & $ 8.08 $ & $ 1.13 $ \\
  14Mns1p07e  & $7.9$ & $52.08$ & $3.71$ & $0.019$ & $21.16$ &  $0.80$ & $1.33 $ &  40m2p07e & $5.51$ & $59.38$  & $5.59$ & $0.033$ & $22.76$ & $ 54.8 $ & $ 1.12 $ \\
 \cline{1-8} \cline{1-8}
 \multicolumn{8}{c|}{ $M_{\rm ZAMS}=30\, M_{\odot}$ Progenitor -exp 1 ( $\rho_{\rm core}R_{\rm star}^{3}=5.280\, M_\odot$ )} & 40m4p07e & $5.51$ & $94.29$  & $2.99$ & $0.028$ & $ 17.97 $ & $-$ & $ 1.06 $  \\
 \hline
     30m1p1e   &  $8.78$ & $96.22$ & $0.91$ & $0.009$ & $0.798$ & $ - $ & $1.004$ & & & & & & & &  \\
  30m1p07e   &  $7.35$ & $115.01$ & $0.91$ & $0.018$ & $1.431$ & $ - $ & $1.004$  & & & & & & & & \\
  30m1p05e   & $6.213$ & $136.07$ & $0.91$ & $0.022$ & $2.498$ & $ - $ & $1.008$  & & & & & & & & \\
 30m1p03e   &  $4.81$ & $175.67$ & $0.91$ & $0.016$ & $28.41$ & $ 4.88 $ & $1.068$ & & & & & & & &  \\
 30m2p03e   &  $4.81$ & $265.28$ & $0.52$ & $0.013$ & $13.84$ & $ - $ & $1.05$ & & & & & & & &  \\
   \hline
     \hline
\end{tabular}
}
\label{tab:Pred_models}
\end{table*}
%

\subsection{Binary evolutionary path}\label{sec:7.2}

{
As we have discussed, in our binary scenario the presence of the CO$_{\rm core}$ provides a natural explanation for the association of long GRBs with type Ic SNe/HNe. The requirement of the CO$_{\rm core}$, at the same time, leads to the formation of compact orbits (few minutes orbital period) needed for the occurrence of high accretion rates onto the NS companion. The HNe, i.e.~the high-velocity SNe associated with GRBs, are explained by the feedback of the energy injected into the SN by the hypercritical accretion \citep{2016ApJ...833..107B} and by the GRB $e^+e^-$ plasma \citep{2018ApJ...852...53R}. This implies that the initial SN explosion/kinetic energy is initially ordinary.
}

{
Once we have established the main binary parameters needed for the explanation of the XRFs and BdHNs following the IGC scenario, it is natural to discuss whether these conditions, namely CO$_{\rm core}$-NS binaries with these features, occur in a given evolutionary path. It is known that massive binaries can evolve into compact-object binaries such as NS-NS and NS-BH. Typical formation scenarios argue that, after the first SN explosion, the compact remnant enters a common-envelope phase with the companion. Such a phase leads to a compaction of the binary orbit. Finally, after the collapse of the companion star, a NS-BH or NS-NS binary is formed if the system remains bound \citep{1999ApJ...526..152F,2012ApJ...759...52D,2014LRR....17....3P}. More recently, a different evolutionary scenario has been proposed in which, after the collapse of the primary star to a NS, namely after the first SN, the binary undergoes a series of mass transfer phases leading to the ejection of both the hydrogen and helium shells of the secondary. This process naturally leads to a binary composed of a CO$_{\rm core}$ and a NS. When the CO$_{\rm core}$ collapses, namely after the second SN in the binary evolutionary path, a compact binary system is formed. The X-ray binary/SN community refer to these systems as ``ultra-stripped'' binaries. Such systems have been also called into cause to explain the population of NS-NS and low-luminosity SNe \citep[see e.g.][]{2013ApJ...778L..23T,2015MNRAS.451.2123T}. The rate of these ultra-stripped binaries are expected to be $0.1$--$1\%$ of the total SN rate \citep{2013ApJ...778L..23T}. Most of the theoretically derived population of these binaries show tight orbital periods in the range $50$--$5000$~h. Ultra-stripped systems have been proposed in these works to dominate the formation channel of NS-NS. In addition, the formation of NS-BH systems was not there considered since 1) they did not find systems where the CO$_{\rm core}$ collapses directly to a BH and 2) they did not consider the possibility of an IGC process in which the BH is formed by the NS companion and not by the collapse of the CO$_{\rm core}$.
}

{
Binary evolutionary paths similar to the above one of ultra-stripped binaries and leading to the tight CO$_{\rm core}$-NS binaries studied here were proposed in \citet{2012ApJ...758L...7R,2015ApJ...812..100B,2015PhRvL.115w1102F}. However, population synthesis analyses scrutinizing the possibility of even tighter binaries than the ones previously considered as well as the physics of the hypercritical accretion process onto the NS companion in the second SN explosion, leading to XRFs and/or BdHNe, have not been yet considered in the literature and deserve a dedicated analysis.
}

\subsection{Occurrence rate density}\label{sec:7.3}

{
The existence of ultra-stripped binaries supports our scenario from the stellar evolution side. Clearly, XRF and BdHN progenitors should be only a small subset that result from the binaries with initial orbital separation and component masses leading to CO$_{\rm core}$-NS binaries with short orbital periods, e.g.~$100$--$1000$~s for the occurrence of BdHNe. This requires fine-tuning both of the CO$_{\rm core}$ mass and the binary orbit. From an astrophysical point of view the IGC scenario is characterized by the BH formation induced by the hypercritical accretion onto the NS companion and the associated GRB emission. Indeed, GRBs are a rare phenomenon and the number of systems approaching the conditions for their occurrence must be low. If we assume that XRFs and BdHNe can be final stages of ultra-stripped binaries, then the percentage of the ultra-stripped population leading to these long GRBs must be very small. The observed occurrence rate of XRFs and BdHNe has been estimated to be $\sim 100$~Gpc$^{-3}$~y$^{-1}$ and $\sim 1$~Gpc$^{-3}$~y$^{-1}$, respectively~\citep{2016ApJ...832..136R}, namely the $0.5\%$ and $0.005\%$ of the Ibc SNe rate, $2\times 10^4$~Gpc$^{-3}$~y$^{-1}$ \citep[see e.g.][]{2007ApJ...657L..73G}. It has been estimated that $(0.1$--$1\%)$ of the SN Ibc could originate from ultra-stripped binaries \citep{2013ApJ...778L..23T}, which would lead to an approximate density rate of $(20$--$200)$~Gpc$^{-3}$~y$^{-1}$. This would imply that a small fraction ($\lesssim 5\%$) of the ultra-stripped population would be needed to explain the BdHNe while, roughly speaking, almost the whole population would be needed to explain the XRFs. These numbers, while waiting for a confirmation by further population synthesis analyses, would suggest that most SNe originated from ultra-stripped binaries should be accompanied by an XRF. It is interesting that the above estimates are consistent with traditional estimates that only $\sim 0.001$--$1\%$ of massive binaries lead to double compact-object binaries \citep[see e.g.][]{1999ApJ...526..152F,2012ApJ...759...52D,2014LRR....17....3P}.

\subsection{Consequences on GRB observations and analysis}\label{sec:7.4}

{
We have recently addressed in \citet{2018ApJ...852...53R} the essential role of X-ray flares in differentiating and act as separatrix between the BdHN model and the ``collapsar-fireball'' model of GRBs \citep{1993ApJ...405..273W}. The gamma-ray spikes in the GRB prompt emission occur at $10^{15}$--$10^{17}$~cm from the source and have Lorentz factor $\Gamma \sim 10^2$--$10^3$. Instead, the analysis of the thermal emission in the X-ray flares in the early (source rest-frame time $t\sim 10^2$~s) afterglow of BdHNe, showed that X-ray flares occur at radii $\sim10^{12}$~cm and expand mildly-relativistically, e.g.~$\Gamma\lesssim 4$ \citep{2018ApJ...852...53R}. These model independent observations are in contrast with an ultra-relativistic expansion all the way from the GRB prompt emission to the afterglow, as traditionally adopted in the majority of the GRB literature in the context of the collapsar-fireball scenario.
}

{
In \citet{2018ApJ...852...53R} we tested whether the BdHN scenario could explain both the ultra-relativistic gamma-ray prompt emission as well as the mildly-relativistic X-ray flare data. Our numerical simulations in \citet{2016ApJ...833..107B} have shown that, at the moment of BH formation and GRB emission, the SN ejecta become highly asymmetric around the collapsing NS. In the direction pointing from the CO$_{\rm core}$ to the accreting NS outwards and lying on the orbital plane, the NS caves a region of low baryonic contamination in which the GRB $e^+e-$ plasma, created once the NS collapses and forms the BH, can expand with high $\Gamma \sim 10^2$--$10^3$ explaining the GRB prompt emission. In the other directions, the GRB $e^+e-$ plasma impacts the SN ejecta at approximately $10^{10}$~cm, evolves carrying a large amount of baryons reaching transparency at radii $10^{12}$~cm with a mildly $\Gamma\lesssim 4$. This result is in clear agreement with the X-ray flares data \citep[see][for details]{2018ApJ...852...53R, 2018EPJWC.16804009M}. A most important prediction of this scenario is that the injection of energy and momentum from the GRB plasma into the ejecta transforms the SN into a HN \citep[see][for a detailed analysis of GRB 151027A]{2017arXiv171205001R}.
}


{
The strong dependence of $P_{\rm max}$ on the initial mass of the NS companion opens the interesting possibility of producing XRFs and BdHNe from binaries with similar short (e.g.~$P\sim $few minutes) orbital periods and CO$_{\rm core}$ properties: while a system with a massive (e.g.~$\gtrsim 2~M_\odot$) NS companion would lead to a BdHN, a system with a lighter (e.g.~$\lesssim 1.4~M_\odot$) NS companion would lead to an XRF. This predicts systems with a similar initial SN, leading to a similar $\nu$NS, but with different GRB prompt and afterglow emission. Clearly being the GRB energetics different, the final SN kinetic energy should be also different being it larger for the BdHNe.
}

{
Besides confirming that XRFs and BdHNe can be produced by these binaries, the present new 3D SPH simulations have also shown new, possibly observable features in GRB light-curves and spectra, e.g.:
}

{
1) the hypercritical accretion occurs both on the NS companion and on the $\nu$NS with a comparable accretion rate, hence roughly doubling the accretion power of the system. 
}

{
2) The above leads to the clear possibility that, under specific conditions, a BH-BH binary can be produced; see e.g. the simulation `30m1p1eb' in Table~\ref{tab:AngularMomentum}. Since the system remains bound (see Table~\ref{tab:inconmodels2}), the two stellar-mass BHs will merge in due time owing to the emission of gravitational waves. However, no electromagnetic emission is expected from such a merger and, in view of the typically large cosmological distances of GRBs, their detection by gravitational-wave detectors such as the ground-based interferometers of the LIGO-Virgo network appears to be difficult.	
}

{
3) The SNe with lower explosion energy create a long-live hypercritical accretion process and produce an enhancement at late times of the accretion rate onto the $\nu$NS. Such a revival of the accretion rate does not exist in the case of single SN namely in absence of the NS companion (see Figure~\ref{fig:25Mzams_Mnsdot_Energy}). In these cases there is a higher probability for the detection of the early phase of an XRF/BdHN by X-ray detectors.
}

{
4) The asymmetric SN explosions lead to a quasi-periodic behavior of the accretion rate which could be detectable by the X-rays instruments (see e.g.~Figure~\ref{fig:25Mzmans_Assymetries}). A possible detection of this feature would be a further test of the binary nature and would pinpoint the orbital period of the binary progenitor.
}

On the other hand, we have evaluated in our simulations whether the binary remains gravitationally bound or it becomes unbound by the SN explosion. Therefore, we are determining the space of initial binary and SN explosion parameters leading to the formation of $\nu$NS-NS or $\nu$NS-BH binaries. {This shows the interesting feature that long GRBs (XRFs and BdHNe) lead to the binary progenitors of short GRBs, hence this topic is relevant for the understanding of their relative density rate and will be further analyzed in forthcoming works.}

\section{Conclusions}
\label{sec:8}

We have performed the first full numerical 3D SPH simulations of the IGC scenario: in a CO$_{\rm core}$-NS binary system, the CO$_{\rm core}$ collapses and explodes in a SN triggering a hypercritical accretion process onto the NS companion. The initial conditions for the simulations were constructed as follows. The CO$_{\rm core}$ stars are evolved using the KEPLER evolution code \citep{heger10} until the conditions for the collapse are met. Then, the stars are exploded with the 1D core-collapse code \citep{1999ApJ...516..892F}. When the forward SN shock reaches the stellar radius, we map the explosion to a 3D-SPH configuration and continue the evolution of the SN expansion with a NS binary companion using the SNSPH code \citep{2006ApJ...643..292F}.

We followed the evolution of the SN ejecta, including their morphological structure, under the action of the gravitational field of both the $\nu$NS and the NS companion. We estimated the accretion rate onto both stars with the aid of Equation~(\ref{eq:MbMns}). The baryonic mass-accretion rates are calculated from the mass of the SPH particles accreted. We have shown that matter circularizes in a disk-like structure around the NS companion (see, e.g., Figure~\ref{fig:25Mzams2Mns}). Therefore, for the angular momentum transfer to the NS, we have adopted that the particles are accreted from the LCO.

We determined the fate of the binary system for a wide parameter space including different CO$_{\rm core}$ masses (see Table~\ref{tab:ProgSN}), orbital periods, SN explosion geometry and energies, {as well as different masses of the NS companion}. We evaluated, for selected NS nuclear EOS, if the accretion process leads the NS either to the mass-shedding limit, or to the secular axisymmetric instability for gravitational collapse to a BH, or to a more massive, fast rotating, but stable NS. We have also analyzed the case of asymmetric SN explosions. We assessed if the binary remains gravitationally bound or it becomes unbound by the explosion. With this information we determined the space of initial binary and SN explosion parameters leading to the formation of $\nu$NS-NS or $\nu$NS-BH binaries.

{
It is worth mentioning some issues not included in the numerical calculations presented here and which deserve to be further studied and/or improved:
}

{
1) We need to improve the resolution of the code to handle the spatial region in the vicinity of the NS companion at scales $10^6$--$10^7$~cm. This will allow to resolve better the structure of the circularized matter (accretion disk) near the NS and evaluate more accurately the angular momentum transfer to the NS (e.g. the value of the angular momentum efficiency parameter, $\chi$). These distances are of the order of the NS Schwarzschild radius, therefore a general relativistic code is needed for this task.
}

{
2) The above will also allow to study the possible outflows from the hypercritical accretion process previously found in the case of fallback accretion \citep{2009ApJ...699..409F} and in which heavy nuclei via r-process nucleosynthesis can be produced \citep{2006ApJ...646L.131F}.
}

{
3) The transformation of the SN into a HN by the impact of the GRB into the ejecta. We are currently performing the hydrodynamical simulations of the interaction of the GRB $e^+e^-$ plasma using a one-dimensional relativistic hydrodynamical module included in the freely available PLUTO \citep[see][for further details]{2018ApJ...852...53R}. We have used in these calculations the SN ejecta profiles at the moment of the NS collapse obtained by \citet{2016ApJ...833..107B}, and evolved them from that instant on under the assumption of homologous expansion.
}

{
4) In view of the above, the present simulations remain accurate/valid up to the instant where the NS reaches the critical mass, hence forming a BH. After that instant the GRB-SN interaction becomes relevant. In the systems when the NS companion does not reach the critical mass, our simulations remain accurate/valid all the way up to the instant where the entire SN ejecta blows past the NS position.
}

We discussed in section~\ref{sec:6} some of the consequences of our simulations in the analysis of GRBs within the context of the IGC scenario. {The simulations have confirmed, extended and improved important previous results. In particular:
}

{1) We showed that long GRBs (XRFs and BdHNe) can be produced by these binaries depending on the binary parameters. We have shown that the main parameters defining the fate of the system are: the initial mass of the NS companion, the orbital period and the SN velocity (or kinetic/explosion energy).}

{2) The accreting NS companion induces high asymmetries in the SN ejecta which are relevant in the GRB analysis.} Recent results on the thermal emission of the X-ray flares in the early (source rest-frame times $t\sim 10^2$~s) afterglow of long GRBs show that they occur at radii $\sim10^{12}$~cm and expand mildly-relativistically with $\Gamma\lesssim 4$. This was shown to be in agreement with the BdHNe of the IGC scenario \citep[see][and section~\ref{sec:6} above for details]{2018ApJ...852...53R}: the $e^+e-$ plasma of the GRB, relativistically expanding from the newborn BH, collides with the SN ejecta at distances of the order of $10^{10}$~cm, to then reach transparency at $10^{12}$~cm with $\Gamma\lesssim 4$. The 3D simulations presented in this work will be essential to explore the dynamics of the $e^+e-$ plasma along all spatial directions and to estimate, as a function of the viewing angle, the light-curve and spectral properties of BdHNe \citep[see e.g.][]{2018ApJ...852...53R, 2018arXiv180305476R, 2017arXiv171205001R}.

{3) One of the most interesting issues is that we have confirmed that some of the systems remain bound after the explosion, implying that XRFs form $\nu$NS-NS binaries and BdHNe form $\nu$NS-BH systems. Therefore, long GRBs (XRFs and BdHNe) produce the binary progenitors of short GRBs, after the shrinking of their orbit until the coalescence by the emission of gravitational waves. The analysis of the number of systems leading to $\nu$NS-NS and $\nu$NS-BH binaries becomes very important for the explanation of the relative occurrence rate of long and short GRBs.}

{4) We have also outlined consequences on the accretion process and its observational features in the case of relatively weak SN explosion energies as well as for intrinsically asymmetric ones. We have addressed possible new observable features in GRB light-curves and spectra, e.g. systems experiencing longer, stronger and quasi-periodic accretion episodes to be possibly observed by X-ray instruments. The observation of this kind of phenomena in the early phases of a GRB would benefit from a new mission operating in soft X-rays like, e.g., THESEUS \citep{2018AdSpR..62..191A}.}

{5) We have also shown that there is also the possibility that some CO$_{\rm core}$-NS binaries could produce BdHNe leading to BH-BH binaries due to the collapse to a BH not only of the NS companion but also of the $\nu$NS due to a massive fallback accretion. The final merger of the two stellar-mass BHs has no electromagnetic counterpart to be detected and its gravitational-wave emission, in view of the large distances of these sources, appears to be extremely weak to be detected by current interferometers such as LIGO-Virgo.}

\acknowledgements
{\bf Acknowledgements:} {We thank the Referee for the comments and suggestions to improve the presentation of our results.} The work of L.B., C.E., and C.F. was partially funded under the auspices of the U.S. Dept. of Energy, and supported by its contract W-7405-ENG-36 to Los Alamos National Laboratory. Simulations at LANL were performed on HPC resources provided under the Institutional Computing program.


\newpage
\appendix

\section{Numerical convergence}
\label{app:1}

In order to evaluate the convergence of our SPH simulation, we have done some numerical experiments varying the number of the particles with which we model the SN ejecta for the different pre-SN progenitors.

We performed simulations with $1$, $1.5$, $2$ and $3$ million particles with different progenitors and with different values of the $\xi$ parameter in Equation~(\ref{eq:CaptureRadius}). We summarize the results of these simulations in Table~\ref{tab:Nparticle}. We compare and contrast the final accreted mass and angular momentum of the $\nu$NS and the NS companion, the final orbital period and eccentricity of the orbit. We also report the relative error of these quantities taking as the reference values the ones of the simulation of about 1 million of particles.

We show in Figure~\ref{fig:25Mzams_Nparticles_rho} profiles of the density on the binary orbital plane and along different directions, taking the NS companion as the center of the reference frame. Figure~\ref{fig:25Mzams_NparticlesMrate} shows the mass-accretion rate on the $\nu$NS and the NS companion. Finally, we plot in Figure~\ref{fig:25Mzams_NparticlesFluxes}, the flux of mass and angular momentum onto the NS companion at two different distances from it: $r=0.02\, R_\odot$ and at $r=R_{\rm cap}$, defined as the maximum capture radius of the NS in each iteration. All these figures corresponds to the simulation of an initial binary system formed by the CO$_{\rm core}$ of the $25\, M_\odot$ progenitor (see Table~\ref{tab:ProgSN}) and a $2\, M_\odot$ NS with an initial orbital period of $\approx 2$~min.

%
\begin{table*}
  \centering
  \caption{Convergence study of the SPH simulation of the IGC scenario.}
 \setlength{\tabcolsep}{1.6pt}
\scriptsize
  \begin{tabular}{c|ccc|cccc|cccc|cccc}
    \hline
    Progenitor & $N$ & $\eta$ & $\xi$ & $m_{\nu ns}$ & ${\rm Er}(m_{\nu ns})$ & $L_{\nu-ns}$ &  ${\rm Er}(L_{\nu ns})$   & $m_{\rm ns}$ & ${\rm Er}(m_{\rm ns})$ & $L_{\rm ns}$ & ${\rm Er}(L_{\rm ns})$ & $p_{\rm orb,f}$ & ${\rm Er}(p_{\rm orb})$ & $e$  & ${\rm Er}(e)$ \\
 $M_{\rm zams}$   & million & & & $(\,M_\odot\,)$ & & $(\, 10^{51}\,{\rm g\, cm^2\, /\, s}\,)$ & & $(\, M_\odot\,)$ & & $(\,10^{51}\,{\rm g\, cm^2\, /\, s}\,)$ & & $(\,{\rm s}\,)$ &  & & \\
     \hline \hline
     \multirow{10}{*}{$25M_\odot$}  & 1.0 & \multirow{4}{*}{$1.0$} & \multirow{4}{*}{$0.1$} & $1.964$ & & $3.478$ &  & $2.078$ & & $3.288$ & & $6298.29$ & & $0.860$ & \\
                      & $1.5$ &       &       & $1.951$ & $0.0066$ & $3.522$ & $0.0127$ & $2.064$ & $0.0067$ & $3.323$ & $0.0106$ & $8274.54$ & $0.3137$ & $0.883$ & $0.0267$ \\
                      & $2.0$ &       &       & $1.943$ & $0.0106$ & $3.455$ & $0.0066$ & $2.065$ & $0.0063$ & $3.250$ & $0.0115$ & $7868.58$ & $0.2493$& $0.880$ & $0.0232$\\
                      & $3.0$ &       &       & $1.935$ & $0.0147$ & $3.482$ & $0.0012$ & $2.051$ & $0.0129$ & $3.292$ & $0.0012$ & $10744.9$ & $0.706$ & $0.897$ & $0.0430$ \\
                      \cline{2-16}
                      & $1.0$   & \multirow{3}{*}{$1.0$} & \multirow{3}{*}{$0.5$} & $1.915$ & & $3.627$ & & $2.126$ & & $3.255$  & & $9843.52$ & & $0.892$ & \\
                      & $1.5$ &       &       & $1.906$ & $0.0046$ & $3.434$ & $0.053$ & $2.099$ & $0.0126$ & $3.118$ & $0.0428$ & $14031.1$ & $0.4254$ & $0.923$ & $0.0348$ \\
                      & $2.0$ &       &       & $1.900$ & $0.0078$ & $3.499$ & $0.035$ & $2.102$ & $0.0112$ & $3.155$ & $0.0307$ & $12871.2$ & $0.3075$ & $0.913$ & $0.0235$ \\
                      \cline{2-16}
                      & $1.0$ & \multirow{3}{*}{$1.0$} & \multirow{3}{*}{$1.0$} & $1.937$ & & $3.669$ & & $2.209$ & & $3.217$ & & $91557.2$ & & $0.979$ & \\
                      & $1.5$ &       &       & $1.926$ & $0.0056$ & $3.119$ & $0.1499$ & $2.189$ & $0.0091$ & $2.745$ & $0.1467$ & $25489.6$ & $0.7215$ & $0.962$ & $0.0174$ \\
                      & $2.0$ &       &       & $1.919$ & $0.0092$ & $3.461$ & $0.0566$ & $2.179$ & $0.0135$ & $3.047$ & $0.0528$ & $145340.6$ & $0.5874$ & $0.985$ & $0.0061$ \\
     \hline \hline
     \multirow{4}{*}{$30M_\odot$ \footnote{$E_{\rm sn}=2.19\times 10^{51}\, {\rm erg}$} }  & 1.0 & \multirow{2}{*}{$2.0$} & \multirow{2}{*}{$0.1$} & $1.783$ & & $4.499$ & & $2.077$ & & $3.861$ & & $-$ & & $1.44$ & \\
                      & $2.0$ &                        &                        & $1.772$ & $0.0061$ & $4.451$ & $0.0106$ & $2.043$ & $0.0164$ & $3.859$ & $0.00051$ & $-$ & & $1.50$ & $0.0416$ \\
                      \cline{2-16}
                      & $1.0$ & \multirow{2}{*}{$2.0$} & \multirow{2}{*}{$1.0$} & $1.781$ & & $4.087$ & & $2.172$ & & $3.351$ & & $-$ & & $1.38$ & \\
                      & $2.0$ &                        &                        & $1.769$ & $0.0067$ & $4.614$ & $0.1289$ & $2.115$ & $0.0262$ &$3.859$ & $0.1550$ & $-$ & & $1.55$ & $0.1096$ \\

     \hline
    $40M_\odot$  & $1.0$ & \multirow{2}{*}{$1.0$}  & \multirow{2}{*}{$0.1$} & $1.875$ & & $4.419$ & & $2.124$ & & $3.902$ & & $-$ &  &$1.84$ & \\
                      & $2.0$ &                         &                        & $1.869$ & $0.0032$ & $4.276$ & $0.0323$ & $2.069$ & $0.0259$ & $3.862$ & $0.0102$ & $-$ & & $2.00$ & $0.0869$ \\
    \hline \hline
  \end{tabular}
  \tablecomments{For each simulation, the four first columns show the progenitor of the CO$_{\rm core}$, the number of particles used in the simulation, the $\eta$ factor that scales the kinetic and internal energy of the SPH particles to mimic a weaker or stronger SN explosions and the $\xi$ parameter that determines the size of the capture radius. In the following columns we show the final mass and angular momentum of the $\nu$NS and the NS companion as well as the orbital period and eccentricity of the final binary system. For each of these values we give the relative errors with respect to the $1$~million particles simulation.}
  \label{tab:Nparticle}
\end{table*}
\begin{figure*}
  \centering
  \includegraphics[width=0.97\hsize,clip]{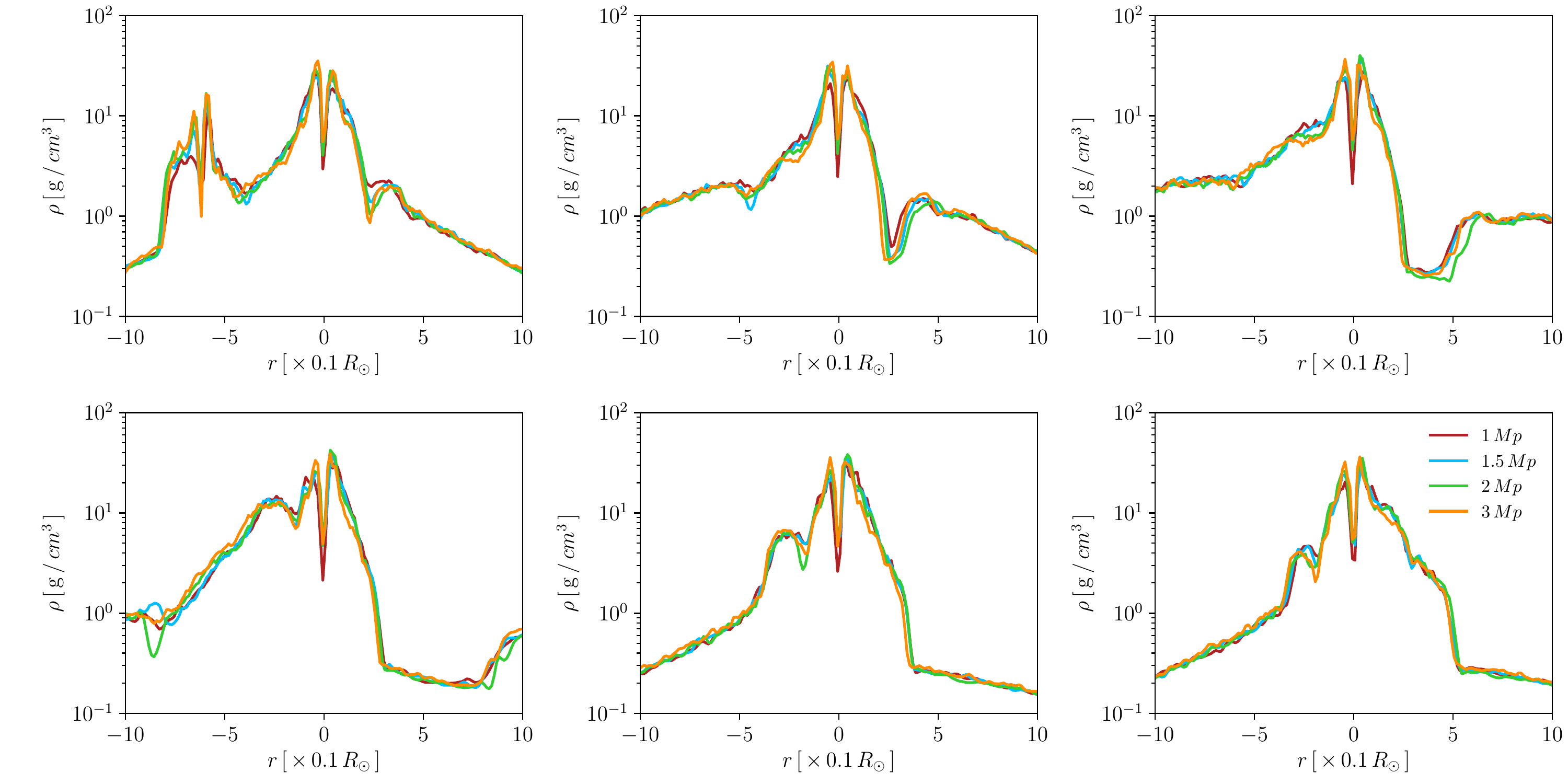}
  \caption{Density along different directions $\theta$ on the orbital binary system plane ($\nu$NS-NS). From left to right of upper panel: $\theta=0.0$, $\pi/6$ and $\pi/3$ and in the bottom panel $\theta=\pi/2$, $2\pi/3$ and $5\pi/6$. The center of the reference system is on the NS position and the $\nu$NS is on the $-x$ axis. The $\theta$ direction if measured from the $+x$ axis. The initial binary system is formed by the CO$_{\rm core}$ of the $M_{\rm zams}=25 \, M_\odot$ progenitor and a $2\, M_\odot$ NS  in an orbital period of about $5$~min. Different colors correspond to different number of particles: $1$~million (red line), $1.5$~million (blue line), $2$~million (green line) and $3$~million (orange line).}\label{fig:25Mzams_Nparticles_rho}
\end{figure*}
\begin{figure*}
  \centering
  \subfigure[$\nu$NS]{\includegraphics[width=0.485\hsize,clip]{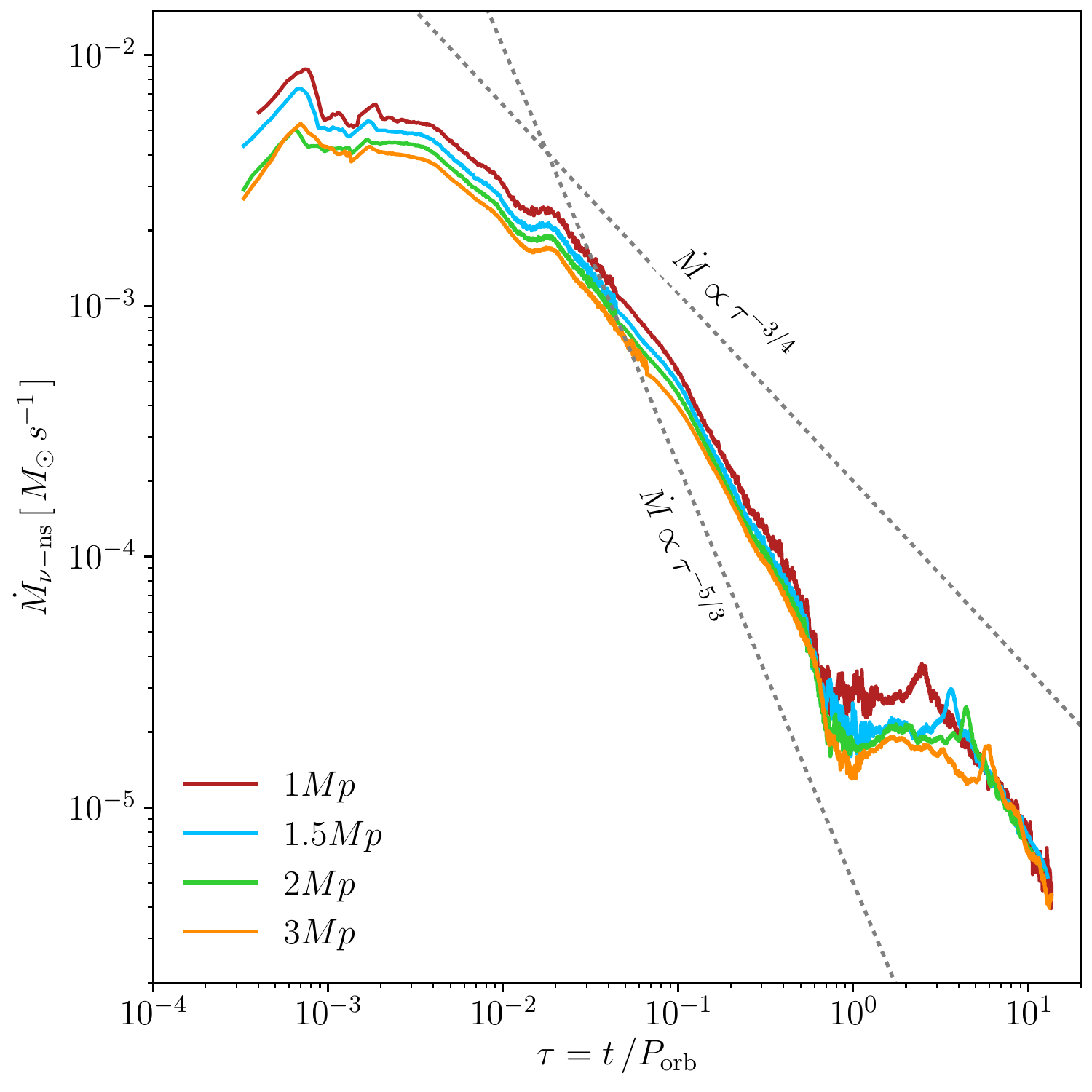}}
  \subfigure[NS]{\includegraphics[width=0.485\hsize,clip]{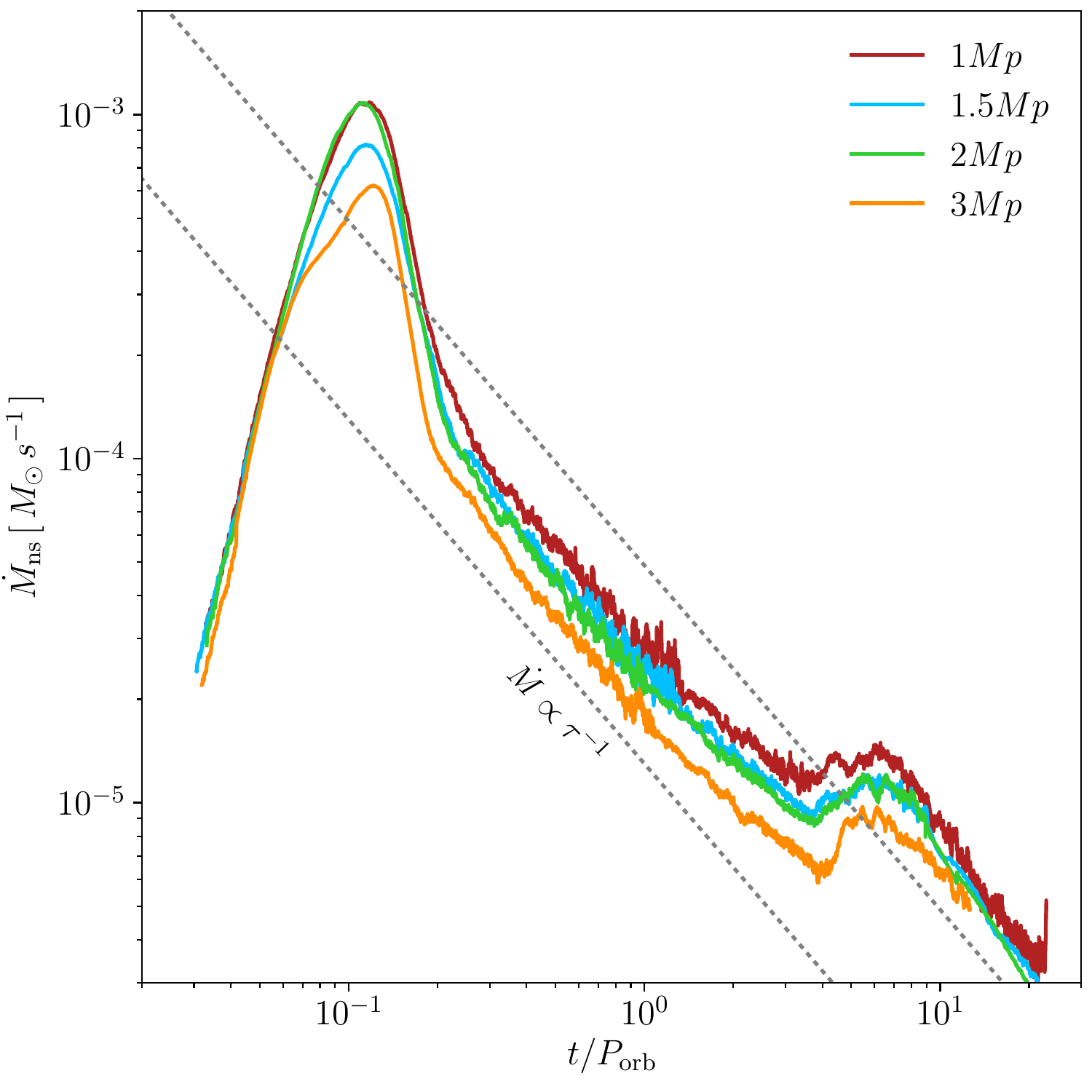}}
  \caption{Mass-accretion rate on the $\nu$NS (left panel) and the NS companion (right panel) for different number of particles modeling the SN expansion in the simulation. The initial binary period is the same as in Figure~\ref{fig:25Mzams_Nparticles_rho}.}\label{fig:25Mzams_NparticlesMrate}
\end{figure*}
\begin{figure*}
  \centering
  \subfigure[Mass flux]{\includegraphics[width=0.485\hsize,clip]{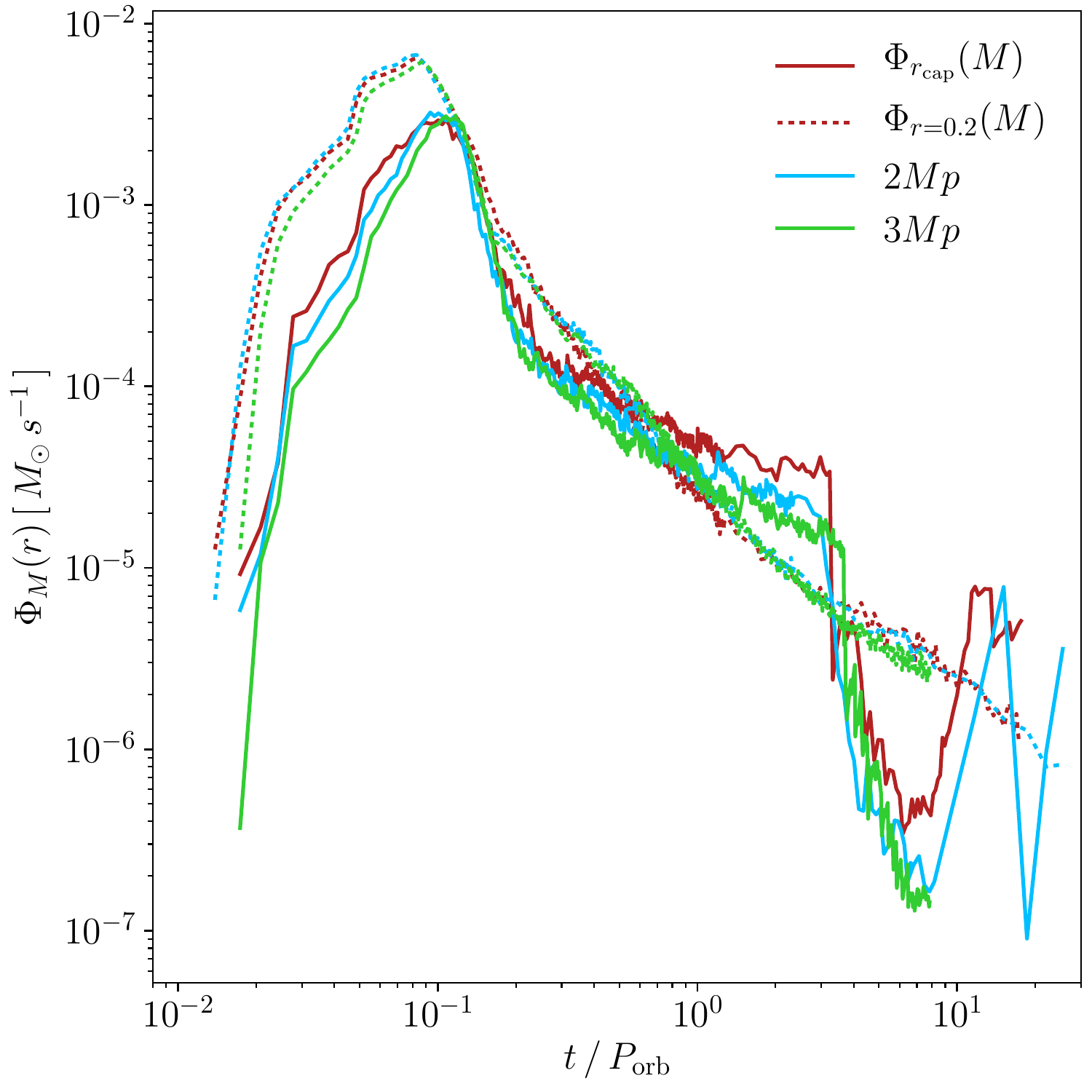}}
  \subfigure[Angular momentum flux]{\includegraphics[width=0.485\hsize,clip]{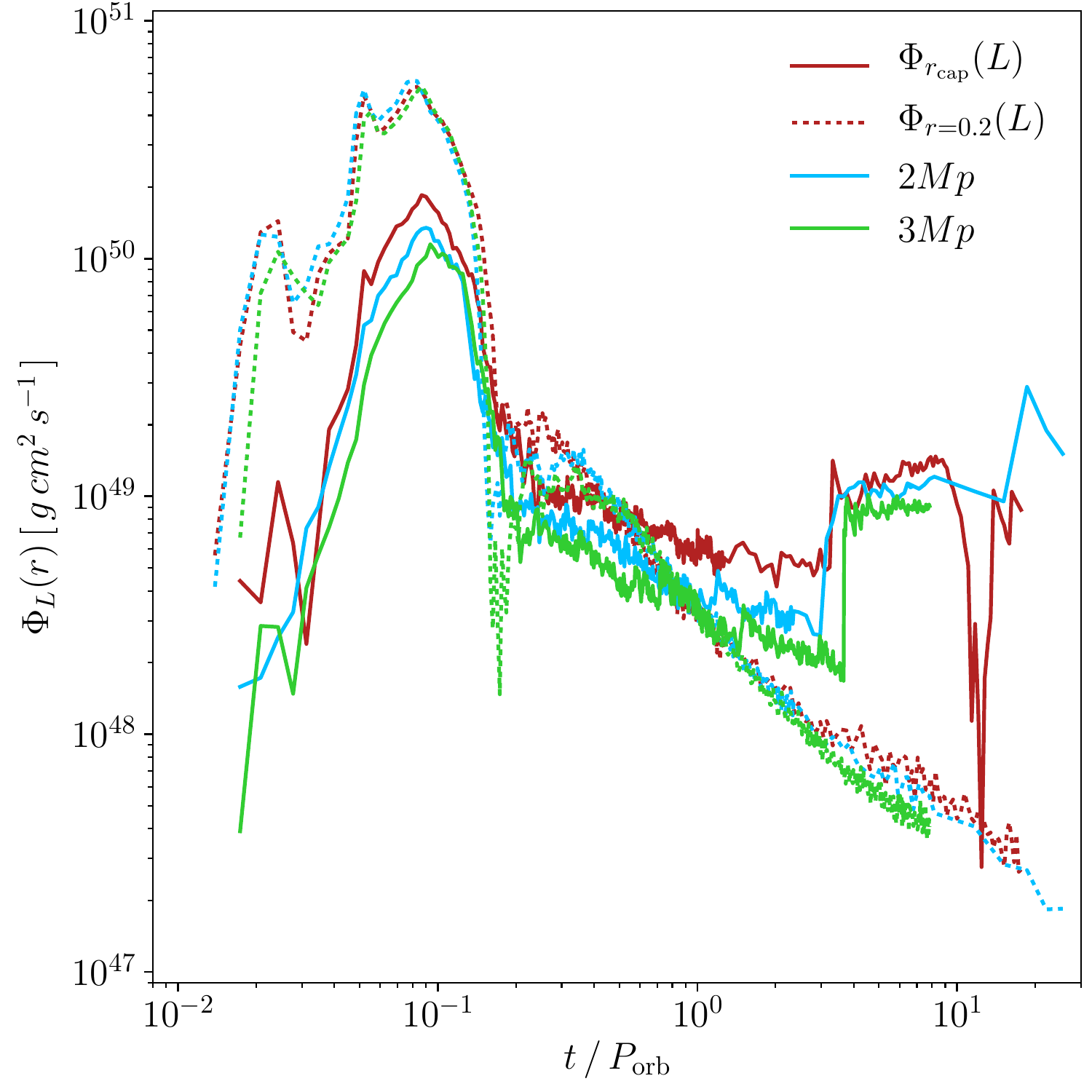}}
  \caption{Mass (left panel) and angular momentum flux (right panel)  thought spheres of radius $r=0.2\, R_\odot$ and $r=R_{\rm cap}$ with the NS in the center. $R_{\rm cap}$ surface is defined as the maximum capture radius between the particles accreted by the NS in each iteration. The initial binary period is the same as in Figure~\ref{fig:25Mzams_Nparticles_rho}.}
  \label{fig:25Mzams_NparticlesFluxes}
\end{figure*}

\end{document}